\documentclass{aa}

\usepackage{graphicx}

\usepackage{txfonts}
\usepackage{xcolor}
\begin{document} 

   \title{{\sc HR-pyPopStar II}: high spectral resolution evolutionary synthesis models low metallicity expansion and the properties of the stellar populations of dwarf galaxies}

\titlerunning{{\sc HR-pyPopStar II:} low metallicity evolutionary synthesis models}
\authorrunning{Mill{\'a}n-Irigoyen et al.}
   \author{I. Mill{\'a}n-Irigoyen
          \inst{1}
          ,
M. Moll{\'a}$^{2}$,
M. Cervi{\~{n}}o$^{3}$,
M.L. Garc{\'{i}}a-Vargas$^{4}$ 
          }

   \institute{ $^{1}$ Max-Planck Institut für Astrophysik, 85741, Garching, Germany \email{imillan@mpa-garching.mpg.de}\\
   $^{2}$ Departamento de Investigaci\'on B\'asica, CIEMAT, Av. Complutense 40, E-28040, Madrid, Spain\\
$^{3}$ Centro de Astrobiolog\'{i´}a (CSIC/INTA), ESAC Campus, Camino Bajo del Castillo s/n, E-28692 Villanueva de la Ca\~{n}ada, Spain\\
$^{4}$ FRACTAL S.L.N.E., C/ Tulip\'{a}n 2, p13, 1A, E-28231, Las Rozas de Madrid, Spain \\
             }

   \date{\today}

  \abstract
   { Low metallicity stellar populations are very abundant in the Universe, either as the remnants of the past history of the Milky Way or similar spiral galaxies, or the young low metallicity stellar populations that are being observed in the local dwarf galaxies or in the high-z objects with low metal content recently found with JWST.}
  { Our goal is to develop new high-spectral-resolution models tailored for low-metallicity environments and apply them to analyse stellar population data, particularly in cases where a significant portion of the stellar content exhibits low metallicity.}
   {We used the state-of-the-art stellar population synthesis code {\sc HR-pyPopStar} with available stellar libraries to create a new set of models focused on low metallicity stellar populations.}
{We have compared the new spectral energy distributions with the previous models of {\sc HR-pyPopStar} for solar metallicity. Once we verified that the spectra, except for the oldest ages that show some differences in the molecular bands of the TiO and G band, are similar, we reanalysed the high resolution data from the globular cluster M~15 by finding a better estimate of its age and metallicity. Finally, we analysed a subsample of mostly star-forming dwarf galaxies from the MaNGA survey we found similar stellar mass-mean stellar metallicity weighted by light to other studies that studied star forming dwarf galaxies and slightly higher mean stellar metallicity than the other works that analysed all types of dwarf galaxies at the same time, but are within error bars.
   }
{}
\maketitle

\section{Introduction}

The spectrophotometric properties of galaxies have been used for decades to study their formation and evolution processes. The analysis of the spectra and its interpretation are usually done using evolutionary synthesis models that create the Spectral Energy Distribution (SED) and the associated data of the so-called Simple Stellar Populations (SSPs). A SSP is a group of stars that formed simultaneously from the same cloud with a fixed Initial Mass Function (IMF), thus, all the stars of a SSP have the same age and initial chemical composition. The SED of a SSP is computed as the sum of all spectra of its individual constituent stars. The synthesis models also contain supplementary information, such as the magnitudes in different spectral bands and their corresponding colours or stellar spectral absorption line indices obtained from those synthetic spectra.

In this way, the observed spectra of a galaxy/galactic region can be decomposed as the spectra of their building blocks, i.e. the SSPs. This technique allows us to obtain information about the studied galaxy/galactic region, e.g. star formation history, metallicity enrichment history, formed stellar mass, current stellar mass, mean stellar metallicity and mean stellar age.
This method of analysis for the stellar spectrum of a galaxy/galactic region is an inverse technique: the decomposition of the spectrum is obtained through a sum of SSPs by searching for the best combination of these that minimizes the residuals. This is used in codes as, for example, {\sc starlight} \citep{cid05}, {\sc vespa} \citep{tojeiro07}, {\sc pipe3d} \citep{Sanchez+16a,Sanchez+16b},  {\sc ppxf} \citep{Cappellari_Emsellem04,Cappellari17}, {\sc firefly} \citep{Wilkinson+2017} or {\sc fado} \citep{gomes18}. 

However, SSP spectra can also be added directly following the results of a Chemical Evolution Model (CEM) applied to the study region, as in \citet{Mancone_Gonzalez2012, Molla14, Millan-Irigoyen20}. The direct method uses the star formation and metallicity enrichment histories predicted by the CEM, which are equivalent to the mass contributions of SSPs of each age and metallicity, necessary to calculate the final theoretical SED of the galaxy/galactic region. 

Irrespective of the analysis method, the spectra of the SSPs used in the analysis need to satisfy some requirements such as a spectral resolution similar to the observed spectrum of the object we want to analyse and a good completeness in ages and metallicities of the possible stellar populations of the object. One of the most important requirements is the spectral resolution of the synthetic spectra of the SSPs, which must be at least similar to the spectral resolution of the observed spectra in order to avoid loss of information in the process of the analysis. 

\begin{table*}
  \begin{center}
	\caption{Summary of some intermediate and high resolution SSP models of the literature, compared to the present work.}
\label{Table:1}
\begin{tabular}{clccrc}
\hline
Reference & Code &  Range (\AA) & $\delta\lambda$\,(\AA) & $ R_{\rm th}$ & Z \\
\hline
 LEB04 & {\sc Pegase-HR} & $4000 \-- 6800$ & 0.55 & 9091 & $0.0002\--0.05$ \\
 GON05 & Sed@      & $3000 \-- 7000$ & 0.30 & 16667 & $0.002\--0.05$ \\
 PER09 & {\sc BASTI} iso.     &$2500 \-- 10500$ & 1.00 & 5000 & $0.0001\--0.04$ \\
 M\&S11 & Maraston & $1000 \-- 25000$ & 0.25 & 20000 & $0.0001\--0.04$ \\
 VAZ16 & {\sc miles}    & $1680 \-- 500000$ & 0.90 & 5556 & $0.0001\-- 0.04$ \\
 CON18 & {\sc fsps}     & $3700 \-- 24000$ & 1.67 & 3000 & $0.00044\--0.028$\\
 CBC20 & {\sc galaxev}  & $3540 \-- 7410 $ & 0.90 & 3000 & $0.0002\--0.030 $\\
Ma20 & Fuel-consumption code& $ 3600-10300$ & 0.8-2.4 & 1800 & $ 0.00007\--0.03$\\ 
XSL & & $3500-24000$ &  0.12-0.8 & 10000 & $ 0.0001-0.03$\\
 MI21 & {\sc pyPopStar} & 91 \-- 24000 & 0.10 & 50000 & $0.004\--0.05$\\
 This work & {\sc pyPopStar} & 2500 \-- 10500 & 0.10 & 50000 & $0.0001\--0.05$\\
\hline
\end{tabular}
\end{center}
\footnotesize{References. LEB04: \citet{LeBorgne+2004};  GON05: \citet{Gonzalez+2005}; PER09: \citet{Percival+2009}; M\&S11: \citet{Maraston_Stromback2011};
VAZ16: \citet{vaz16}; CON18: \citet{conroy18}; CBC20: \citet{Coelho+2020}; Ma20: \citet{Maraston+2020}; MI21: \citet{Millan-Irigoyen+21}; XSL: \citet{Verro+2022}. }
\end{table*}

If the synthetic spectra have a lower resolution than the data, it would be necessary to degrade the observed spectra \citep[see][for a detailed explanation of the different definitions of spectral resolution]{Millan-Irigoyen+21}. Thus, given the resolutions of state-of-the-art and future instruments that are partially or completely devoted to extragalactic studies in very large telescopes, such as MEGARA (GTC), WEAVE (WHT), VLT/MOONS (VLT), or MOSAIC (ELT), it is necessary to have a base of SSPs with a similar spectral resolving power of the instruments ($R$ in the range $18\,000 - 20\,000$).

Another important requirement for the synthesis models is a wide coverage in both ages and metallicities to consider all the possible environments. The majority of galaxies in the Universe have star-formation histories with stellar populations with very different ages and metallicities coexisting.

There are a large number of works devoted to the synthesis of SSPs. Some of them are low-resolution models \citep{CER1994, Fioc_Rocca_Volmerange1997, Kodama_Arimoto1997, lei99, bc03,Maraston2005,fritze06,elst09, Conroy_Gunn_White2009, Maraston+2009, Fioc_Rocca_Volmerange2019}, some of them use empirical libraries \citep{Conroy_VanDokkum2012, vazdekis2015, vaz16,Maraston+2020}. Some of the well-known medium to high resolution synthesis models are those of \citet{LeBorgne+2004, Gonzalez+2005, Coelho+2007, Percival+2009, Maraston_Stromback2011, Conroy_VanDokkum2012, lei14, vaz16, conroy18, Coelho+2020}. 

In Table~\ref{Table:1}, we summarize the current evolutionary synthesis models from the literature with medium or high spectral resolution (considering only those with $R_{th} \ge 1800$). The columns indicate: (1) the reference of each model, (2) the synthesis code used to compute the SED, (3) the wavelength range, (4) the wavelength step $\delta\lambda$ at $\lambda = 5000$\,\AA, (5) the value $R_{\mathrm{th}}=\lambda/\delta\lambda$, as defined for theoretical models, at that wavelength, and (6) metallicity range of each model.

{\sc PopStar} models \citep[][hereinafter MOL09, MAN10, GV13, respectively]{molla09, mman10,GarciaVargas_Molla_MartinManjon2013}, based on previous works by \citet{garcia-vargas94,garcia-vargas95, garcia-vargas98}, were specially computed to study young stellar clusters in regions of star formation, such as the {\sc STB99} models \citep{lei99}. However, PopStar models also included old stellar populations (age $> 2.0$\,Gyr) in their model set. Starting from this base, we have recently developed in \citet[][ hereinafter MI21]{Millan-Irigoyen+21} an updated python version of {\sc PopStar} models called {\sc HR-pyPopStar} by producing high-resolution SSP models ( $R_{\textrm{th,5000}}= 50\,000$) for 106 ages in the range $\log{\tau}= [5.0,10.18]$ and 4 metallicities $Z=0.004$, 0.008, 0.02 and 0.05.

Low-metallicity stellar populations constitute a significant part of the stellar content of local dwarf galaxies given their relatively inefficient SFH. Given the amount of dwarf galaxies found in huge-volume surveys such as Mapping Nearby Galaxies at APO \citep[MaNGA,][]{Bundy+2015} or Dark Energy Spectroscopic Instrument (DESI) \citep{DESI2025}, low-metallicity models that include young stellar populations are essential to properly study the stellar content of these galaxies. Thus, the development of low-metallicity high-wavelength-resolution models, which also encompasses the maximum possible range of ages, is important to study correctly this kind of system.

The main aim of the present work is to update our {\sc HR-pyPopStar} models by now including the library of stellar models from \citet{mun05}, the new PoWR O and B stellar library and the PHOENIX stellar library for cool stars. This allows us to make SSP models with lower metallicities than the set of models of the MI21 version, by including two more sets for $Z=0.0001$ and $Z=0.0004$, while maintaining the same spectral resolution as in those previous models.

The paper is structured as follows: we describe the model in Section ~\ref{Sec:Model}, mainly introducing the stellar libraries used in this work and explaining the differences from our previous work. The resulting SEDs, the derived magnitudes and D4000s, the comparison with our previous models, and the comparison with other authors are described in Section~\ref{Section:Results}. In Section ~\ref{sec:Low_z_analysis}, we analyse the spectra of the low metallicity globular cluster M~15, and a survey of dwarf galaxies from the MaNGA sample, and we compared them with the results of different authors. 
Our conclusions are recapitulated in Section~\ref{Sec:Conclusions}.

\section{Model Description}
\label{Sec:Model}

The main ingredients of any population synthesis code are the isochrones, which trace the stellar parameters of each star present in the SSP at a given age, the spectral library, or collection of the spectra for each type of star, and the IMF. 

We used here the same set of isochrones from the Padova group as in MI21, \citep{Bressan_Bertelli_Chiosi1993, Fagotto+1994a, Fagotto+1994b, Girardi+1996}. However, instead of only using 4 metallicities, as in MI21, $Z = 0.004$, 0.008, 0.02 and 0.05, we added the isochrones of the two lowest metallicities $Z =  0.0001$, and 0.0004.

\begin{table}
\small
\label{Table:2}
\caption{Summary of the stellar libraries used in this work for each isochrone of our models.}
\begin{center}
\begin{tabular}{cccccc}
\hline
Isochrone & WR & O $\&$ B & Mun05 & \textsc{phoenix} & Rauch03\\
\hline
0.0001 & subSMC & subSMC & 0.0001 & 0.0001 &  0.002\\
0.0004 & subSMC & subSMC & 0.0004 & 0.0004 &  0.002\\
0.004 & SMC & SMC & 0.004 & 0.004 &  0.002\\
0.008 & LMC & LMC & 0.008 & 0.008 &  0.02\\
0.02 & solar & solar & 0.02 & 0.02 &  0.02\\
0.05 & solar & solar & 0.05 & 0.05 & 0.02 \\
\hline 
\end{tabular}

\end{center}
\end{table}

\begin{table}
\caption{Summary of the differences in $T_{\rm eff}$ coverage and number of stars in the new and old models of O and B stars from PoWR group.}
\label{Table:3}
\normalsize
\begin{center}
\begin{tabular}{ccc}
\hline
Z & $\rm N_{\star}$ Old PoWR & $\rm N_{\star}$ New PoWR \\
\hline
solar & 200  & 174\\
LMC & 191 & 200 \\
SMC & 192 & 240 \\
\end{tabular}

\end{center}
\end{table}

\subsection{Stellar libraries}\label{subsec:Stellar_libraries}

Regarding the stellar libraries, we have the following stellar atmosphere classification: 1) Hot stars (O and B) with $T_{\rm eff} > 25000$\,K; 2) Wolf-Rayet (WR) stars, defined using the surface abundance in the stellar atmosphere of C, N, O and H; 3) "cool" stars from A to G spectral types with $T_{\rm eff} \le 25000$\,K; 4) post- Asymptotic Giant Branch (AGB)/Planetary Nebulae (PNe) stars; and 5) cool K and M stars.

The stellar libraries used in the population synthesis models and the metallicity range that each one has are summarized in Table \ref{Table:2}. We explain each library and the change with respect to MI21. 

In the case of WR and post-AGB/PNe stars, we used the same libraries as in MI21, which cover the metallicity range from 0.002 to 0.05. 
\vspace{0.3cm}

\subsubsection{O and B type stars}
For the O- and B-type stars, we used the new O- and B-stellar atmosphere models from PoWR. They computed the stellar atmospheres using the mass-loss recipe by \citet{Vink+2001} divided by three, a mass-loss recipe that is more in agreement with current estimates of the stellar mass-loss. They created models for 4 metallicities: solar, LMC, SMC and sub-SMC ($1/31 Z_{\odot}$). We checked the spectra of all the models individually and removed the few ones that have non-physical emission line profiles that seem to be errors in the computation of the models and could affect the emission lines of the synthesized spectra.
The comparison of the number of stars in each metallicity range for old and new versions of the models is described in Table~\ref{Table:3}.

\subsubsection{Spectral type A to G type stars}\label{subsubsec:}

For A to G type stars, we substituted the stellar library from \citet[][hereinafter C14]{Coelho2014} used in MI21 by the stellar models from \citet[][hereinafter MUN05]{mun05}. These authors gave a set of synthetic spectra based on the code of \citet{Kurucz2013} that covers the wavelength range from 2500 to 10\,500\,\AA. These models are available for different combinations of stellar parameters, in particular effective temperature $T_\mathrm{eff}$ in the range \mbox{$3500\le T_{\mathrm{eff}} \le 47\,500$\,K}, with steps of \mbox{250\,K} up to \mbox{10\,000\,K} , with progressively larger steps in $ T_{\rm eff}$ for hotter stars, gravity in the range \mbox{$0.0\le \log{g}\le 5.0$}, metallicity in the interval \mbox{$-2.5\le[\mathrm{M/H}]\le 0.5$}, $\alpha$-enhancement abundance from \mbox{$[\alpha/\mathrm{Fe}]=0.0$ and $+0.4$}, three values of micro-turbulence velocity \mbox{$\xi=1$, 2, and 4\,km\,s$^{-1}$}, and stellar rotation in the range \mbox{$0\le V_{\mathrm{rot}}\le 500$\,km\,s$^{-1}$}. We have selected models with \mbox{$\mathrm [\alpha/\mathrm{Fe}]=0.0$},\, \mbox{$\xi=2$\,km\,s$^{-1}$} and \mbox{$V_{\mathrm{rot}}=0$\,km\,s$^{-1}$}. Given that Kurucz models are not adequate to model hot stars where the approximations considered in them do not hold, we are going to include only the stars of this library MUN05 with \mbox{$T_{\mathrm{eff}} \le 25\,000$\,K}. These models were originally computed at several spectral resolving powers $\mathrm{R} =$ 20\,000, 11\,500 (to simulate Gaia \citep{Gaia18}), 8500 (to simulate RAVE), 2000 (to simulate SDSS/SLOAN), and an uniform dispersion of 1 and~\mbox{10\,\AA\,pix$^{-1}$}. For this particular work, we used the models with $R=20\,000$. 

{\begin{figure}
\hspace{-0.7cm}
\begin{center}
\includegraphics[width=0.475\textwidth,angle=0]{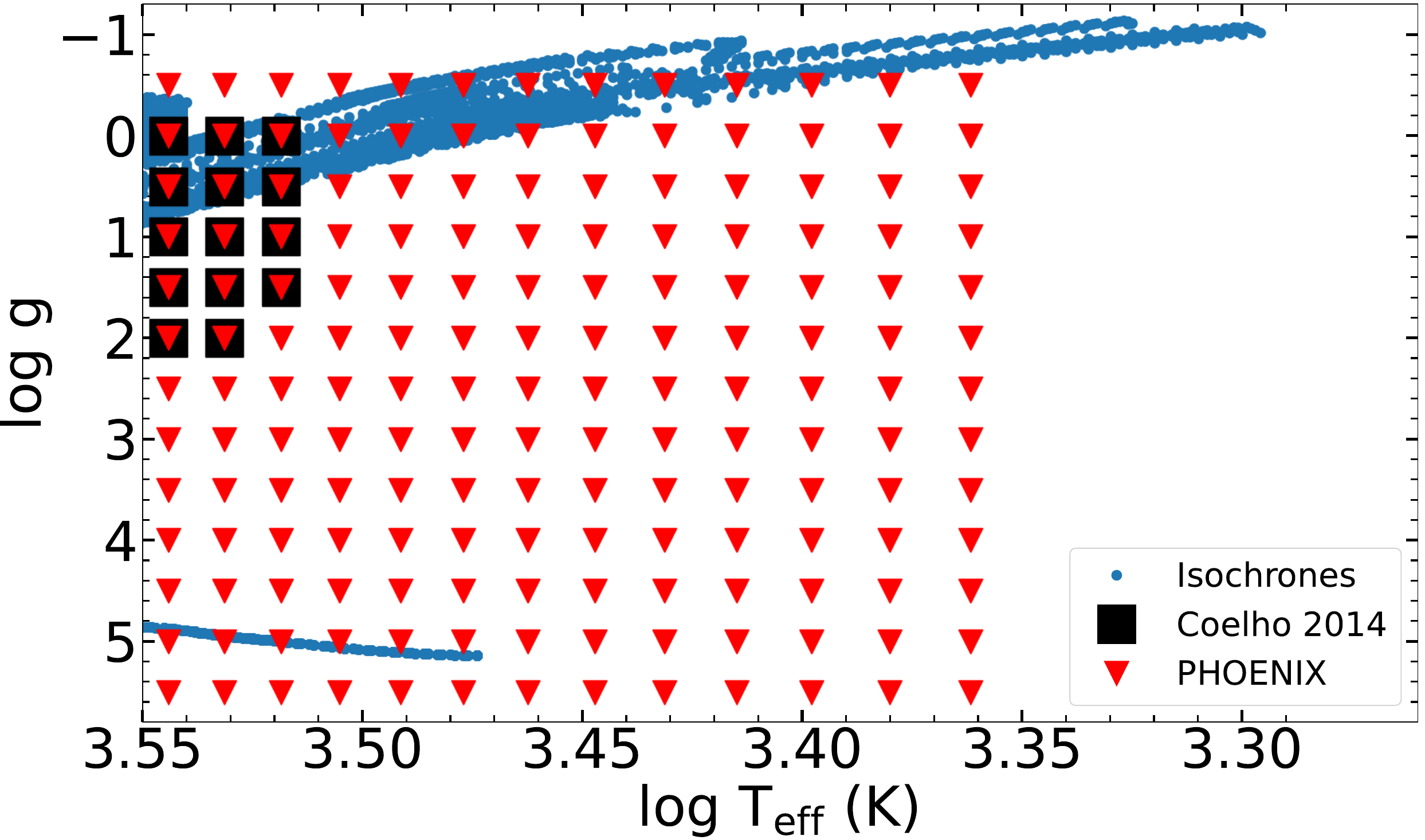}
\caption{Comparison of the coverage of the stellar libraries of C14, black squares, and PHOENIX, red triangles, for the low temperature stars, $T_{\rm eff}<4000 \,{\rm K}$. Blue points represent the isochrones for all ages and solar metallicity.}
\label{fig:Coverage}
\end{center}
\end{figure}
}

\subsubsection{K and M type stars}\label{subsubsec:cool_stars}

We have included models for cool stars, spectral types K and M, using the PHOENIX atmospheres models \citep{Husser+2013} not used in MI21. These models simulate the 1-D stellar atmosphere instead of using plane-parallel geometry. Moreover, they have an up-to-date list of atomic and molecular lines and a new equation of state. This allows to model more accurately cool giant stars, especially in the red and NIR parts of the spectra, where molecular bands are prominent. The PHOENIX library has a wider coverage of $T_{\rm eff}$ and $\log g$ on the HR diagram than the C14 library at low temperature. The comparison of coverage between C14 at low temperature and PHOENIX libraries with respect to the isochrones used in this work is shown in Fig. ~\ref{fig:Coverage}. 

Since PHOENIX model atmospheres are computed with 1D spherical geometry, they have one more degree of freedom than plane-parallel models and require the stellar mass to determine the effective radius. Thus, for each model of the PHOENIX library that have $T_{\rm eff}<4000$\,K, we obtain the stellar radius from the header of the FITS file.
Then, we computed the bolometric luminosity by integrating the spectra across its full wavelength range, $500 - 55,000$\, \AA, and this computed stellar radius. We compared the luminosity we obtain with the bolometric luminosity also provided by the FITS header of each stellar atmosphere model. 

We found that for giant stars ($\log{g} < 2.0$), the luminosities obtained this way were significantly lower than the bolometric luminosities of each model, in some cases accounting for just half of the bolometric luminosity. These discrepancies exceed the expected differences between the bolometric luminosity and the flux integrated over UV, optical and NIR bands, and they affect the fluxes of cool giant stars. As we use plane-parallel models for stars with $T_{\rm eff} < 25,000\,{\rm K}$ and want to preserve the luminosity of each star, we normalize the flux of giant stars, to make their integrated luminosity in the whole range $500-55000$\,\AA\ match the bolometric luminosity of its model.

\subsection{Initial mass function}

In the case of the IMF, we have used the same four ones as in MI21: \citet{Salpeter1955}, \citet{Ferrini_Penco_Palla_1990}, \citet[][with a slope $\alpha=-2.7$ for massive stars]{Kroupa2002}, and \citet{Chabrier2003}, hereinafter referred to as SAL, FER, KRO and CHA, respectively. In the following, we will show the results using the CHA IMF which is similar to the KRO IMF with $\alpha = 2.3$.
\\

\subsection{ Stellar Population synthesis}

The spectrum of a SSP along an isochrone of a given age, $\tau_i$, and metallicity, $Z$, is computed with the following equation:
\begin{equation}
    F_{\lambda, \mathrm{SSP}}(Z,\tau_i) = \sum_{j=1}^{\mathrm{N_{tot\,lib}}(Z)} w_j(Z,\tau_i) \times  {F_{\lambda,\,j}^\mathrm{lib}}(Z), 
\end{equation}
where $\mathrm{N_{tot\,lib}}(Z)$ is the number of elements in the stellar library with a given metallicity $Z$, $F_{\lambda,\,j}^\mathrm{lib}(Z)$ is the monochromatic flux of the element $j$ of the library with metallicity $Z$, and $w_{j}(Z,\tau_i)$ is the weight of each library element $j$ for each point $i$ of the isochrone of the corresponding metallicity $Z$ and age.

The weight of each point $j$ of the isochrone has a contribution given by the IMF and another one related to the correction of the luminosities of each point. The contribution of the IMF in follows the equation:

\begin{equation}
w_\mathrm{IMF}(m_j, Z, \tau_i) =
\int_{m_{\mathrm{low},j}(Z, \tau_i)}^{m_{\mathrm{up},j}(Z,\tau_i)}\ \phi(m)dm,
\end{equation}
\noindent where $m_{\mathrm{low},j}(Z,\tau_i)$ and $m_{\mathrm{up},j}(Z,\tau_i)$ define the mass intervals along the isochrone that share the same element $j$ in the stellar library.

Furthermore, we have to correct the weights for each point of the isochrone $j$ using the ratio between the bolometric luminosity of the isochrone, $L_{\mathrm{iso},j}$, and the integrated luminosity of the stellar library model $i$. Thus, the total weight $w_{j}(Z,\tau_i)$ is:

\begin{equation}
    w_j(Z,\tau_i) = w_\mathrm{IMF}(m_j, Z, \tau_i)\ 
     \frac{L_\mathrm{iso}(m_i,Z,\tau_i)}
    {L_{j}^{\mathrm{lib}}(Z)},
    \label{eq:}
\end{equation}
where $L_{j}^{\mathrm{lib}}(Z)$ is the integral over wavelength of ${F_{\lambda,\,j}^\mathrm{lib}}(Z)$ in the stellar library, which can be pre-computed or obtained from the parameters  of the library. For plane-parallel non-expanding atmosphere models such as those from MUN05 or \cite{Rauch2003}, ${F_{\lambda,\,j}^\mathrm{lib}}(Z)$ is equivalent to the flux at the surface per surface unit $\mathrm{cm}^2$, so we can calculate directly the luminosity for a given $T_\mathrm{eff}$ as:
\begin{equation}
L_\mathrm{bol}/R^2 = \sigma_\mathrm{SB} T_\mathrm{eff}^4, 
\end{equation}

\noindent being $\sigma_\mathrm{SB}$ the Stefan-Boltzmann constant.
This method is adequate for the MUN05 stellar library, whose spectral range from 2500 to 10500\,\AA, is too short to obtain the bolometric luminosity by integrating their spectra.

We have used the same assignment of an element $j$ to each element $i$ of the isochrone that was used in MI21. For groups 1), 3) and 4) we find the closest stellar model in $T_{\rm eff}$ and $\log{g}$; For WR stars we use the mass loss and the radius at a given optical depth (see details in MI21). 
The theoretical stellar libraries used in this work, see \ref{subsec:Stellar_libraries}, have dense enough binning in the HR-diagram to not produce problems in the synthesis. Regarding the old stellar population where cold stars dominate, the PHOENIX library coverage for the range 2300-4000\,K in the $T_{\rm eff}$-$\log(g)$ diagram is given by steps of 100\,K for $T_{\rm eff}$, and of 0.5 for $\log{g}$.

\begin{figure}
\hspace{-0.5cm}
\begin{center}
\includegraphics[width=0.48\textwidth,angle=0]{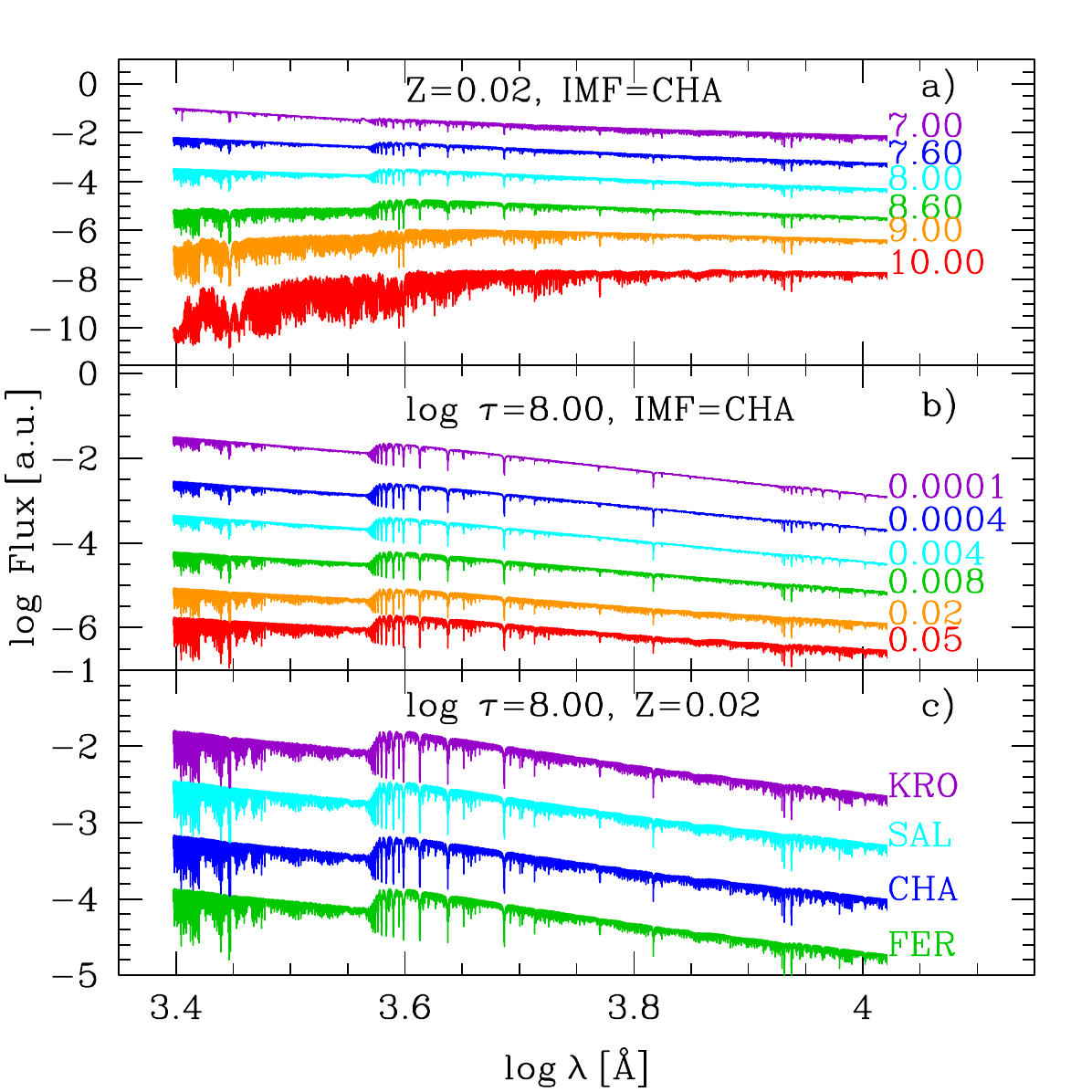}
\caption{SEDs obtained with this new version of the code {\sc HR-pyPopStar}. We show a comparison among SEDs with a) different ages ($\tau$) for the same metallicity $Z = 0.02$ and CHA IMF; b) different metallicities for a given age log $\tau$ = 8.00 and CHA IMF; and c) different IMFs for a given age log $\tau$ = 8.00 and metallicity $Z = 0.02$.}
\label{fig:SED-MUN}
\end{center}
\end{figure}

\begin{figure*}
    \centering
\includegraphics[width=0.48\textwidth]{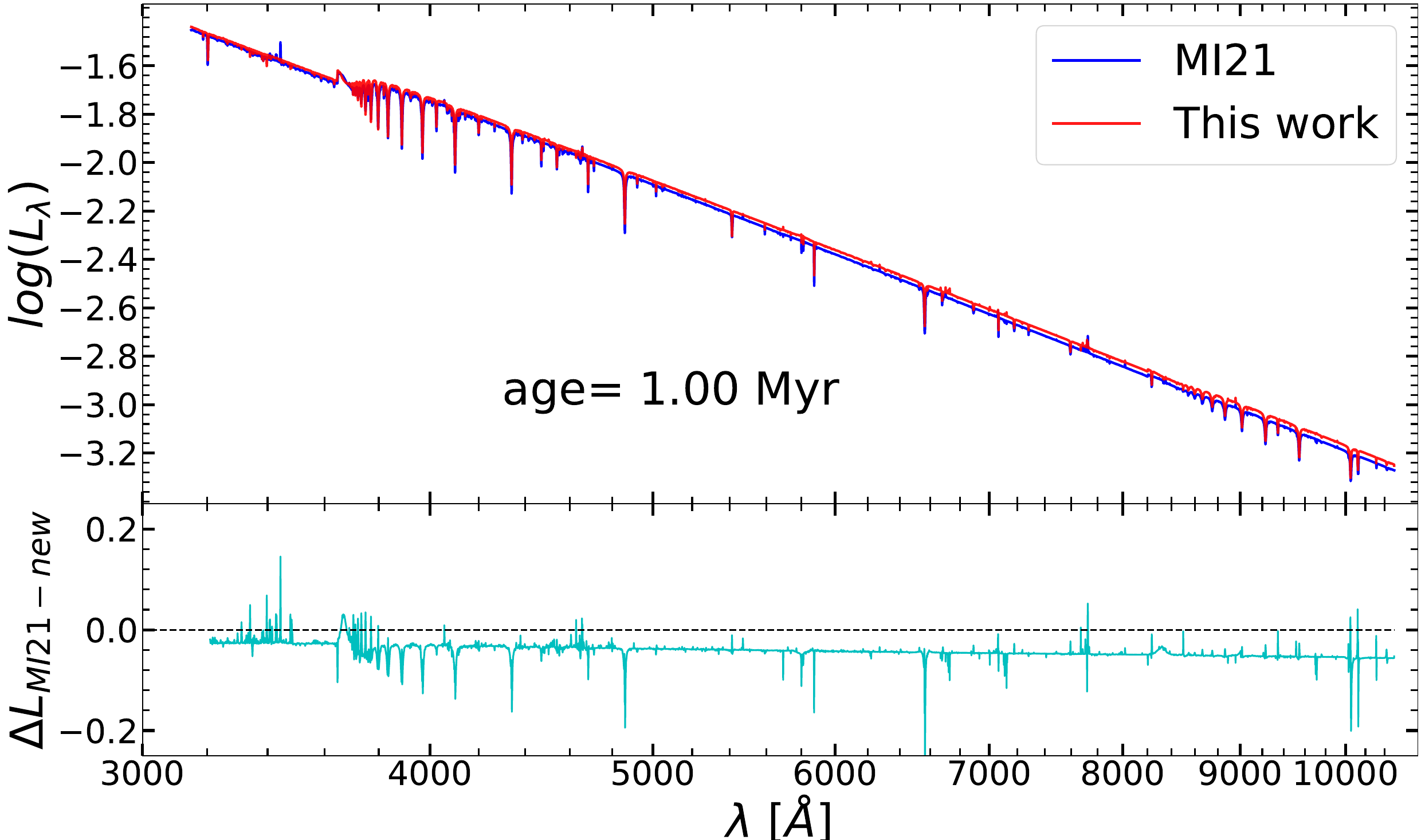}
\includegraphics[width=0.48\textwidth]{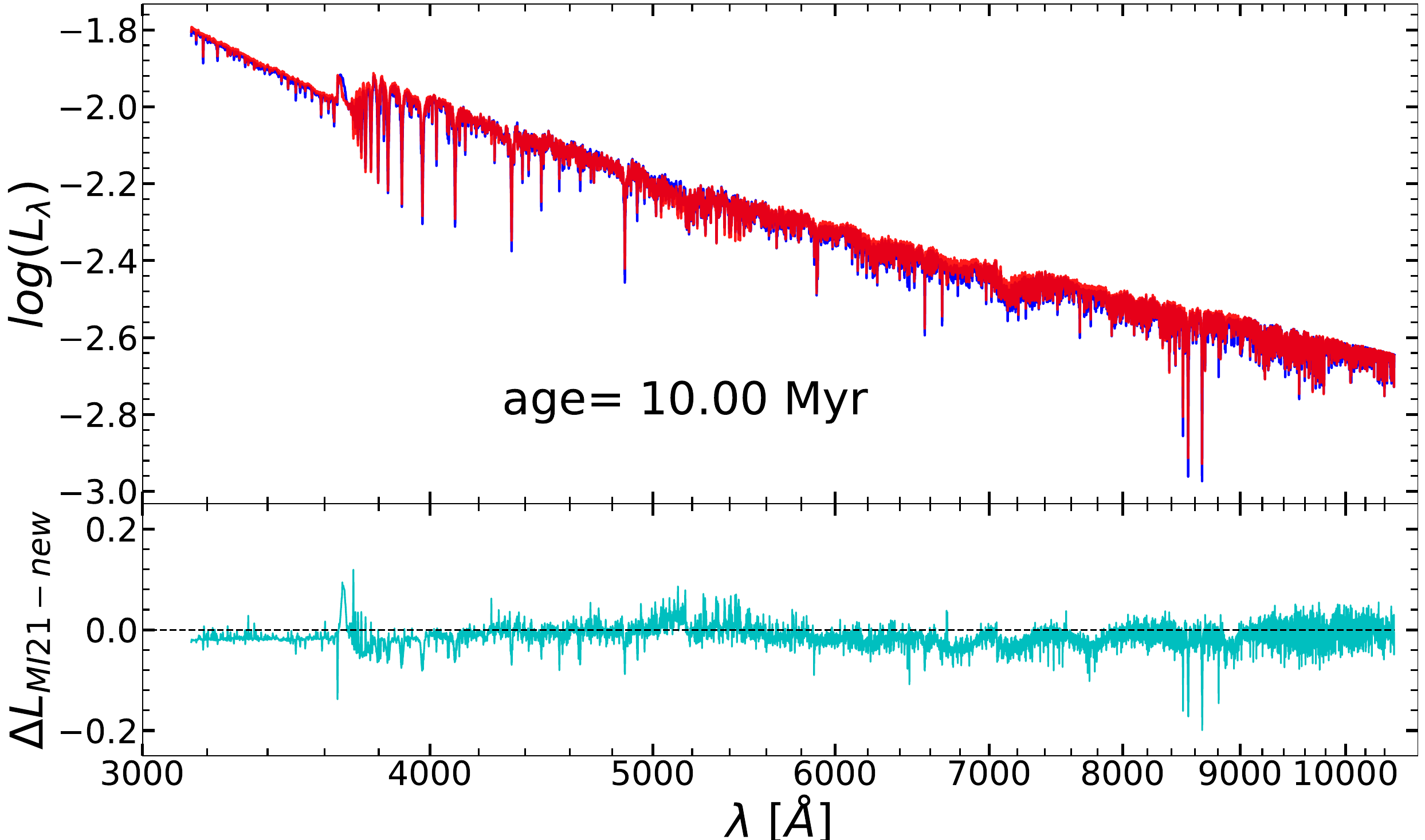}
\includegraphics[width=0.48\textwidth]{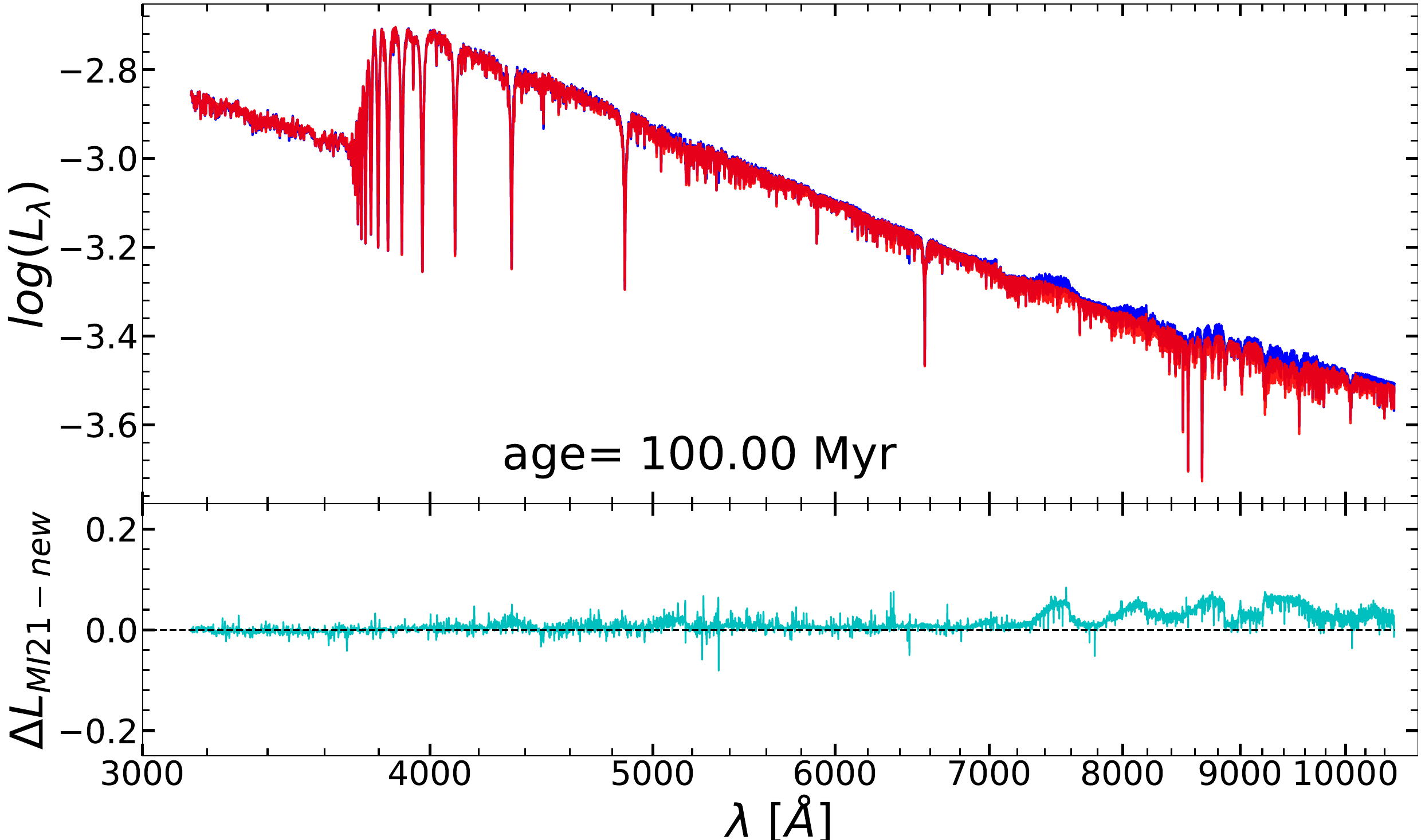}
\includegraphics[width=0.48\textwidth]{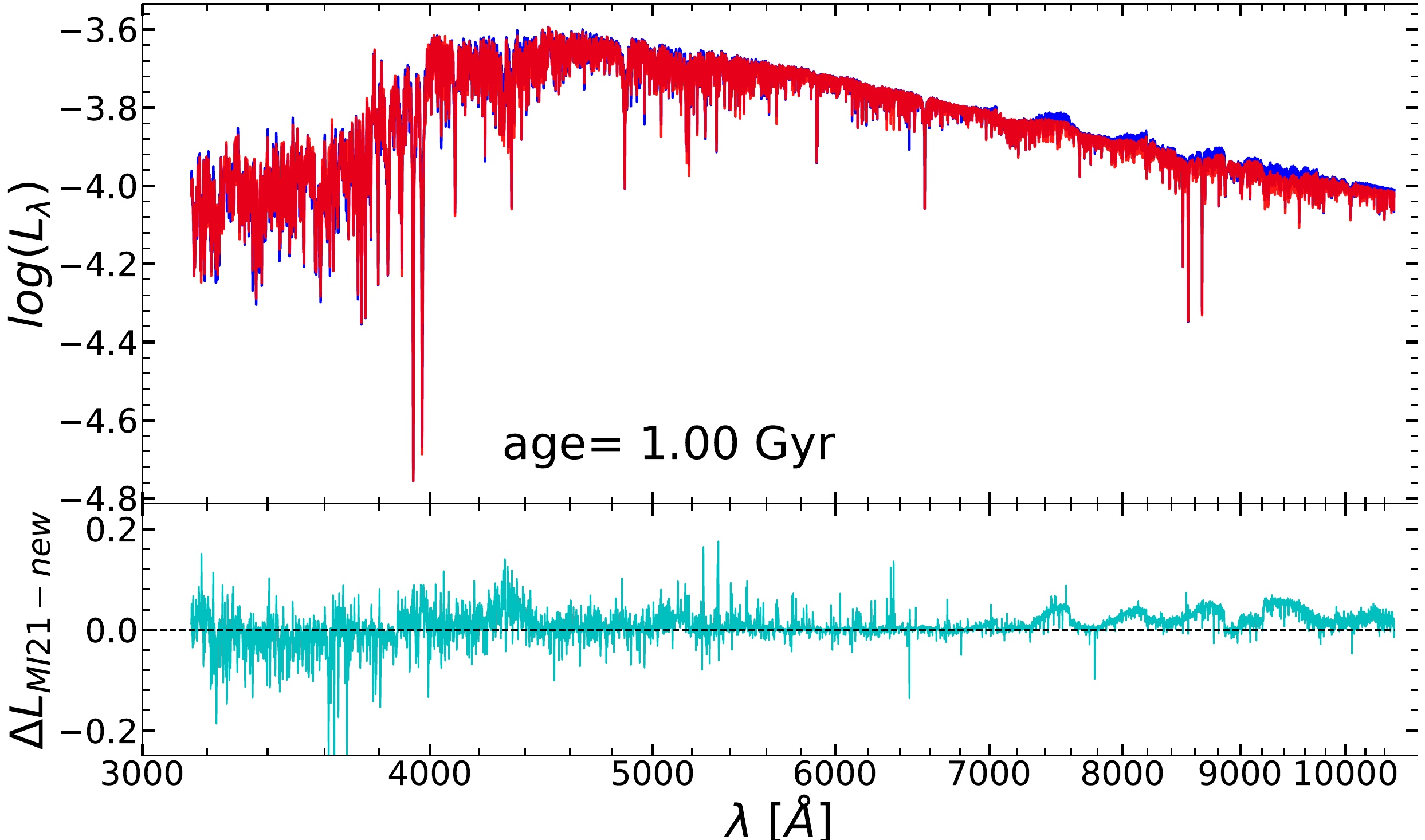}
\includegraphics[width=0.48\textwidth]{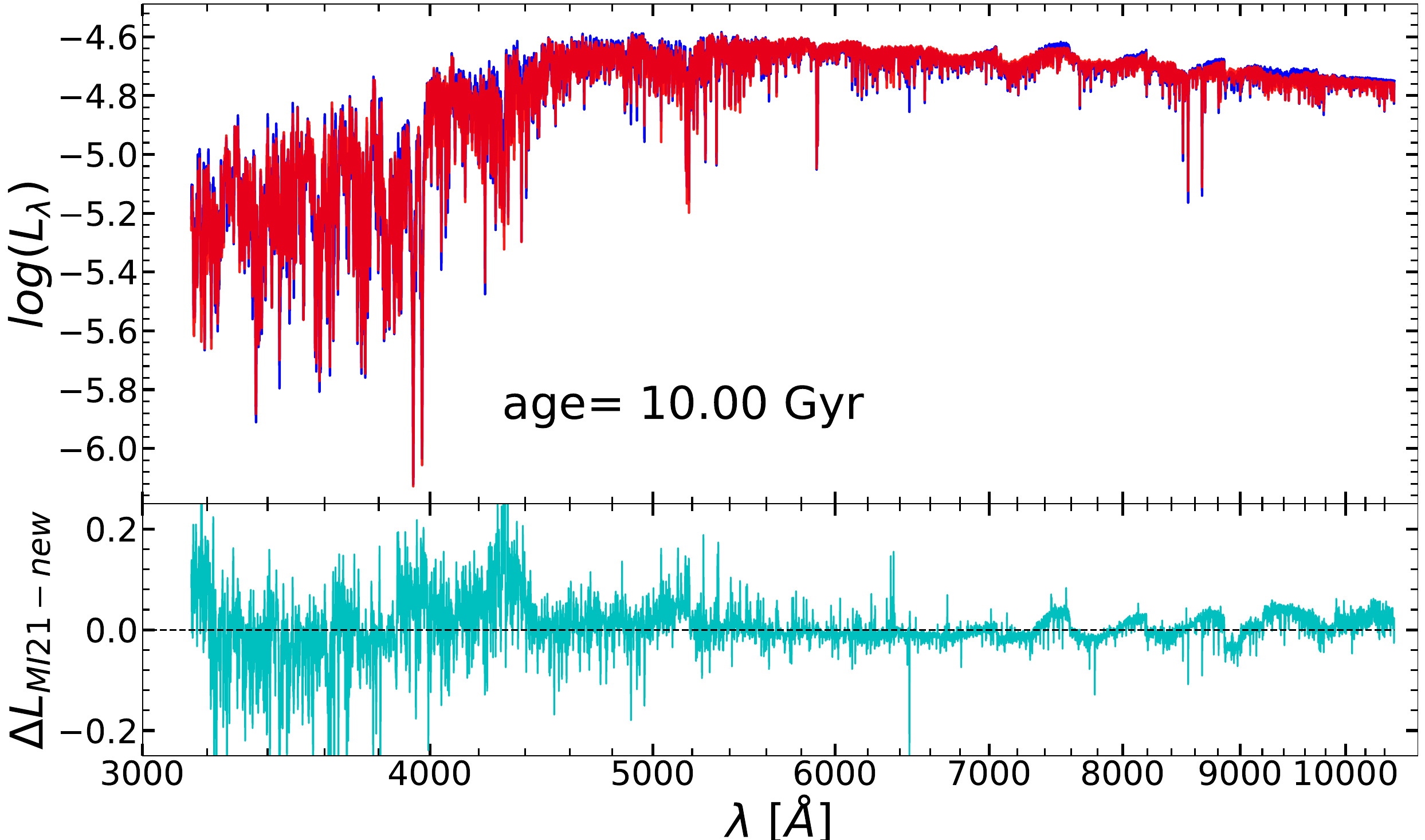}
\caption{Comparison between the old version of {\sc HR-pyPopStar} using the stellar library of C14 (MI21), in blue line, and the present models using MUN05 +PHOENIX stellar library, in red line, for $Z=0.02$ and CHA IMF. The bottom part of every subfigure shows the residuals $\Delta L_{MI21-new}= \frac{L_{MI21}-L_{new}}{L_{MI21}}$.
}
\label{fig:Munari_Coelho1}
\end{figure*}
\section{Results}\label{Section:Results}
\subsection{Spectral Energy distributions: comparison with previous models}

We have produced the complete set of updated {\sc HR-pyPopStar} models using the 4 IMFs (SAL, FER, KRO and CHA), with 106 ages \footnote{The SAL IMF produces a shorter number of ages because lowest mass considered is $1 \ M_{\odot}$}, and the six  metallicities $ Z=$0.0001, 0.0004, 0.004, 0.008, 0.02 and 0.05.  The wavelength range is [2500\,\AA\ -- 10\,500\,\AA] and the wavelength step is  $\delta\lambda = 0.1$\,\AA. The resulting SEDs are publicly available in the web page of {\sc HR-pyPopStar}\footnote{See in \url{https://www.fractal-es.com/PopStar}}.

Fig.~\ref{fig:SED-MUN} shows the SEDs with this version of {\sc HR-pyPopStar} for: a) six selected ages at $Z=0.02$, and the IMF from CHA;  b) six different values of the metallicity, for a given age $\tau=100$\,Myr and the IMF from CHA; and c) four different IMFs with the same age $\tau=100$\,Myr, and metallicity $Z=0.02$. Each SED has an arbitrary shift in logarithmic scale for the sake of clarity.

Then, we have compared the effect of using different stellar libraries in the synthesis models, by comparing the new SEDs created for this paper with the ones from MI21 for metallicity $Z=0.02$ and the IMF of CHA for some specific ages: 1, 10, 100, 1000 and 10000\,Myr in Fig.~\ref{fig:Munari_Coelho1}.

For the age of 1\,Myr, the new models have between a 2 and a 3\% more luminosity than the MI21 ones. Moreover, there are notable changes in specific lines, such as the Balmer series and some other lines that even changed from emission to absorption and vice versa. These differences are attributed to the new models of O and B stars from PoWR group.

\begin{table*}
\label{Table:4}
\caption{Magnitudes in broad band filters for the CHA IMF, solar abundances and different ages in column 3. Columns 4 to 7 are the Johnson-Cousins-Glass system magnitudes U, B, V and R in the Vega system, while columns 8 to 12 refer to the magnitudes $u$, $g$, $r$, $i$ and $z$ for the SDSS filters AB system. The whole table is available in electronic format.
}
\begin{center}
\resizebox{16cm}{!}{
\begin{tabular}{cclccccccccc}
\hline
IMF & Z & $\log t$ & U  & B & V & R & $u$ & $g$ & $r$ & $i$ & $z$\\
\hline
CHA & 0.02 & 5.00 & -0.340 & 0.674 & 0.905 & 1.031 & 0.773 & 1.122 & 1.559 & 1.967 &  2.247 \\
CHA & 0.02 & 5.48 & -0.380 & 0.635 & 0.868 & 0.995 & 0.732 & 1.084 & 1.522 & 1.931 & 2.213 \\
CHA & 0.02 & 5.70 & -0.422 & 0.594 & 0.826 & 0.953 & 0.691 & 1.044 & 1.480 & 1.889 & 2.171 \\
CHA & 0.02 & 5.85 & -0.465 & 0.550 & 0.781 & 0.907 & 0.647 & 0.999 & 1.435 & 1.843 & 2.124 \\
CHA & 0.02 & 6.00 & -0.526 & 0.493 & 0.726 & 0.853 & 0.586 &  0.943 & 1.380 & 1.789 & 2.071 \\
CHA & 0.02 & 6.10 & -0.635 & 0.387 & 0.619 & 0.745 & 0.477 & 0.836 & 1.273 & 1.682 & 1.964 \\
\hline 
\end{tabular}
}
\end{center}

\end{table*}

In the case of 10\,Myr, the new models have a flux 3\% higher than MI21 in the NIR and in the molecular bands. This is attributed to the different modelling of the red supergiants in both atmosphere models in the NIR, even though they are very similar in the optical model. At these ages, the contribution of red supergiant stars to the spectra of the population increases with the wavelength having a significant contribution in the NIR \citep[see][]{CER1994}

For 100\,Myr, the new models have a lower flux in the TiO molecular bands of the NIR than MI21. This is caused by the new stellar atmosphere models of PHOENIX for the same reasons as for 10\,Myr, but more focused in the molecular bands. PHOENIX spectra have higher opacities than the C14 models, and thus their spectra in that region are less luminous in that wavelength range. Moreover, the PHOENIX stellar library has a better coverage of $T_{\rm eff}$ in the H-R diagram, by actually including stars as cold as 2300\,K, whereas in C14 the coolest star is 3300\,K. This is especially relevant for intermediate and old ages, since there is a significant fraction of stars cooler than 3300\,K.
\\

\begin{figure*}
\hspace{-0.5cm}
\begin{center}
\includegraphics[width=0.32\textwidth]{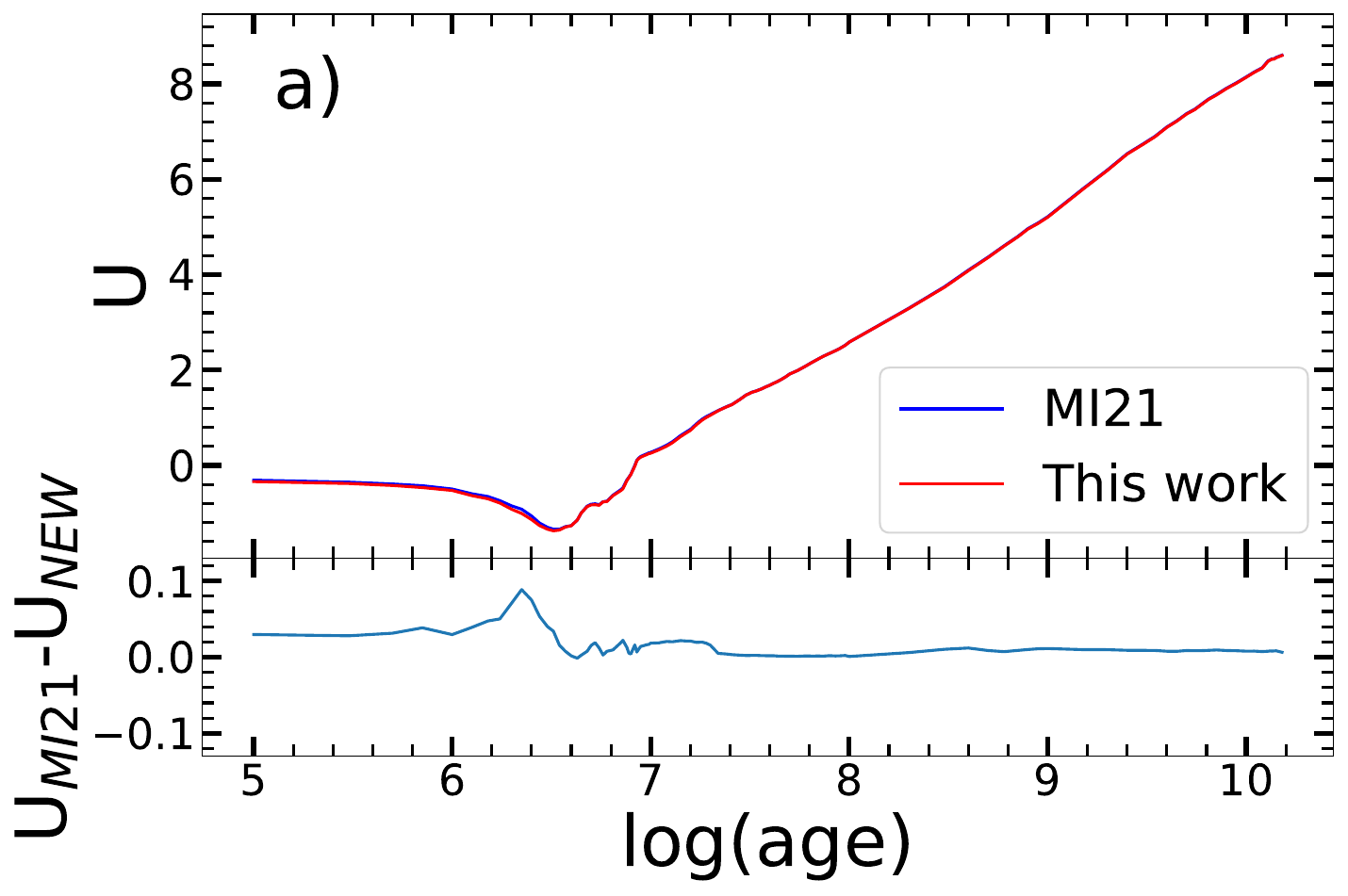}
\includegraphics[width=0.32\textwidth]{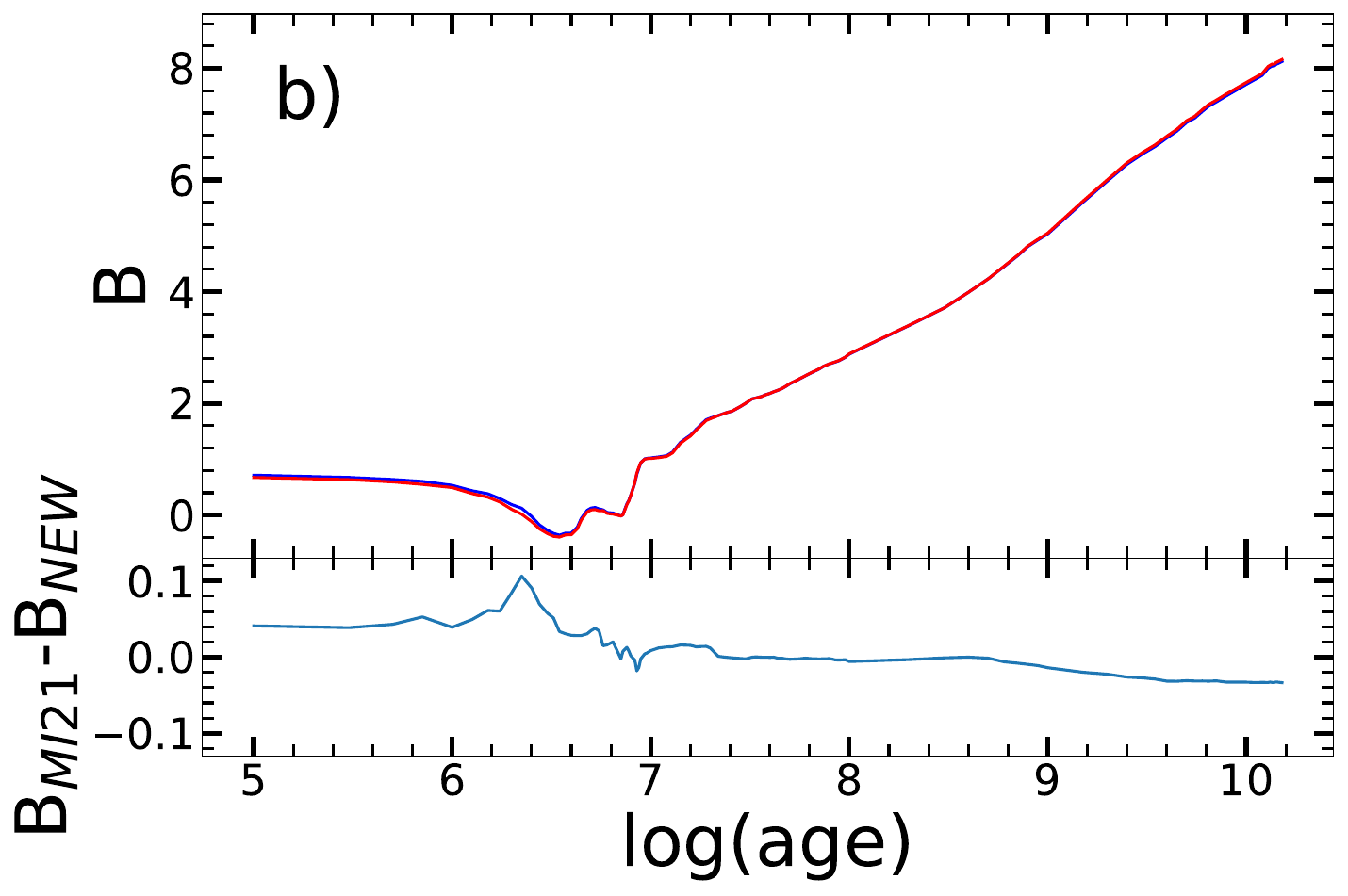}
\includegraphics[width=0.32\textwidth]{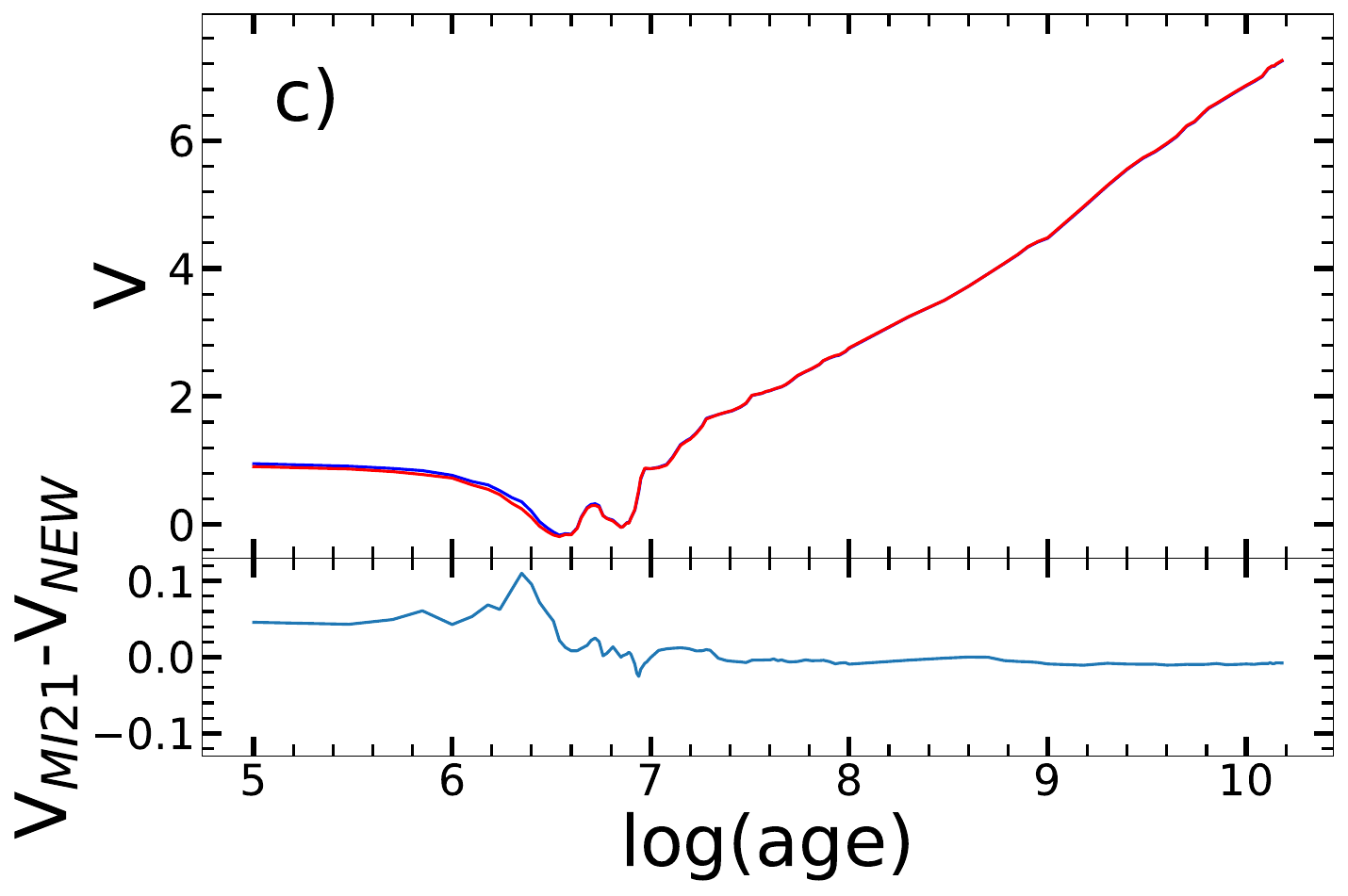}
\includegraphics[width=0.32\textwidth]{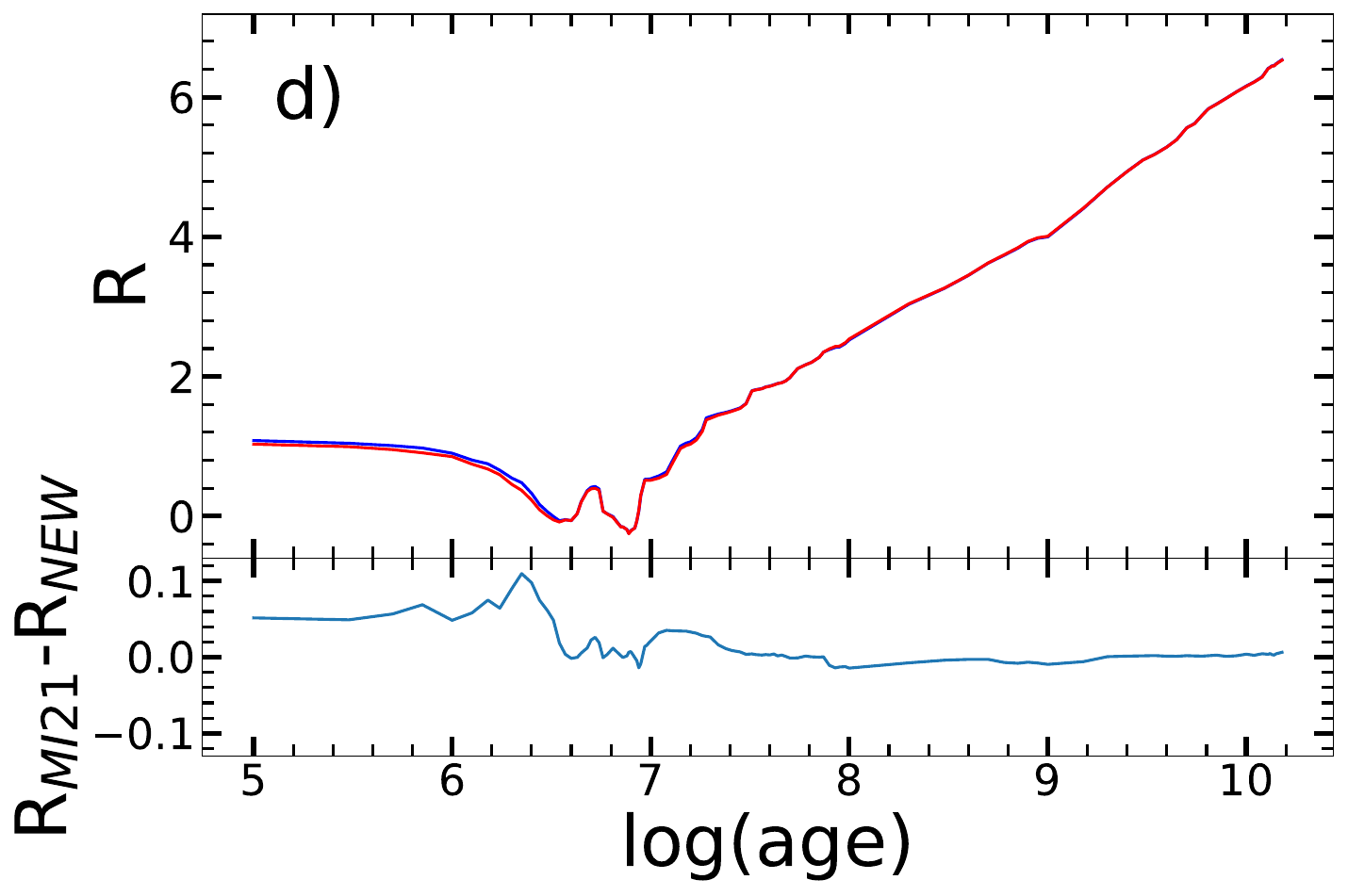}
\includegraphics[width=0.32\textwidth]{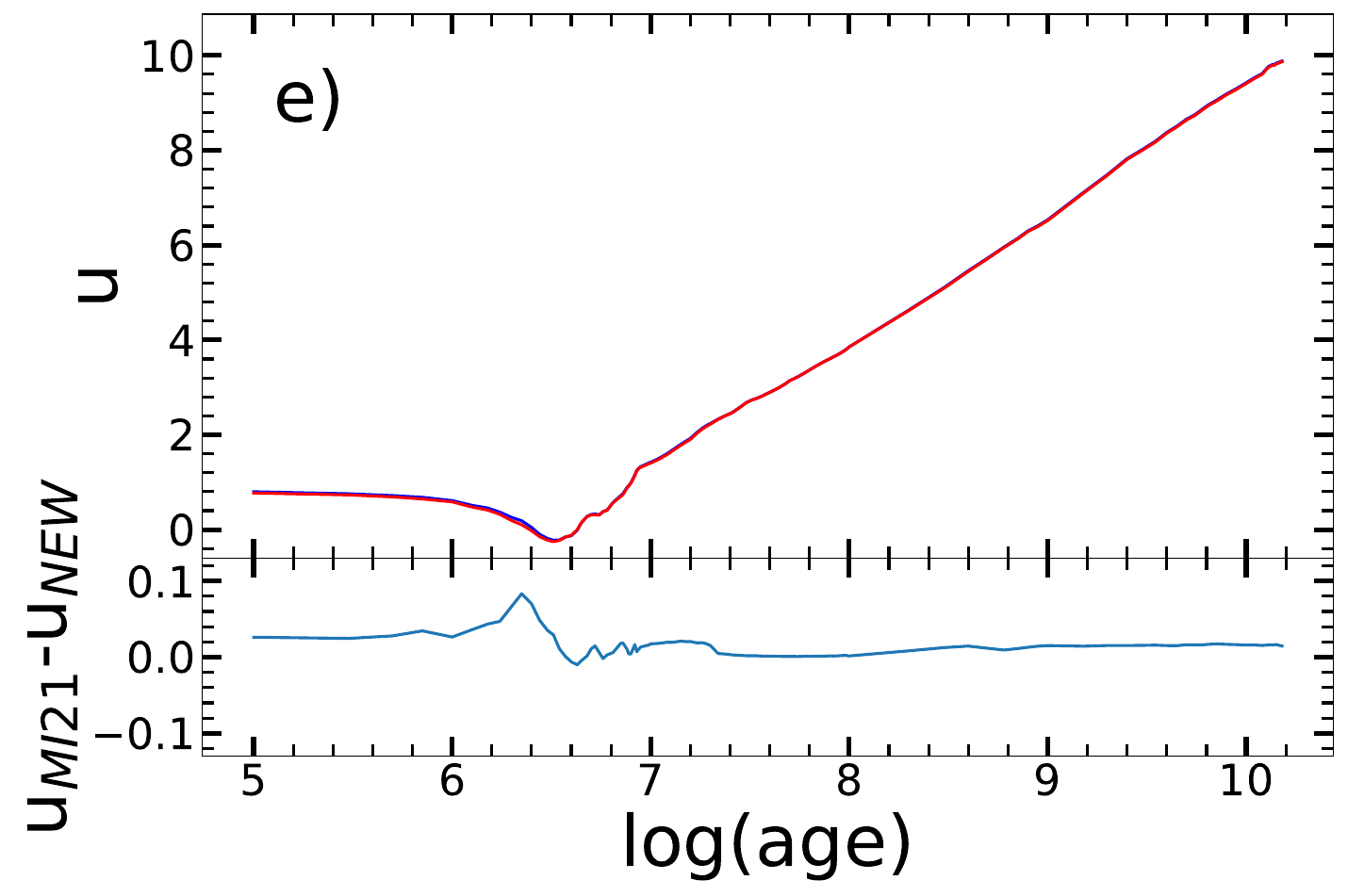}
\includegraphics[width=0.32\textwidth]{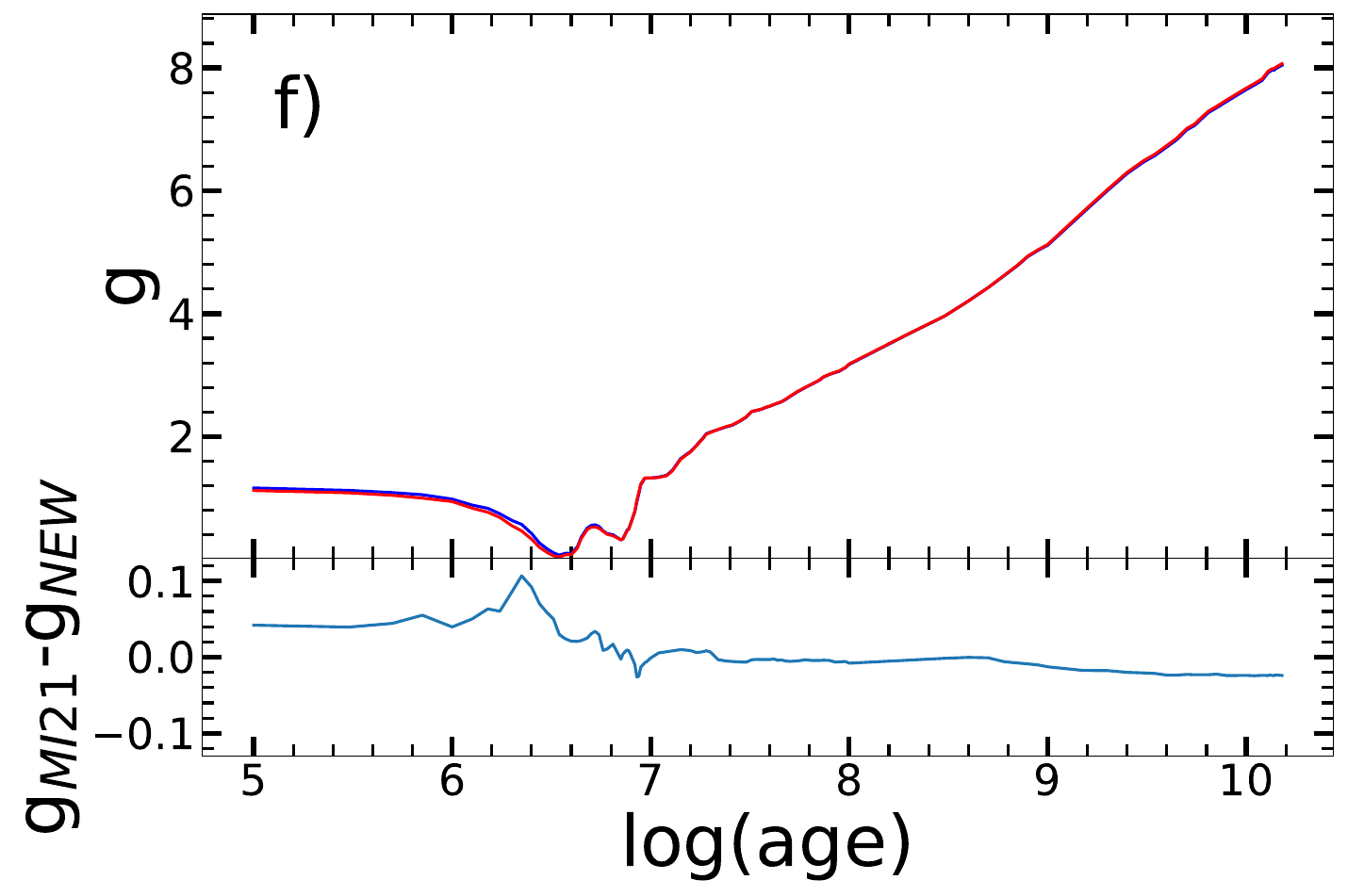}
\includegraphics[width=0.32\textwidth]{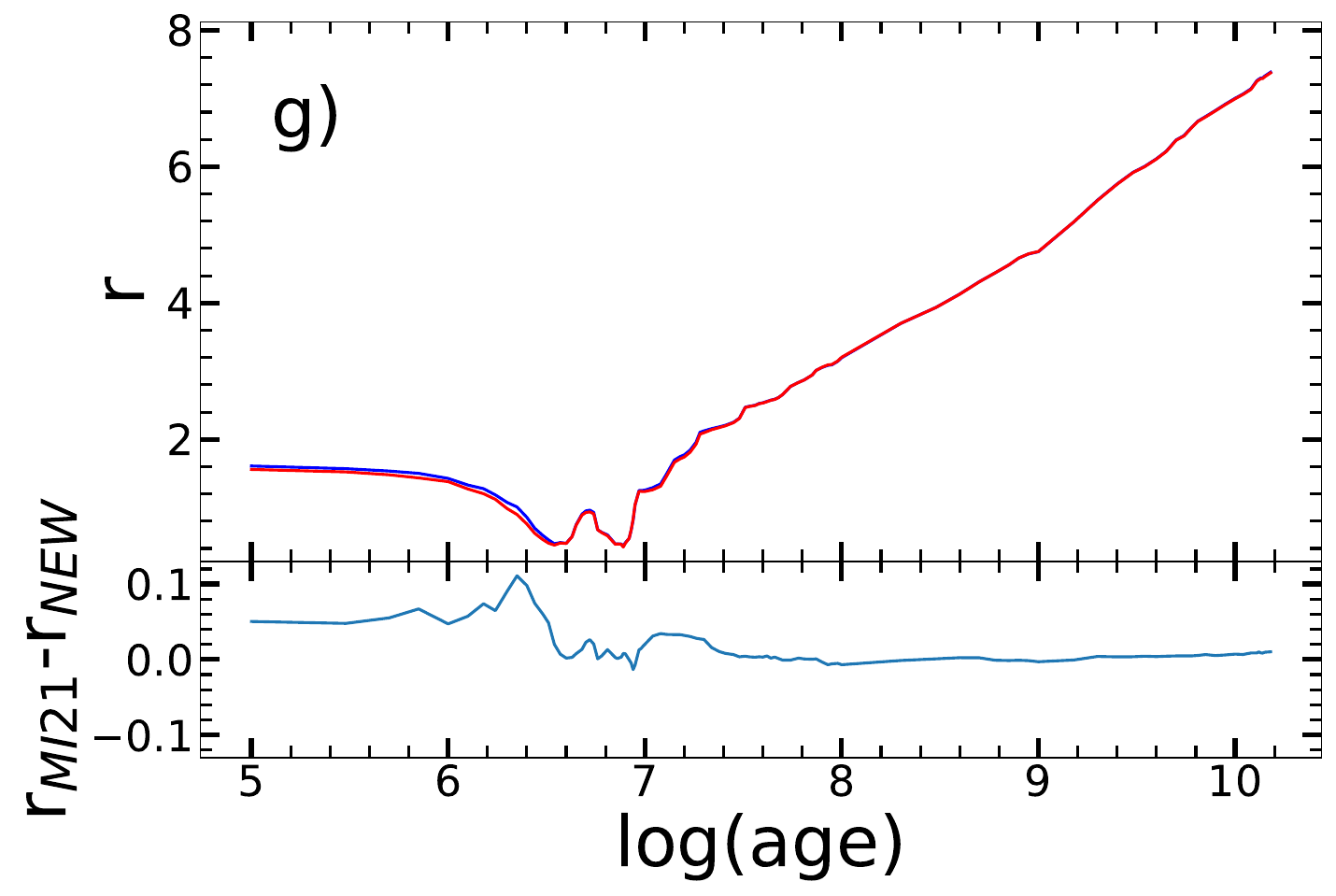}
\includegraphics[width=0.32\textwidth]{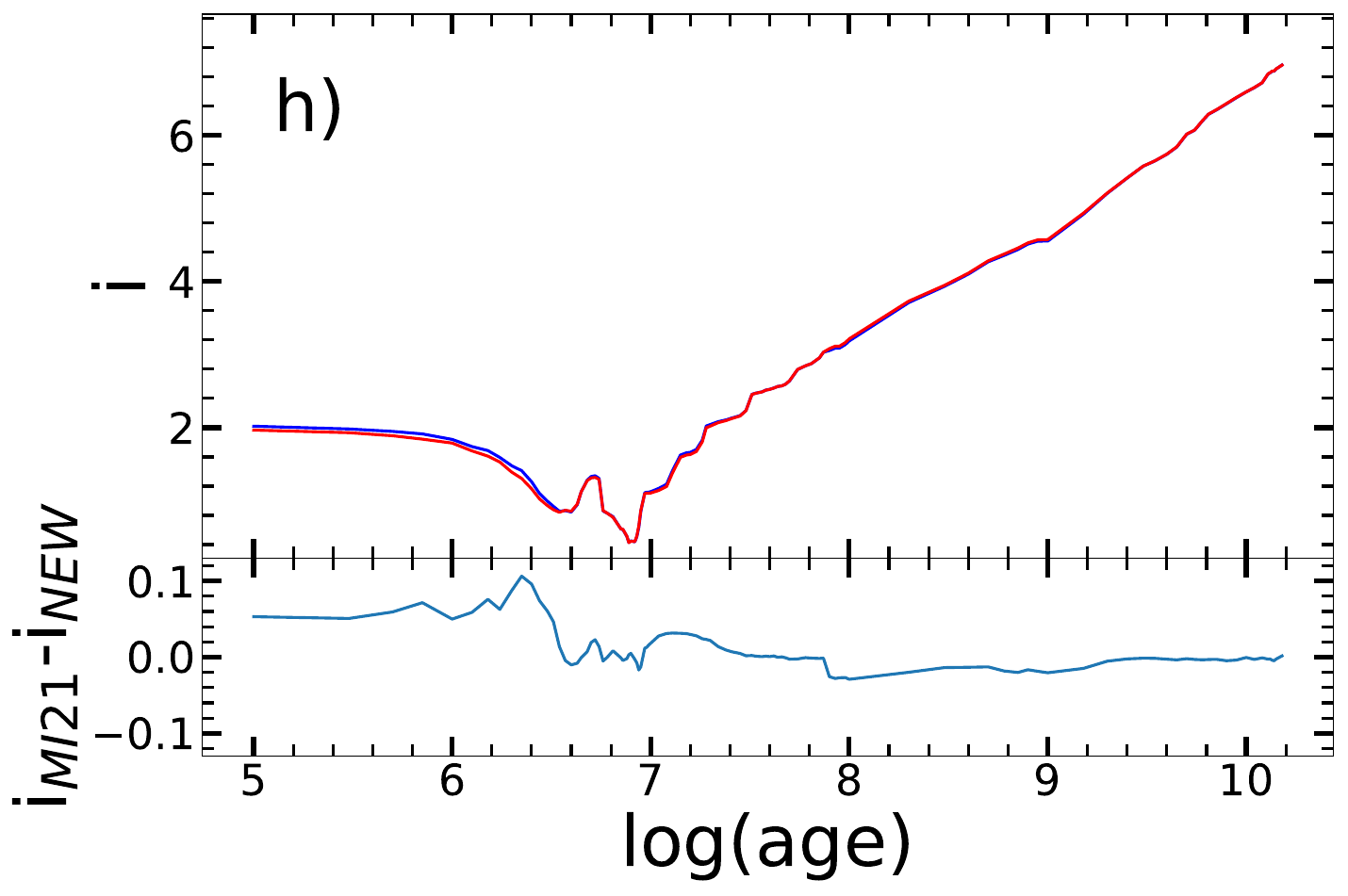}
\includegraphics[width=0.32\textwidth]{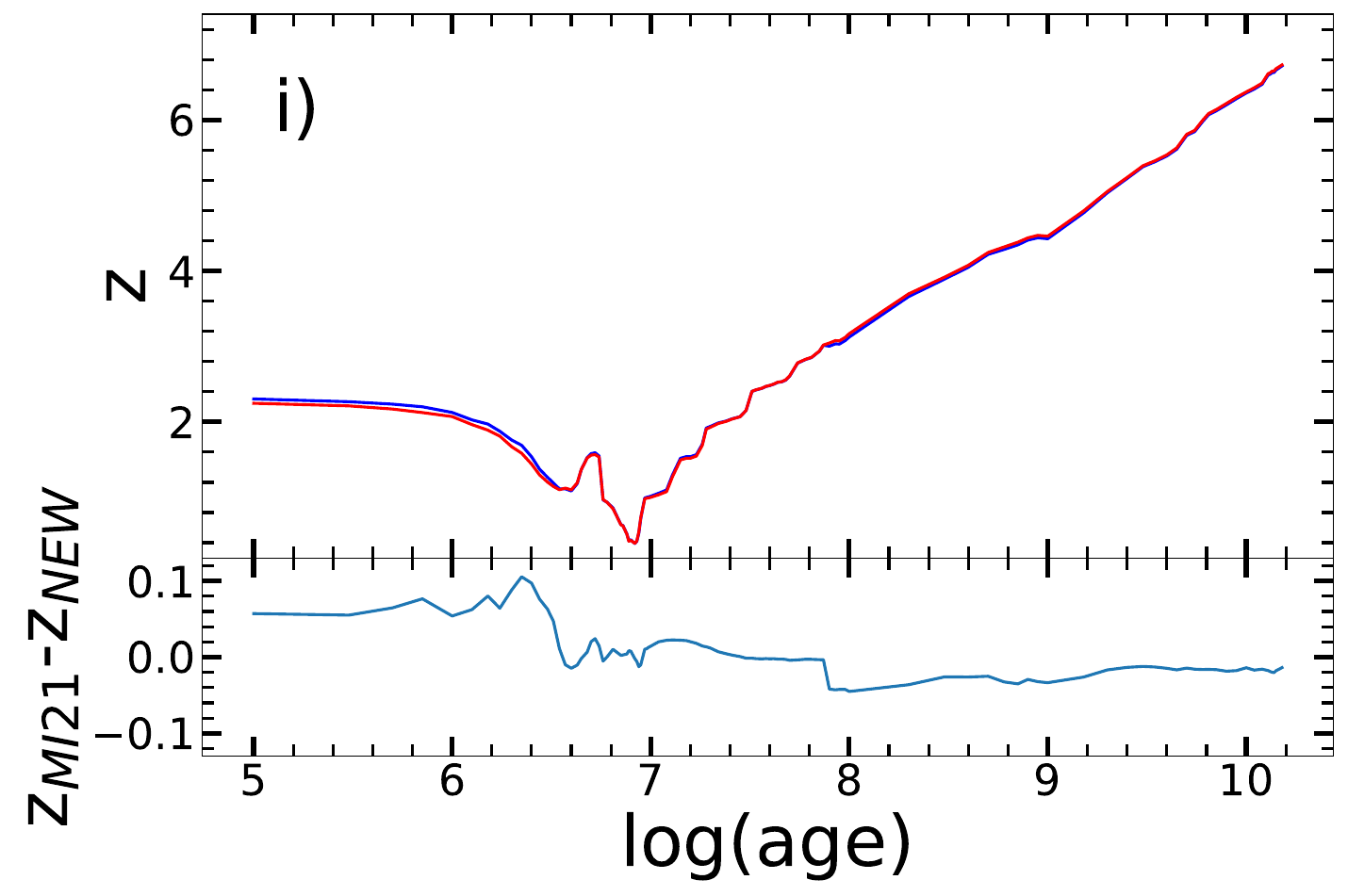}
\caption{Evolution of magnitudes a) U, b) B, c) V, d) R, e) u, f) g, g) r, h) i, and i) z with age: comparison  of results between MI21, blue, and this work, red, with residuals in the bottom panel for CHA IMF and $Z=0.02$.
}
\label{Fig:magnitudes}
\end{center}
\end{figure*}

Finally, for the 1\,Gyr and 10\,Gyr ages, there are two main differences between the new models and MI21. In the bluer part of the spectra, the new models have less flux in the G band area $\sim 4300$ \AA, more prominent for 10\,Gyr than for 1\,Gyr. This is caused by the different modelling of the CH molecular absorption bands in the C14 and PHOENIX libraries. 

\subsection{Derived properties}

\normalsize

\subsubsection{Magnitudes and colours}

We also computed some magnitudes in bands that are within our wavelength range: U, B, V and R, in the Johnson system and $u$, $g$, $r$, $i$ and $z$ in the SDSS system. 
In Table~\ref{Table:4} we give these magnitudes for all ages, metallicities and IMFs in electronic format. Here we show as an example the values for six ages, solar metallicity and the CHA IMF. The evolution of the 9 magnitudes is shown in Fig.~\ref{Fig:magnitudes}, with a magnitude in each panel, as labelled. 
\begin{figure*}
\hspace{-0.5cm}
\begin{center}
\includegraphics[width=0.495\textwidth]{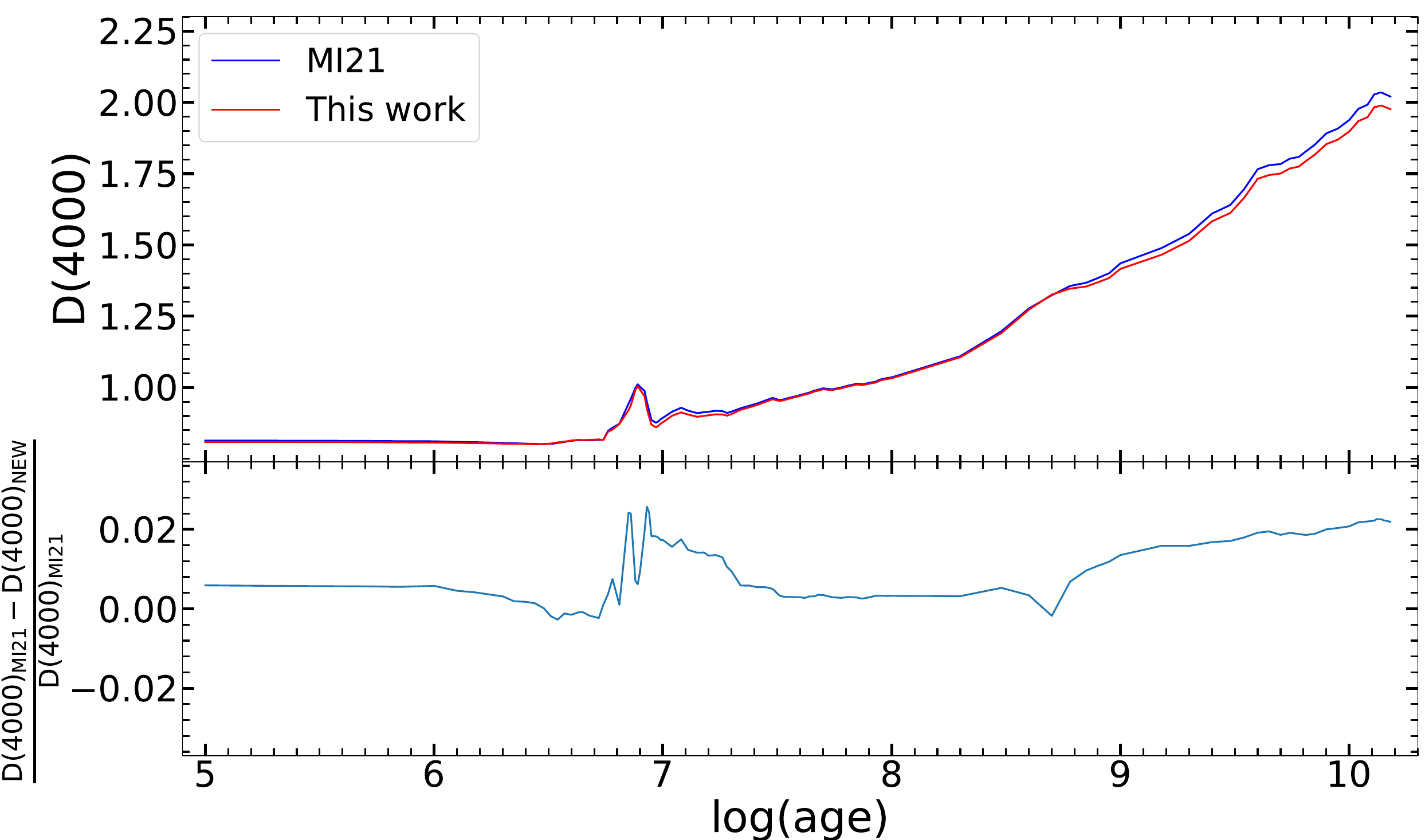}
\includegraphics[width=0.495\textwidth]{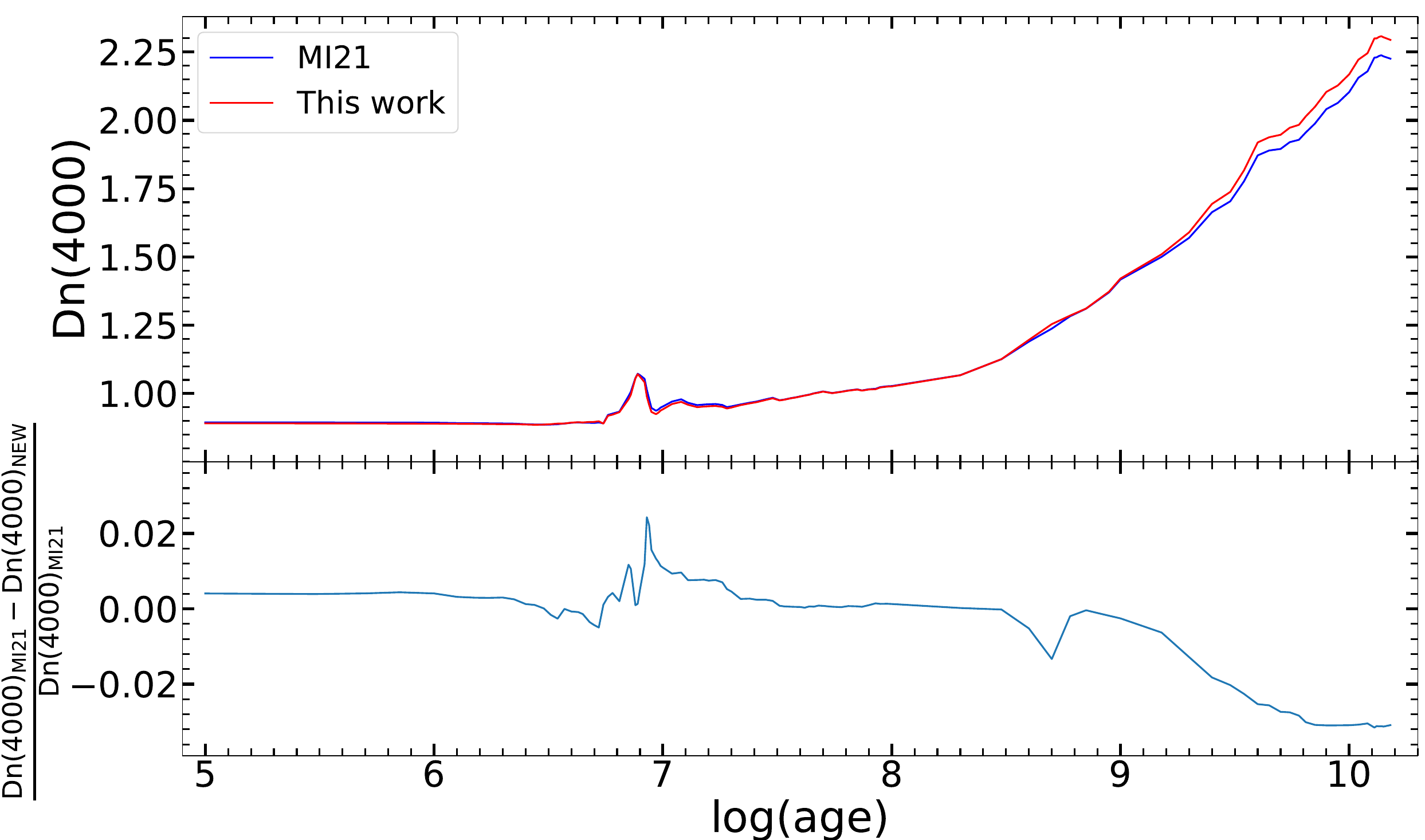}
\caption{D4000 and Dn4000 break time evolution comparison for between MI21 in blue line and this work red line for $Z=0.02$.}
\label{Fig:D4000}
\end{center}
\end{figure*}

There are differences between the old and the new models versions for all magnitudes. For young ages, the magnitudes of the new models are 0.04-0.06\,dex higher, thus less luminous than MI21. Then, the new models magnitudes decrease until they reach a maximum difference at around at 2-3\, Myr, that is 0.12\,dex smaller than MI21 for redder (R, r, i and z ) bands and  0.09\,dex smaller than MI21 for the bluer (U,B,V, u and g) ones, respectively. Then, both models are very similar until they reach ages around 100\,Myr, when the new models have smaller magnitudes than MI21, around 0.03 dex, for the blue and bigger magnitudes than MI21, also around 0.03, for the red magnitudes.

\subsubsection{4000 \AA\, break}
We have also compared the indices D4000 \citep{Bruzual1983} and Dn4000 \citep{Balogh+1999,Kauffmann+2003} of the 4000\,\AA\ break, which are known to be proxies of the age of the stellar population. In Fig.~\ref{Fig:D4000}, we show the comparison for $Z=0.02$ between MI21 and the present work models. In general, both models agree very well, except in two age ranges: 1) 10-30\,Myr, where the newest models are a $ ~2$\% smaller than the MI21; and 2) when the age is older than 4\,Gyr, where the newest models are a $3.5\%$ higher  than the previous ones in the case of Dn4000. In both cases, the difference is due to the cool stars spectral libraries used for them, C14 for MI21 and PHOENIX in this work, that have small differences in the area before the break, mainly in the CN band 3883 area that could be attributed to the modelling of the molecular lines done in each atmosphere model. The comparison for $Z=0.004$ is in ~\ref{appendix}. The general trends of both libraries are maintained by using the MUN05 +PHOENIX+new PoWR models. The difference in the D4000 index does not have a great effect on estimating the age using the new D4000 breaks. 

\subsection{Calcium triplet}

\begin{figure}
\hspace{-0.5cm}
\begin{center}
\includegraphics[width=0.495\textwidth]{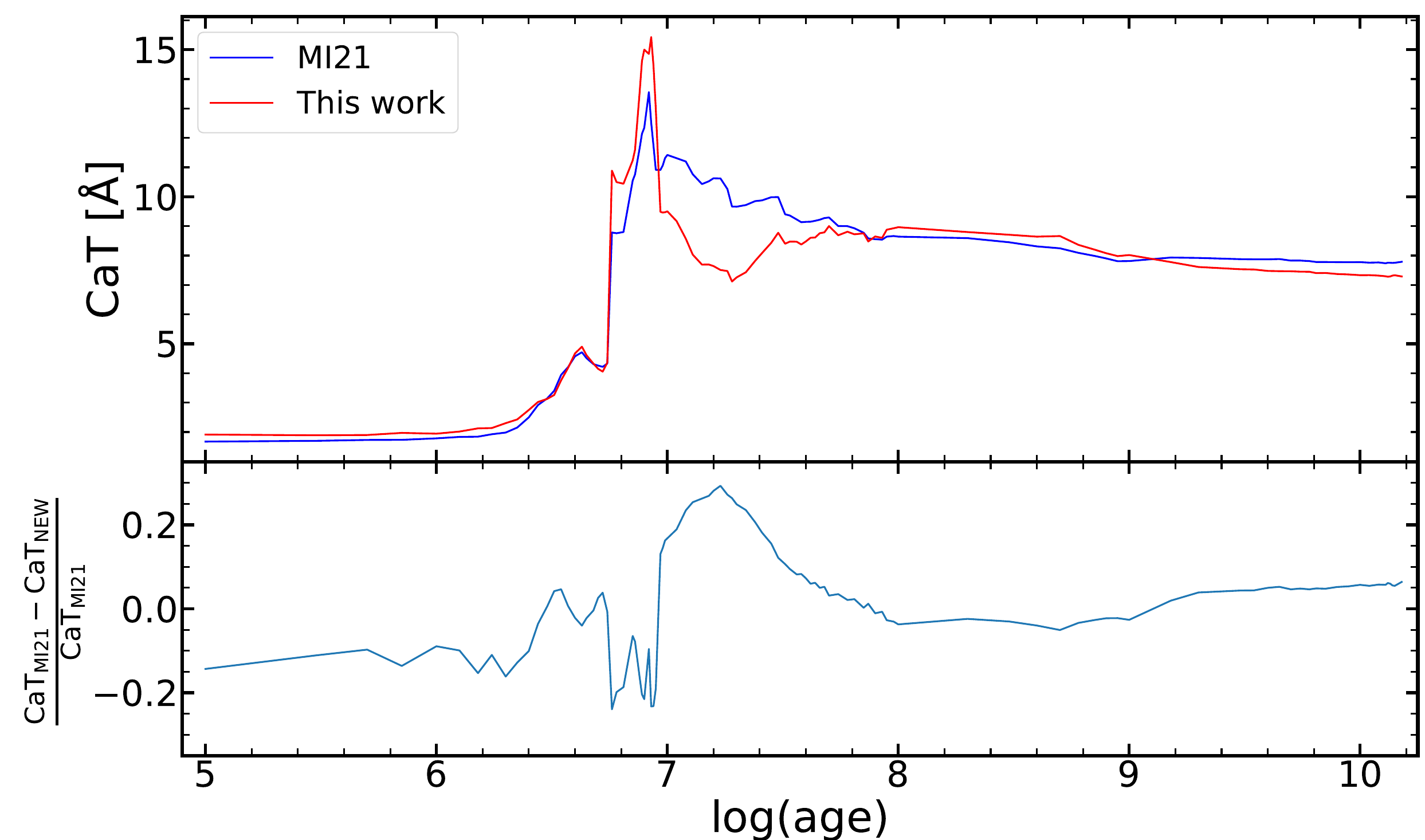}
\caption{Calcium triplet time evolution comparison for between MI21 in blue line and this work red line for $Z=0.02$.}
\label{Fig:CaT_Z02}
\end{center}
\end{figure}

We have also compared the Calcium triplet (CaT) indices using the continuum definition of  \cite{Cenarro+2001} and the line definition of \cite{GV20}. The comparison between the new models and MI21 for $Z=0.02$ is shown in Fig.~\ref{Fig:CaT_Z02}. For the youngest ages, (0.1-10 Myr) new models have higher values than the old ones with the biggest difference at the peak at 10 Myr with a difference of $20\%$. In the ages from 10 Myr to 100 Myr, the CaT of the new models are smaller than the MI21 models, this is specially significant for the ages around 20 Myr (log(age)=7.3). In the case of intermediate ages between 100 Myr and 1 Gyr, the CaT index of the new models are slightly higher $\sim 3\%$, than the ones of the old models. Finally, for ages older than 1 Gyr, the CaT of the new models have smaller value than the MI21 models, with a maximum difference in that age range of $\sim 7\%$.

The differences in the CaT are caused by the change of the new PoWR models for the young ages with a secondary contribution, specially for ages between 10 and 30 Myr, of the change from C14 atmospheres to the PHOENIX atmospheres for red supergiants. The comparison for the lowest metallicity is shown in \ref{subsec_app:CaT}.

\subsection{Comparison with other SSP models}\label{subsec:other_SSP}

\begin{figure}
\begin{center}
\includegraphics[width=0.46\textwidth]{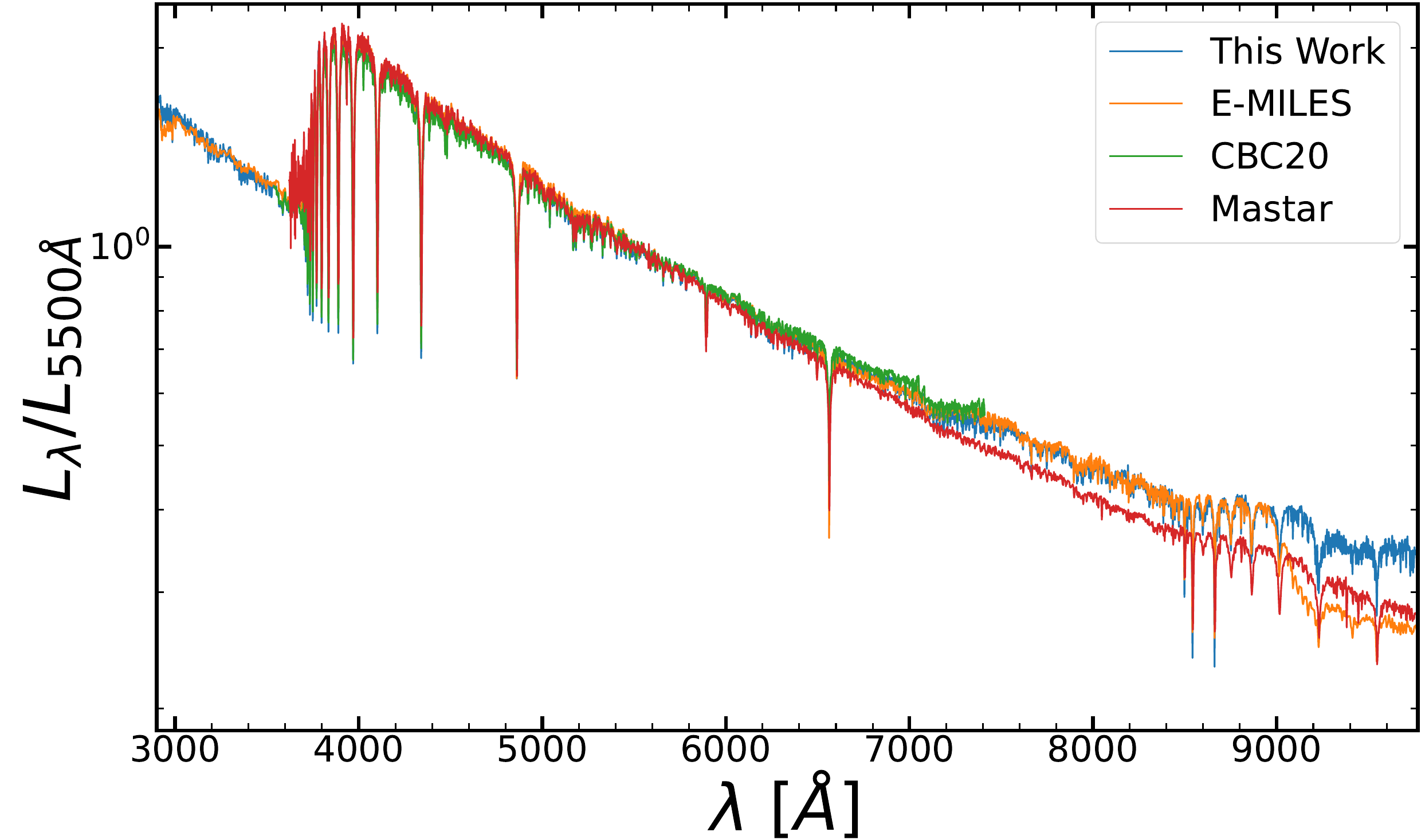}
\includegraphics[width=0.46\textwidth]{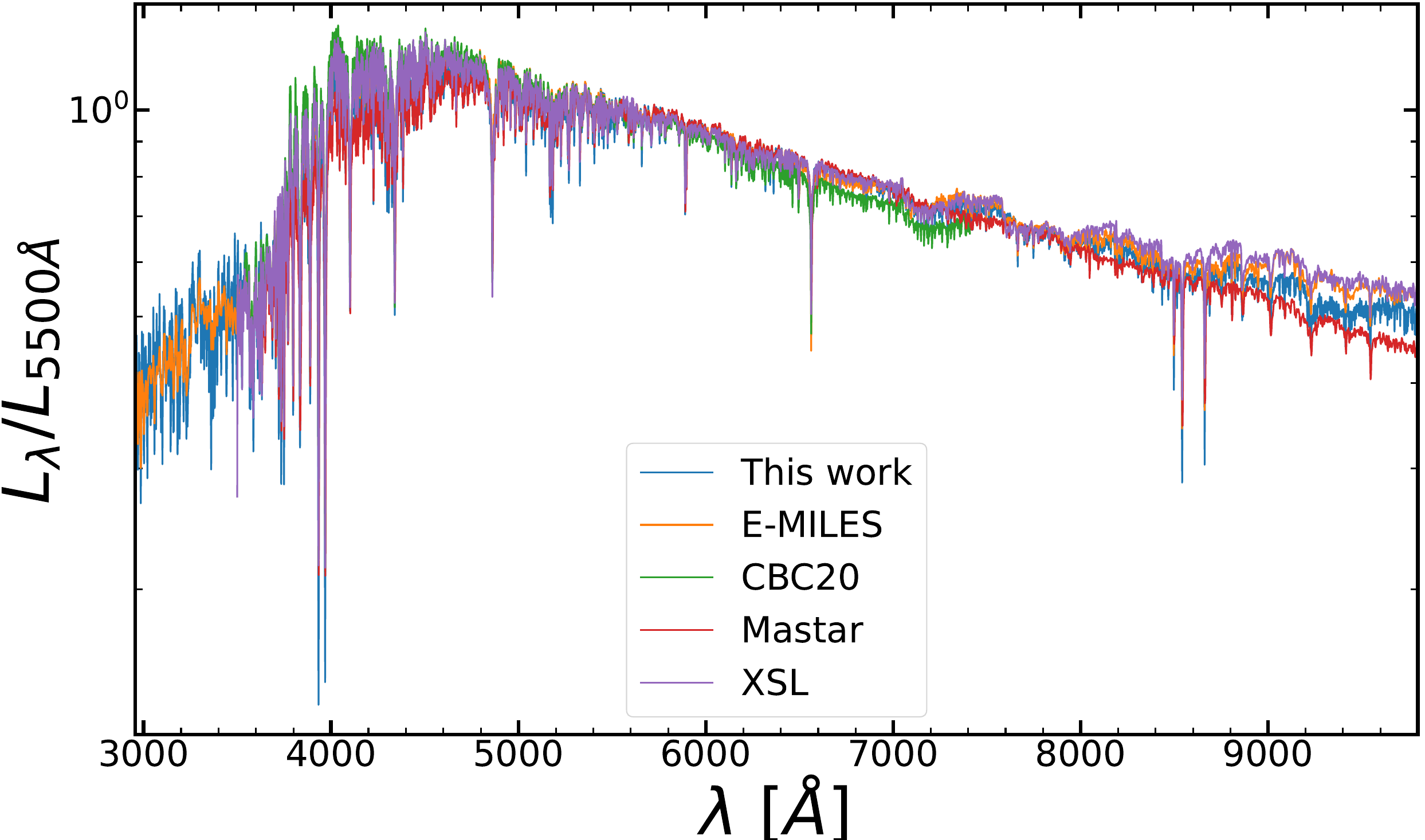}
\includegraphics[width=0.46\textwidth]{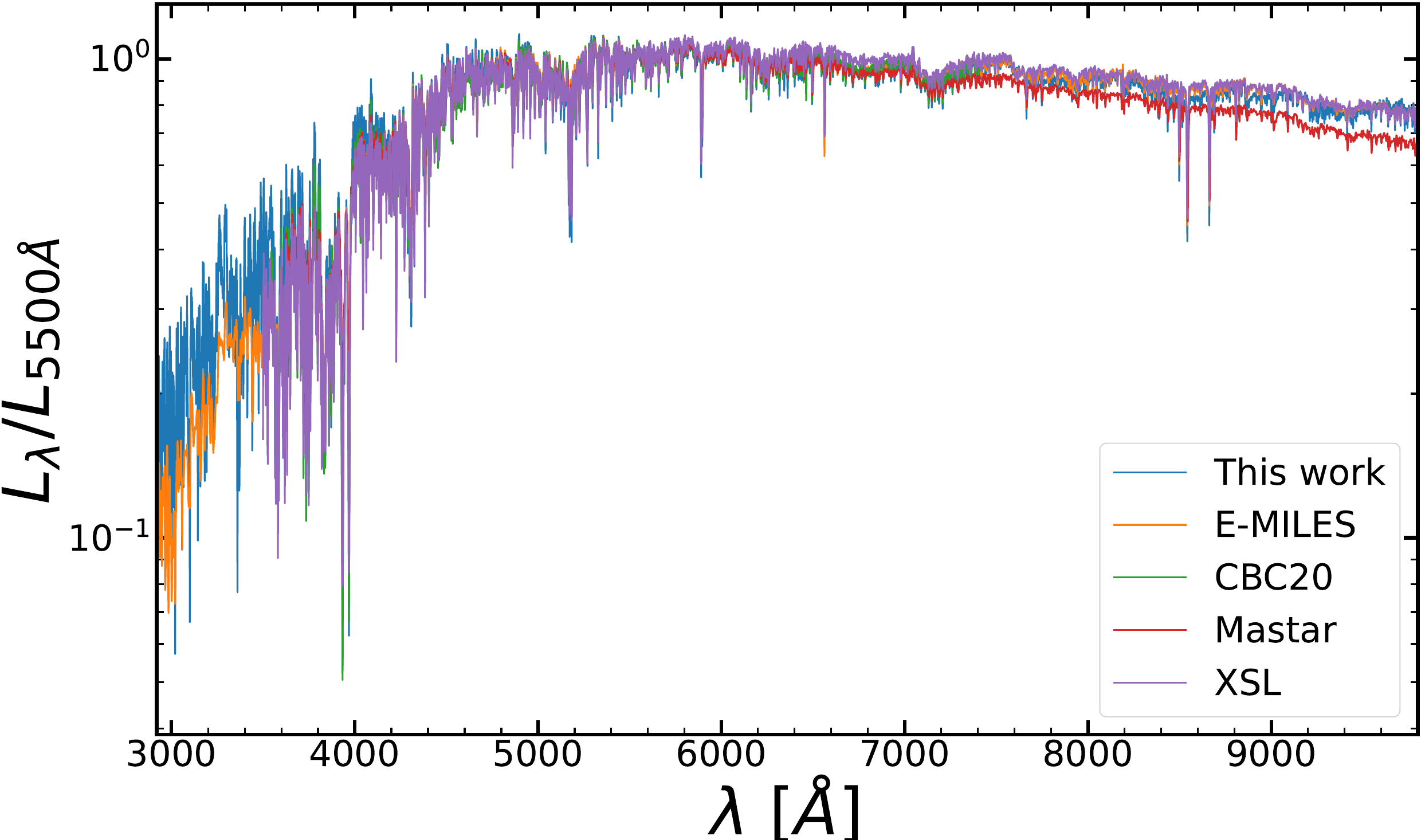}
\caption{Comparison of the E-MILES (orange), CBC20 (green), MaStar (red), XSL (purple) and our new models (blue) for solar metallicity and 100 Myr, left, 1 Gyr, centre, and 10 Gyr, right.
}
\label{fig:Other_models}
\end{center}
\end{figure}

We compared our new models with other stellar population synthesis models, both theoretical, CBC20, and empirical, E-MILES, MaStar (Ma20) and XSL. For XSL and E-MILES, we used the models computed with the Padova 2000 isochrones to make a more similar comparison and see only the effects of the stellar libraries. We compared the models of solar metallicity for 100\,Myr, 1\,Gyr and 10\,Gyr. We have normalized all the spectra to a 10\,\AA\ band around 5500\,\AA, where there is no prominent feature. The comparison of the models is shown in Fig. \ref{fig:Other_models}. All models agree in the optical region except CBC20 that has lower fluxes in the red part. However, in the NIR each model has a very different spectrum. 

For 100 Myr, E-MILES and our models agree until $\sim 9000$\,\AA\ where E-MILES models suddenly drop whereas our models decrease slowly. MaStar models has lower fluxes in the NIR than the rest of the models. 

In the case of 1\,Gyr, XSL has the highest fluxes and very strong molecular bands in the NIR, E-MILES has smaller fluxes in the NIR in general and in the TiO molecular band around 8600\,\AA\ in particular. Our models are similar to those of E-MILES until $\sim 8800$,\AA, where our models have lower fluxes than those of E-MILES. Finally, MaStar has fluxes lower than those of the rest of the models. 

Finally, for 10\,Gyr, XSL, E-MILES and our models are very similar except in the region $8800-9100$\,\AA\ where our models have slightly lower fluxes than XSL and E-MILES.  

The differences between XSL and E-MILES compared with our models are mostly attributed to the different libraries they include, since the same isochrones are used for the models. However, the differences with MaStar models could also be due to the fact that they are using a different algorithm, and different stellar tracks \citep[the Geneva group tracks][]{Strom+2012}.  

\section{Analysis of stellar populations of low metallicity systems}\label{sec:Low_z_analysis}
\subsection{High resolution observations of the globular cluster M15}

After quantifying the difference between the models from MI21 and the new ones of this work, we have tested the new models with observations that have a similar spectral resolution. Similarly to the analysis performed in MI21, we have analysed the spectra of the globular cluster M15 which is a classical example of an old and metal-poor stellar population.
\begin{figure}
\hspace{-0.5cm}
\includegraphics[width=0.495\textwidth]{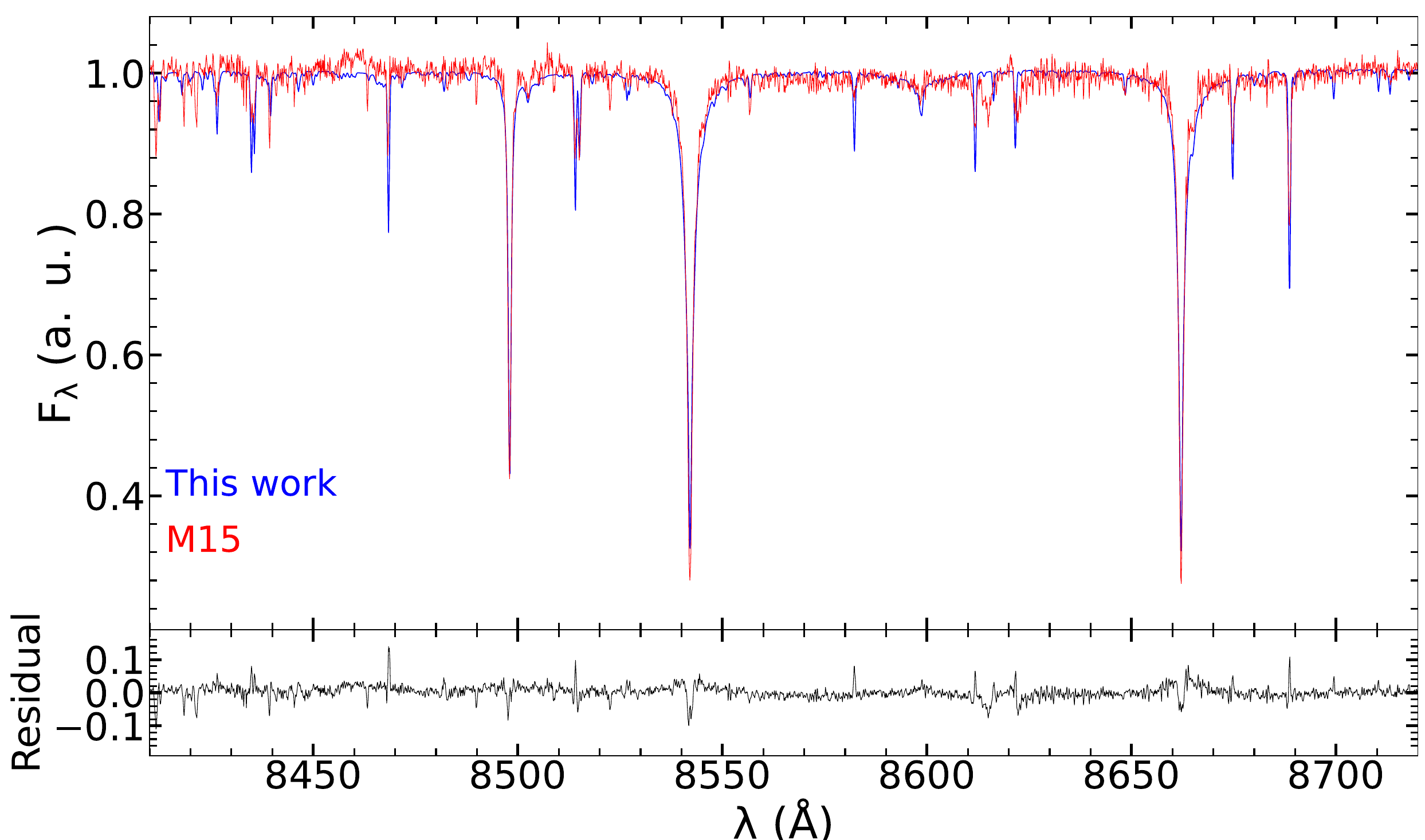}
\caption{The spectra of M15 from MEGARA observations in red, and the best model from {\sc HR-pyPopStar} fitting the data in blue.}

\label{fig:M15_analysis}
\end{figure}

In order to do that, we have used the spectrum of the globular cluster M15, observed with MEGARA in the HR-I setup (see MI21 for more specific details). This spectrum has very high S/N in the CaT area ($\sim$100 S/N per \AA). We have performed a $\chi^2$ method to find the best model that fits the observed spectrum and we have obtained an age of $\tau =  12.8$\,Gyr and a metallicity $Z=0.0004$, which corresponds to  $\log{(Z/\mathrm{Z}_{\sun})}=-1.7$, by using the solar abundance of $\mathrm{Z}_{\sun}=0.02$ \citep{Grevesse_Sauval_1998}. These results are better estimates than the ones obtained in MI21, with a $ \chi^{2}_{\rm new}=5.14$, smaller than the $\chi^{2}_{\rm old}=6.54$. Moreover, the new results are closer to recent estimates from the literature by \citet{OMalley17}, who found $\tau = 12.5 \pm 1.3$\,Gyr and $[\rm Fe/H]=-2.33$, obtained by analysing colour magnitude diagrams with GAIA data of key stars in the globular cluster. In the previously fitted model, there were too many metallic lines that couldn't be found in the observed spectrum even with the lowest metallicity modelled in MI21 $Z = 0.004$, whereas the new model fit better these observed features in the spectrum. This shows the great advantage of having available low metallicity models to be able to reproduce these kind of systems. We present the new best fit model in Fig.~\ref{fig:M15_analysis}, where the observed spectrum is shown with the red line, while the best fitting model, corresponding to the indicated parameters, is the blue line. 

\subsection{Analysis of the stellar populations of dwarf galaxies }

We used these new models to analyse the stellar populations of dwarf galaxies where the low metallicity stars have a significant contribution to their total stellar population. Dwarf galaxies have a wide range of morphological types, Star Formation Histories (SFH), etc. However, they have generally lower metallicities than normal massive galaxies. Therefore, they are the ideal examples to see the capabilities of the new generation low metallicity SSPs for the analysis of stellar populations. In this case, we have chosen the dwarf galaxies observed by MaNGA \citep{Bundy+2015} that were studied in the MaNDALA Value Added Catalogue \citep{Cano-Diaz+2022}. These galaxies were selected to have stellar masses lower than $\log M(\mathrm{M}_\odot) < 9.1$. These observations have good coverage in the wavelength range [3650 \,\AA --10300 \,\AA] with $R \sim 2100$. 

\begin{figure*}
\begin{center}
\hspace{-0.5cm}
\includegraphics[width=0.795\textwidth]{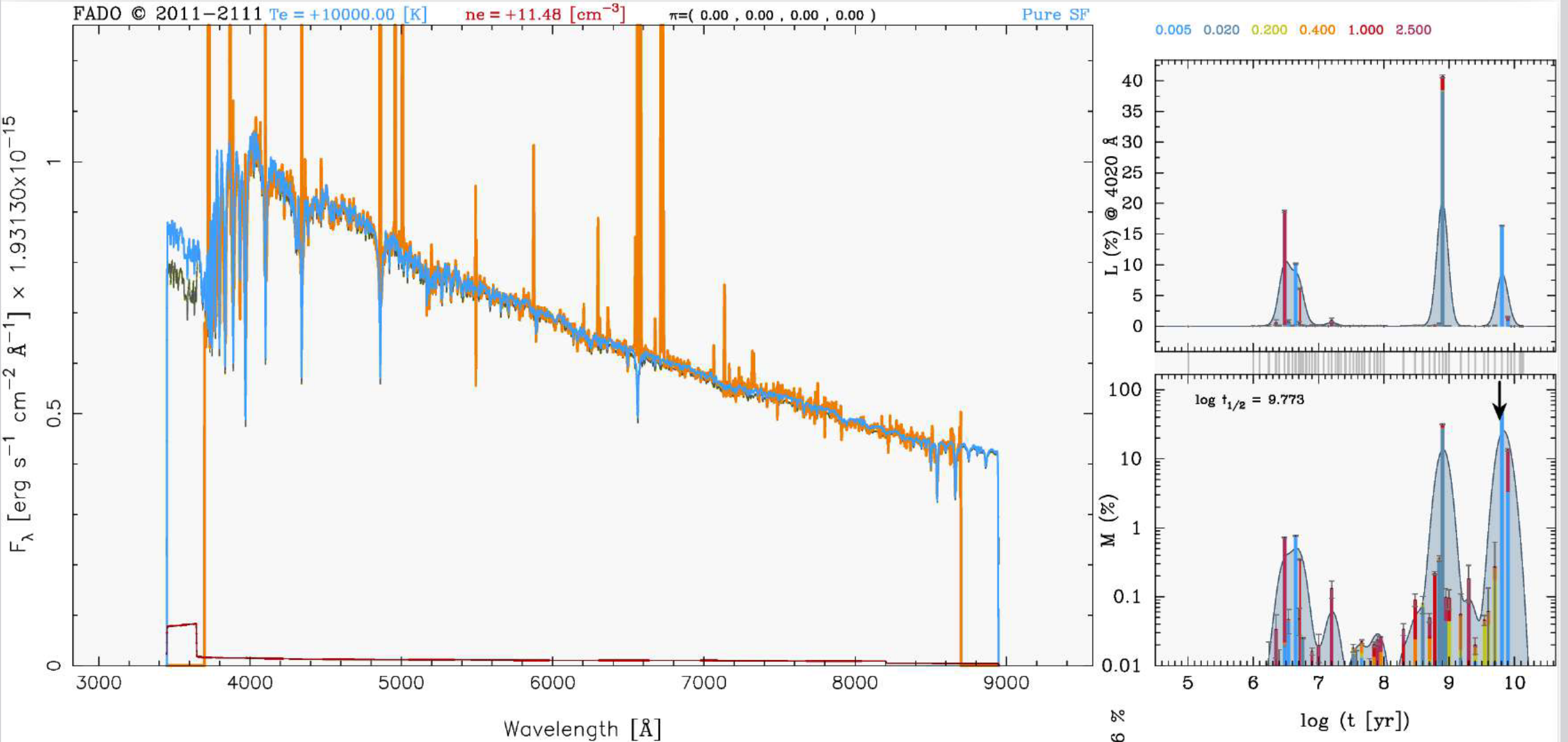}
\caption{ Example of one of the full spectral fitting fits of FADO of the dwarf galaxy 8596-6103. In the left plot, the orange line is the observed spectra, the blue line the fitted stellar spectra and the green line the fitted nebular continuum. On the right top image is the contribution of each SSP to the fit in luminosity at $\lambda = 4020$\,\AA. On the right bottom panel is the star formation history of the galaxy given by the contribution in mass of each SSP to the fit. }
\label{fig:Example_FADO}
\end{center}
\end{figure*}

We have extracted the spectra of the 131 galaxies adding the spaxels of the IFU up to 1 effective radius, $R_\mathrm{e}$. Then, we have made a selection cut of the spectra based on the criteria of having a $S/N$ greater than 6 in $\rm H_{\alpha}$, $\rm H_{\beta}$, $\rm H_{\gamma}$, Mg I b triplet region and the Calcium triplet (CaT) region. After applying the $S/N$ cut, we obtain a total of 31 galaxies from the original sample. We have analysed all the observed galaxy spectra using the full spectral fitting code {\sc FADO} \citep{gomes18} and the newly computed SSP using the CHA IMF. {\sc FADO} has been already tested with the previous version of {\sc HR-pyPopStar} in MI21. During the analysis, we masked the region around 5570\,\AA\ to avoid the contamination from telluric lines. 
Regarding the full spectral fitting analysis, we have run the code 10 times with different initial random seeds for each galaxy, in order to avoid local minima in the final solution and we have averaged over the different runs. In Fig.~\ref{fig:Example_FADO}, we show an example of one of the fit for one of the dwarf galaxy from FADO.

In addition to the star formation and chemical enrichment history, FADO gives as outputs the mean stellar age, mean stellar metallicity (both weighted by mass and light) and the total formed and current stellar masses. In Table~\ref{Table:4}, we summarized the averaged values of the stellar populations properties of the galaxies obtained from the full spectral fitting, the whole table is available in electronic format.

\normalsize
\begin{table}
\caption{Properties of the stellar populations of the sample obtained by FADO using the new SSP models. The whole table is available in electronic format.}
\normalsize
\resizebox{9cm}{!}{
\begin{tabular}{clcccc}
\hline
 MaNGA Id & $\log M_{*}$ & $\left< \log({\rm \tau})\right>_{L}$ & $\left<\log( \tau)\right>_{M}$ & $\left<\log(Z/Z_{\odot})\right>_{L} $  & $\left<\log(Z/Z_{\odot})\right>_{M}$ \\
\hline
7815-6101 & 9.11 & 8.58 & 9.66 & -0.29 & -0.41 \\
7992-12703 & 8.75 & 8.44 & 9.17 & -0.36 & -0.27 \\
8135-6101 & 8.42 & 8.62 & 9.51 & -0.38 & -0.23 \\
8150-3701 & 6.77 & 8.03 & 9.61 & -0.64 & -0.86 \\
8152-12702 & 8.14 & 9.01 & 9.58 & -1.01 & -1.26 \\
8449-6101 & 8.83 & 8.11 & 8.93 & -0.29 & -0.29 \\
\hline 
\end{tabular}
}

\label{Table:5}
\end{table}
\normalsize

After computing the mean stellar metallicities and stellar masses, we have considered the mass-mean metallicity and the mass-mean age relations of the galaxies of the sample.
There have been several works that have tried to explore the mass-metallicity relation of dwarf galaxies in the past using different methods and different subsamples. Some of the work devoted to study all the galaxies also traced at least part of the range of dwarf galaxies. Thus, \citet{Gallazzi+2005} analysed up to 200\,000 galaxies from the SDSS DR2 survey \citep{Abazajian+2004} using spectral indices sensitive to age and metallicity and reconstructing the star formation histories of the galaxies; \citet{Panter+2008} used 300\,000 galaxies from the SDSS DR3 \citep{Abazajian+2005} using a similar analysis than the one from \citet{Gallazzi+2005} to reconstruct the SFH of the galaxies; \citet{Kirby+2013} studied red giants in local dwarf spheroidal and irregular dwarf galaxies in the Local Group using indices to obtain the mean age and metallicity of these dwarfs galaxies; \citet{Kudritzki+2016} studied blue supergiant stars in several Local Group dwarf galaxies using low spectral resolution spectra; \citet{Zahid+2017} stacked the spectra of 90\,000 star-forming galaxies from SDSS DR7, DR10 and DR11 \citep{Abazajian+2009,Alam+2015} in different stellar mass bins creating combinations of star formation bursts and chemical evolution models to fit the stellar absorption lines; and more recently, \citet{Sextl+2023} performed a similar work to \citet{Zahid+2017}, but distinguishing the galaxies between star forming and passive galaxies and a third group that considers all of them to find the mass-metallicity relation of these three groups. In general, there are hints that star-forming dwarf galaxies have on average higher metallicity than their passive counterparts. 

\begin{figure}
\begin{center}
\includegraphics[width=0.4999\textwidth]{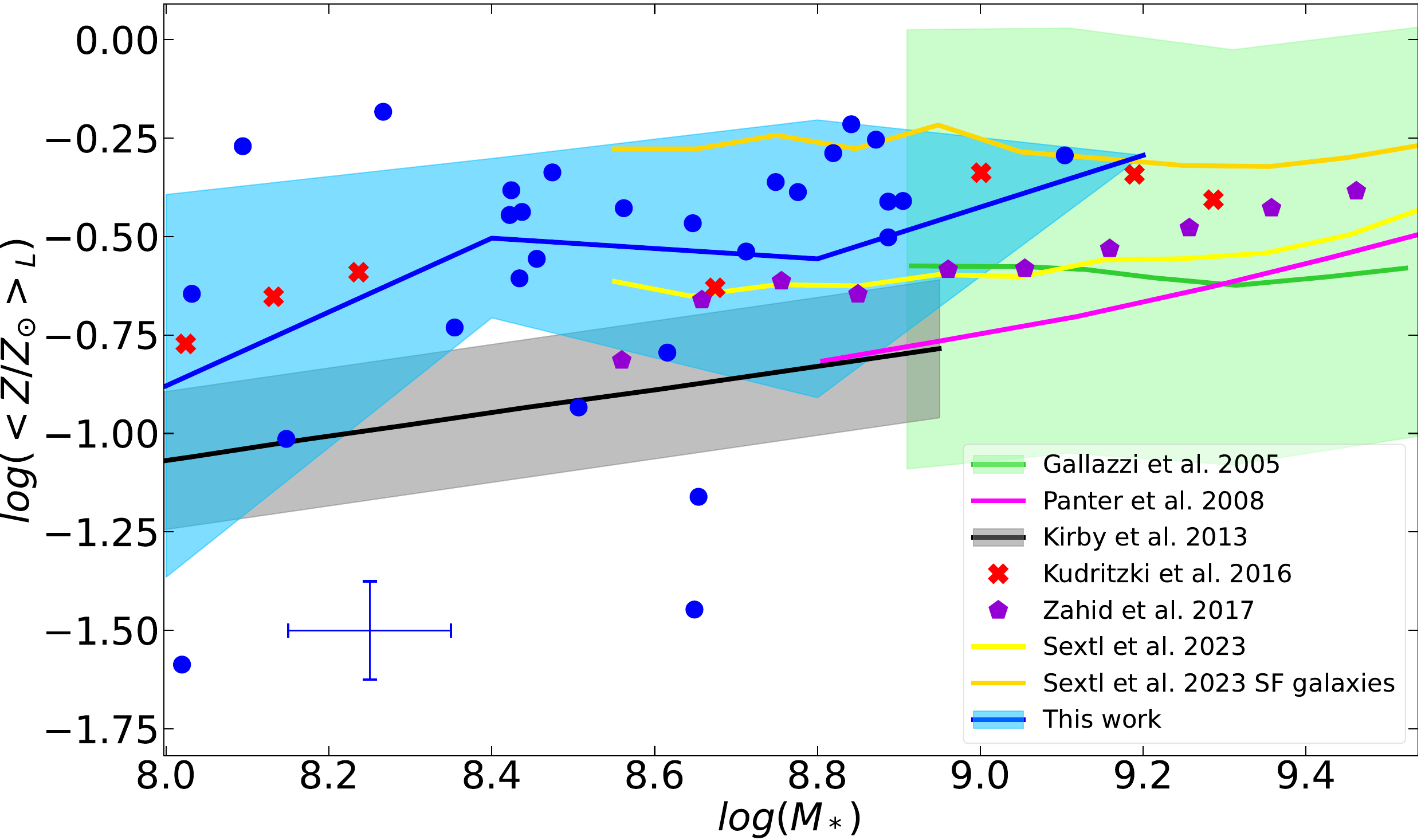}
\caption{Current stellar mass and mean stellar metallicity weighted by light relation. In blue circles our result with the mean error in the bottom, green line is from \citet{Gallazzi+2005}, the magenta line is from \citet{Panter+2008}, the black line is from \citet{Kirby+2013}, red Xs are from \citet{Kudritzki+2016}, violet pentagons are from \citet{Zahid+2017} and the orange and yellow lines are the star-forming regions and the mean of all galaxies from \citet{Sextl+2023}. The blue dot are the results from our work and the blue cross in the bottom represents the mean error of our analysis.}
\label{fig:Mass_metallicity}
\end{center}
\end{figure}

In Fig.~\ref{fig:Mass_metallicity}, we show our results of the current stellar mass and the mean stellar metallicity weighted by light as blue dots. We have also binned the results with mass bins of 0.4\,dex to show the trends of the results in a more comprehensive way. We compare our results with those found by
\citet{Gallazzi+2005, Panter+2008, Kirby+2013, Kudritzki+2016, Zahid+2017, Sextl+2023}.

Our results follow similar trends to \citet{Gallazzi+2005} and \citet{Sextl+2023} for the mean of all galaxies of SDSS, but our models have slightly higher mean stellar metallicity weighted by light, $\left<Z\right>_{L}$, than both of them. In particular, our models show values $\left<Z\right>_{L}$ between the subsample of SF and the mean of all galaxies from \citet{Sextl+2023}. Our models also have a similar mass-metallicity relation of studies of local star-forming galaxies from \citet{Kudritzki+2016,Zahid+2017}. The star-forming galaxies tend to show higher $\left<Z\right>_{L}$ than other works done with dwarf spheroidals \citep{Kirby+2013}, or take all the dwarf galaxies as a whole such as \citet{Gallazzi+2005,Panter+2008}. In fact, \citet{Sextl+2023} showed that star-forming galaxies tend to have higher $\left<Z\right>_{L}$ than the no-star-forming galaxies. Regarding our sample, 29 of 31 galaxies, a $93.5\%$, of the sample are star-forming galaxies.     

\subsection{Comparisons with MaNDALA Value Added Catalogue}

\begin{figure*}
\begin{center}
\includegraphics[width=0.495\textwidth]{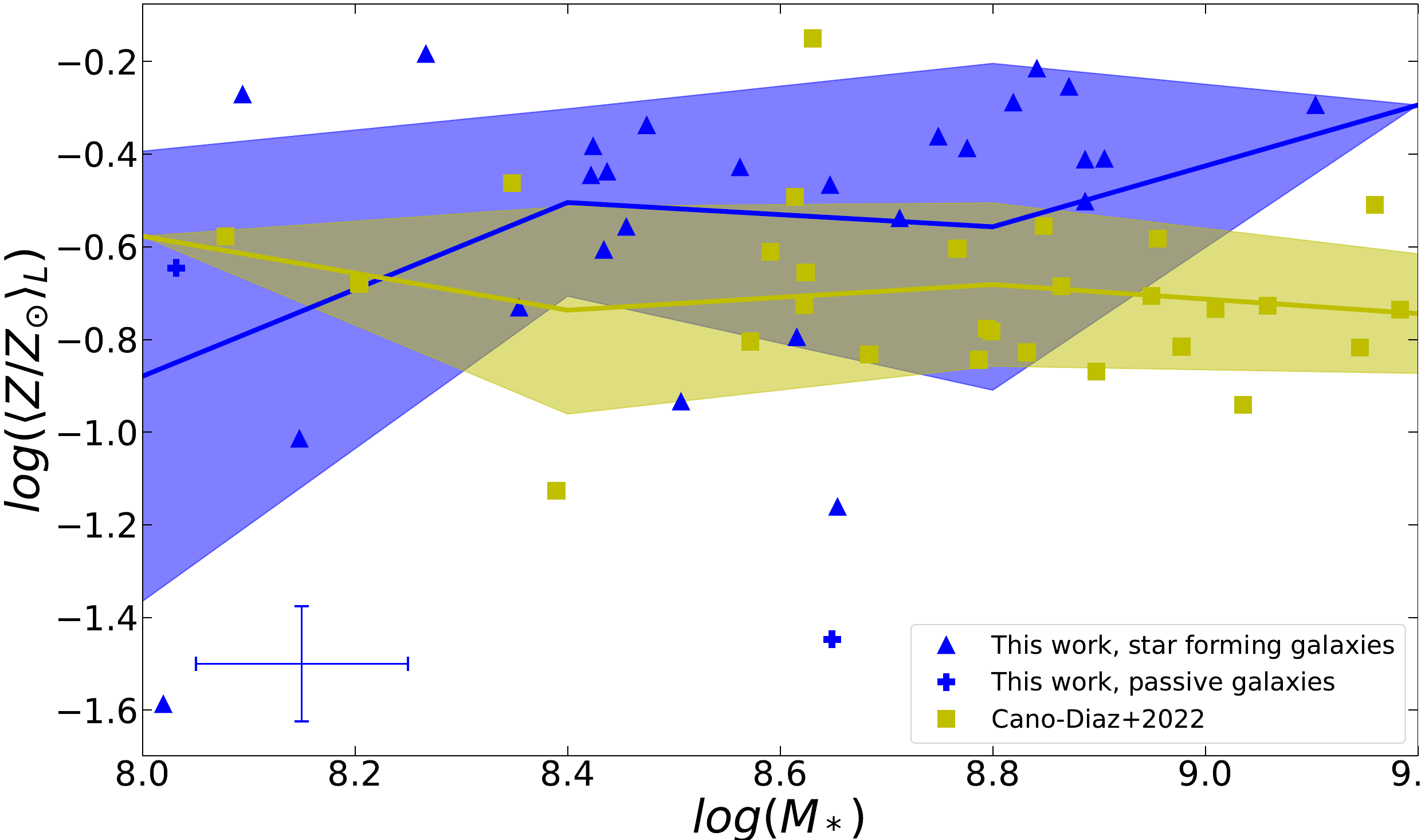}
\includegraphics[width=0.495\textwidth]{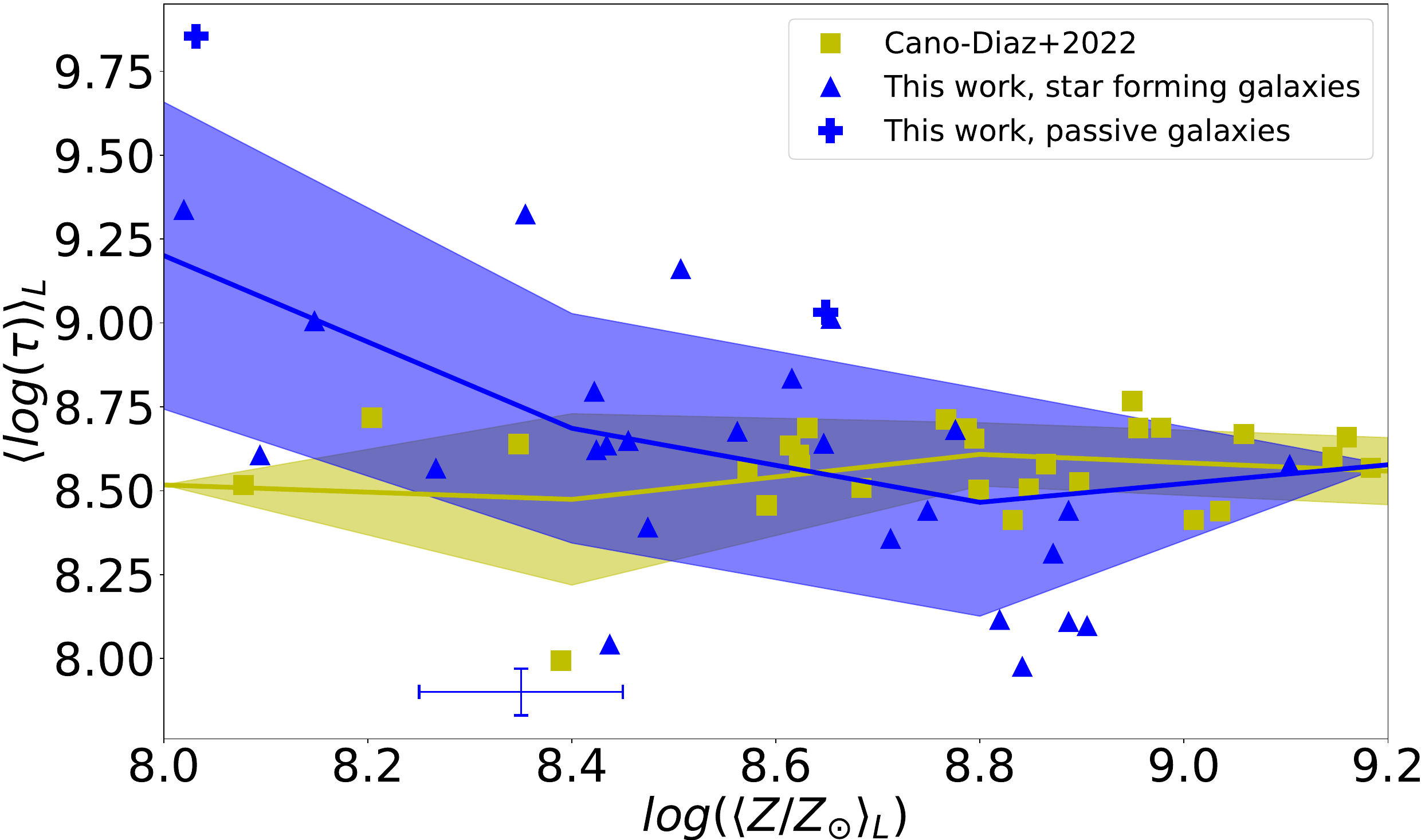}
\caption{Relation of a) the mean stellar metallicity as $\left<Z\right>_{L}$ and b) the mean stellar age, as $\log\left<\tau\right>_{L}$, {\sl vs } the stellar mass as $\log M_{*}$. Green squares are the results from \citet{Cano-Diaz+2022}, blue  triangles for star forming and blue crosses for passive galaxies are the results from this work and the blue cross in the bottom left shows the mean error of this work in stellar mass, mean stellar age and mean stellar metallicity.}
\label{fig:Mass_metallicity_age_Mandala}
\end{center}
\end{figure*}

\begin{figure}
\includegraphics[width=0.495\textwidth]{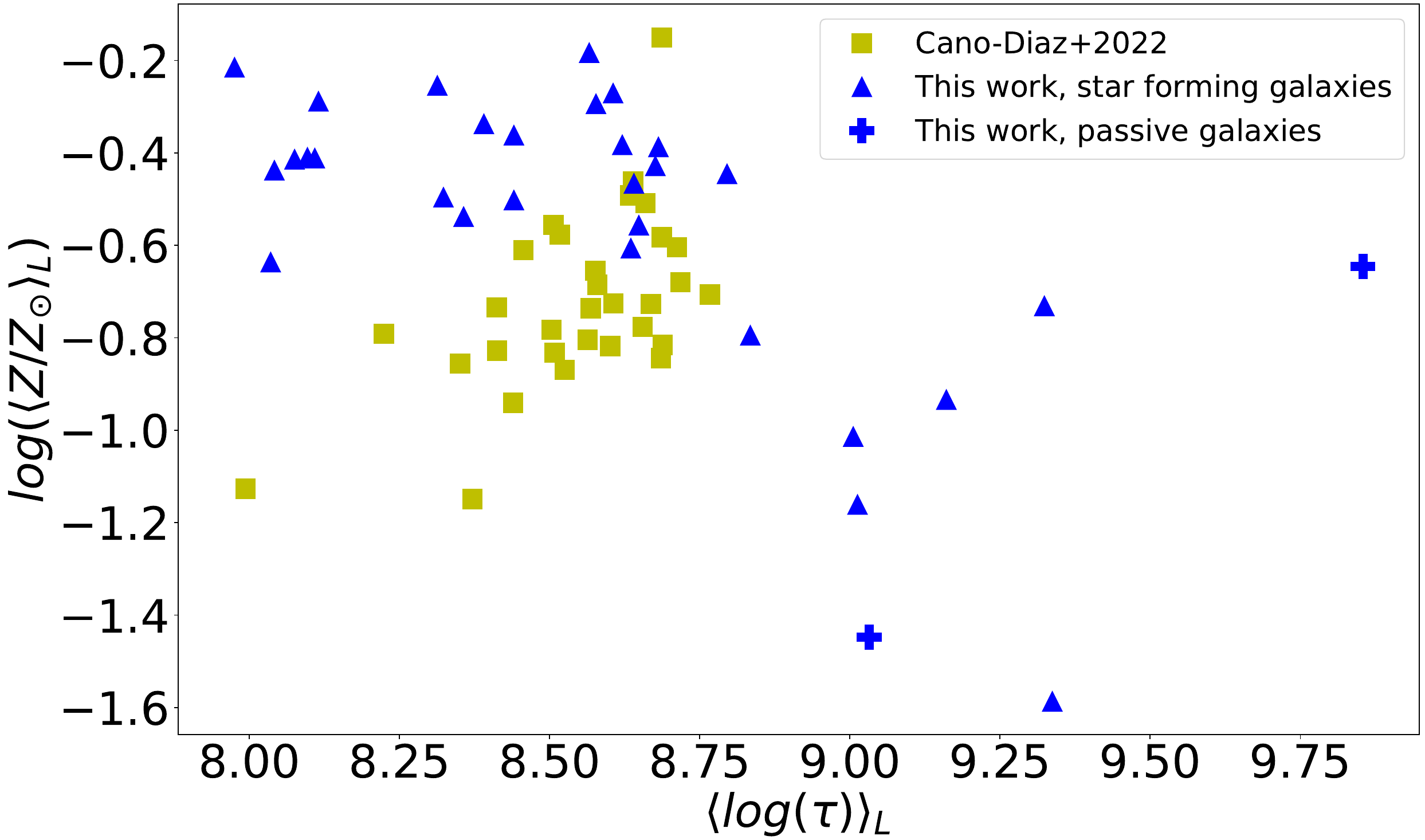}
\caption{Comparison of the mean of the age and the mean stellar metallicity of \citet{Cano-Diaz+2022}, green squares, and this work, blue triangles for star forming and blue crosses for passive galaxies.}
\label{fig:age_metallicity}
\end{figure}

\begin{figure*}
\begin{center}
\includegraphics[width=0.495\textwidth]{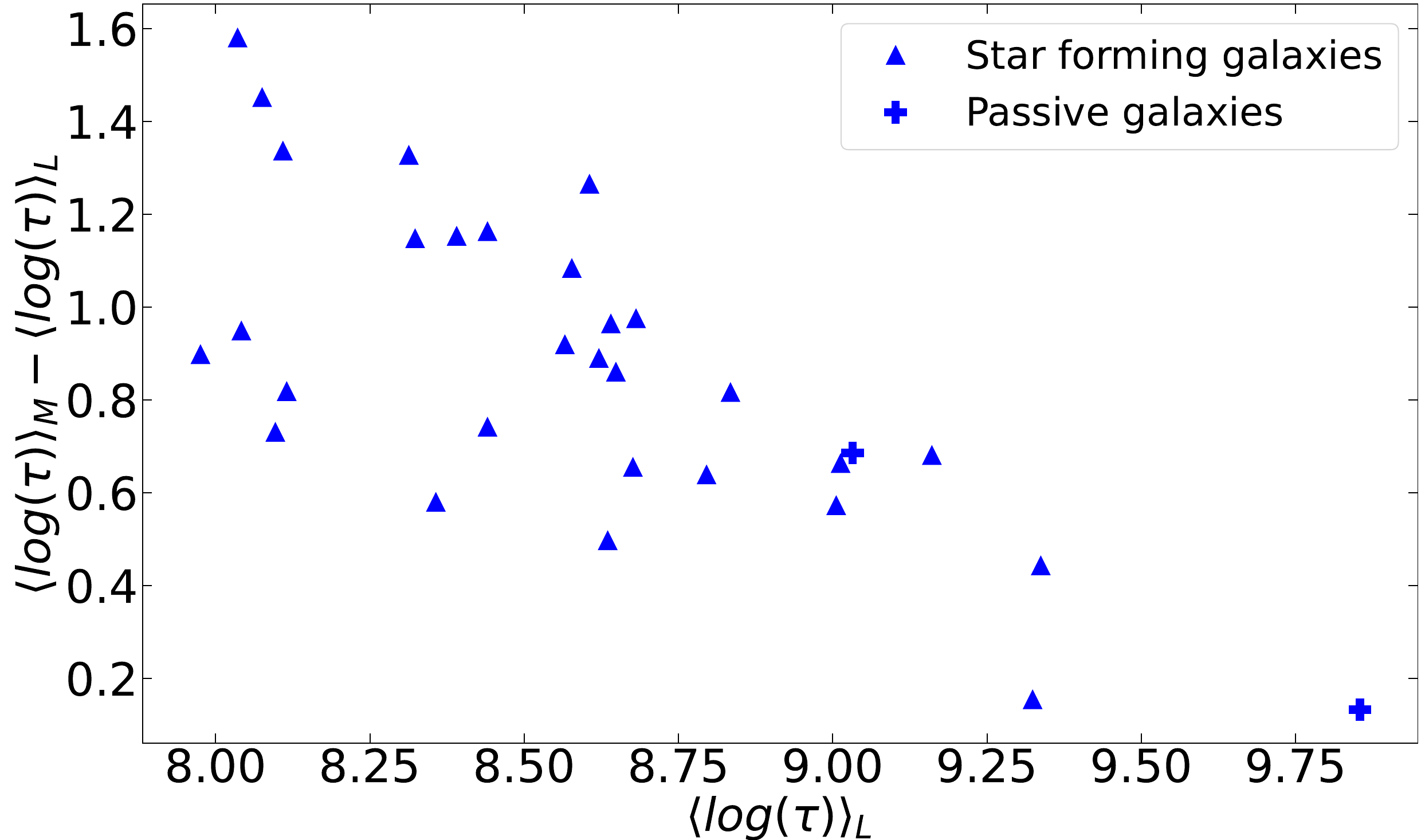}
\includegraphics[width=0.495\textwidth]{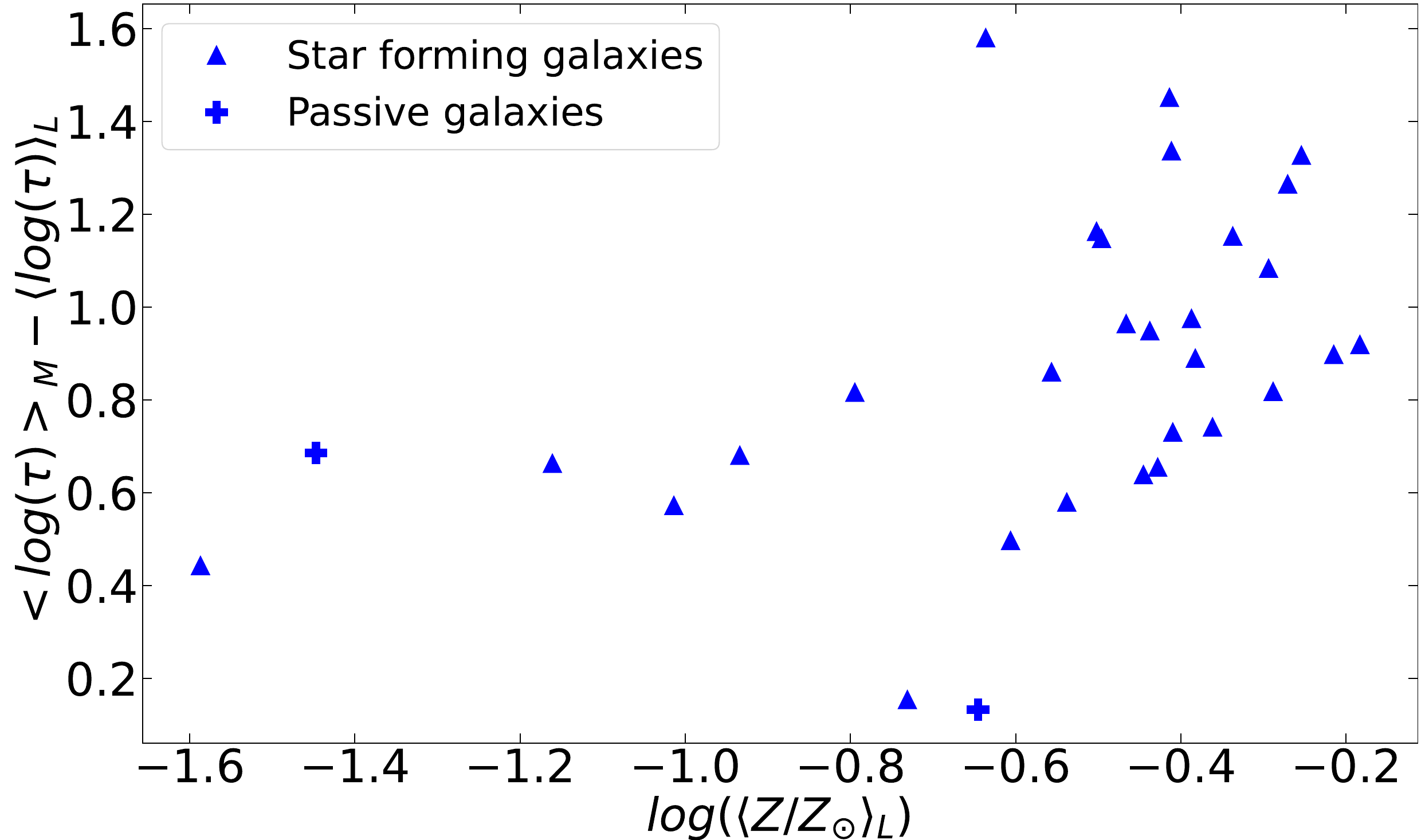}
\caption{Relation of the $\log\left<\tau\right>_{L}-\log\left<\tau\right>_{M}$ {\sl vs } a) the mean stellar age weighted by light, as $\log\left<\tau\right>_{L}$, and the mean stellar metallicity as $\log\left<Z\right>_{L}$, the star forming galaxies are represented by blue triangles and passive galaxies as blue crosses.}
\label{fig:ageM-ageL_ratio}
\end{center}
\end{figure*} 

Finally, we have made a direct comparison of our results of the 31 analysed dwarf galaxies and the results obtained by \citet[][hereinafter CD22]{Cano-Diaz+2022} for the same sample. 

In Fig.~\ref{fig:Mass_metallicity_age_Mandala}, panel a) we show the comparison of the results of the analysis within 1\,$R_e$ from CD22 and this work of the mean stellar metallicity weighted by light {\sl vs} the current stellar mass. We also show the binned results using bins of 0.4\,dex. Our models have a slightly higher $\left<Z\right>_{L}$ than CD22 for the same mass bin, but both models agree within error bars, except for the bin of the highest mass where our models have clearly more metallicity $\left<Z\right>_{L}$ than CD22. In general, our models have an increasing $\left<Z\right>_{L}$ with increasing mass, whereas CD22 have no trend of $\left<Z\right>_{L}$.

Similarly, in panel b), we show the relation between the mean age weighted by light and the stellar mass and the binned results using the same stellar mass bins as in panel a). We find younger ages $\left<\log (\tau)\right>_{L}$ than CD22 for the highest mass and older ages $\left<\log (\tau)\right>_{L}$ than CD22 for the low stellar masses. Our analysis shows a small trend of lower mean stellar age weighted by light $\left<\log (\tau)\right>_{L}$ for higher masses. On the contrary, CD22 does not find any trend of age $\left<\log (\tau)\right>_{L}$ with the mass.

Finally, in Fig.~\ref{fig:age_metallicity} we show the age metallicity relation of these galaxies. In our results, there seem to be two distinct stellar populations: one population predominantly young that has high mean stellar metallicity and another older stellar population with lower mean stellar metallicity. In the case of CD22, they seem to have one cloud with similar properties of stellar populations for all galaxies.

This suggests that these new models could be useful to disentangle the age-metallicity degeneracy and thus obtain more accurate estimates of the stellar population parameters. However, we have too little spectra of galaxies to have robust statistics and to have a solid conclusion, and more observations are necessary to have any final conclusion.

These two distinct populations that we find in the age-metallicity relation have very different star formation histories. On one hand, the population of old age and low metallicities has star formation histories with one or two star formation bursts some few Gyr ago, and a stopped star formation after that, so their star formation is now very low, the old and low metallicity stellar population being the only population at present. However, the young-age and high-metallicity population group had a burst of intermediate age, followed by a recent burst, now showing a high SFR.  

In order to show this difference in the SFH, we use the difference between the averaged by light and averaged by mass ages: $\left<\log(\tau)\right>_{M}$ and $\left<\log(\tau)\right>_{L}$. $\left<\log(\tau)\right>_{L}$ is dominated by the youngest stellar populations, while $\left<\log(\tau)\right>_{M}$ is dominated by the oldest stellar populations. Thus, the difference can be used as a tracer of how dispersed in time are the star formation bursts, if there have been recent star formation events, the difference between both quantities will be important, but if the star formation stopped long ago the difference will be insignificant. In Fig.~\ref{fig:ageM-ageL_ratio}, we show a) the $\left<\log(\tau)\right>_{L}$ {\it vs} $\left<\log(\tau)\right>_{M}-\left<\log(\tau)\right>_{L}$ and b) the $\log(\left<Z\right>_{L})$ {\it vs} $\left<\log(\tau)\right>_{M}-\left<\log(\tau)\right>_{L}$, it can be observed that low metallicity and old age population  from Fig.\ref{fig:age_metallicity} tend to have small values $\left<\log(\tau)\right>_{M}-\left<\log(\tau)\right>_{L}$, so they have a very low fraction of young stellar population. By implying, that way, that the SFR of the galaxies with just the old stellar population dropped heavily after the low metallicity population was formed.

\citet{Sextl+2023} also found that the proportion of young to old is higher in dwarf galaxies and depends strongly on the SFR. Moreover, they also found that the metallicity of the young stellar population was significantly higher than the one of old stellar population in low mass systems in comparison with high stellar mass galaxies. This can explain the diversity on the properties of the stellar populations of the analysed dwarf galaxies. 

There are differences in the mean stellar metallicity weighted by light and stellar mass relation, and mean stellar age weighted by light and the stellar mass relation of CD22 and this work. There could be several causes behind these differences: the different SSP models used, the different full spectral fitting used codes and also the analysis method. 
First, the SSP models of CD22 and ours are different. CD22 used a version of the Bruzual and Charlot code (Charlot, in prep) that included the MaStar \citep{Yan+2019} stellar templates for cold stars and theoretical spectra \citep{Leitherer+2010} for hot stars, and they used the IMF of \citet{Salpeter1955}. We have showed in ~\ref{subsec:other_SSP} that our models follow similar trends to other SSP models that use empirical libraries and very similar isochrones to our models, such as, XSL and EMILES, except in the reddest part of the NIR, $\lambda>9000$, but this region is excluded of the analysis. Thus, there should not be any systematic effect in our models.

Another possible cause of this difference is the full spectral fitting code used. We have used FADO, whereas CD22 used pyPipe3D. These two codes are very different and explore the parameter space in a very different way: FADO uses a genetic algorithm to make a random walk through the parameter space more efficiently, and pyPipe3D uses "non-lineal models" to fit the extinction, the redshift and the velocity dispersion and then analyses the parameter space. However, a full comparison of the effect of using a different full-spectral fitting code in the analysis of the stellar populations is beyond the scope of this paper.

Finally, another possible cause of the difference is the analysis method. On one hand, we stacked the spectra for 1\,$R_e$, and then we analysed the resulting spectra to obtain the stellar properties at 1\,$R_e$. On the other hand, CD22 created a map of the stellar population properties of the galaxy, by stacking the result of the map to obtain those properties. These two methods are not equivalent and given the low S/N of these systems some possible differences could arise when comparing their results.

\section{Conclusions}
\label{Sec:Conclusions} 

We have made a new set of {\sc HR-pyPopStar} models with a wider range of metallicities, basically reaching lower metallicities than in our first version of the models. Using the stellar libraries from citet{mun05}, \citet{Husser+2013} and the new version of PoWR models for O and B stars, a set of high wavelength-resolution theoretical spectral energy distributions for SSPs from 2500\,\AA\ to 10\,500\,\AA\ with a constant wavelength step of $\delta\lambda = 0.1$\,\AA\ has been computed. We have used isochrones from the Padova group \citep{Bressan_Bertelli_Chiosi1993, Fagotto+1994a, Fagotto+1994b,Girardi+1996}, for 106 different ages in the range of $t=0.1$\,Myr to 15\,Gyr, six metallicities $Z = 0.0001$, 0.0004 0.004, 0.008, 0.02 and 0.05, and four IMFs from \citet{Salpeter1955}, \citet{Ferrini_Penco_Palla_1990}, \citet[][with exponent $-2.7$ for the massive star range]{Kroupa2002} and \citet{Chabrier2003}. 

We have checked the effect of using these libraries by comparing the SEDs obtained by MI21 and the ones obtained in these new models for solar metallicity and CHA IMF. We found some differences, specially in the young ages, due to the update of the O and B stars from PoWR, and the old ages populations, due to the use of the {\sc PHOENIX} stellar spectra. Regarding the old population, the differences are specially in the NIR and in molecular bands, such as, G-band and TiO bands, which are attributed to the use of this {\sc PHOENIX} library for cool stars (in MI21 we used C14). The {\sc PHOENIX} library has a more complete list of molecular lines has spherical geometry instead of plane-parallel geometry, which reproduces better this kind of stars. Moreover, the coverage in temperature of the HR diagram is more complete in the {\sc PHOENIX} library.

The magnitudes remain similar in both old and new models, with average differences smaller than 0.06\,dex, the highest difference being 0.12\,dex for the youngest ages, where we have updated the models for O and B stars. We have also checked the change in the D4000 and the Dn4000 for $Z=0.02$ and CHA IMF, finding small differences in D4000 (the maximum difference is a 2\%) and in Dn4000 (the maximum difference is a 3.5\%), that are caused by the different modelling of the CN bands in C14 and PHOENIX libraries.

We have also compared our resulting SEDs with E-MILES, MaStar, CBC20 and XSL. For E-MILES and XSL we choose the Padova00 isochrones that are very similar to the isochrones that we use. All models agree well in the optical, but each SSP model is different in the NIR. XSL, E-MILES and our models agree on the continuum level, while each one has its own strength for the TiO molecular bands.

In order to test low metallicity HR SSP models, we have used our new models to analyse different set of stellar populations data. First, we have analysed high-resolution data of the globular cluster M~15 using data from HR-I of MEGARA. We have made an analysis of $\chi^2$ to obtain the best estimate of the age and the metallicity. We have found an older age and a lower metallicity than the previous results we found in \citet{Millan-Irigoyen+21}. We have found an age of $\tau = 12.8$\,Gyr and the metallicity $Z=0.0004$. These results are similar to other works from the literature that studied the same object using other techniques as this one from \citet{OMalley17}, who found $\tau = 12.5 \pm 1.3$\,Gyr and $[Fe/H] = -2.33$. 

Finally, we have analysed the stellar populations of dwarf galaxies from the MaNGA sample up to redshift 0.15, where the new low metallicity models are expected to contribute significantly to their stellar content. After obtaining the spectra at $1\,R_e$, we have made a selection of the spectra based on the signal-to-noise ratio of key areas for the stellar population analysis, $H_{\alpha}$, $H_{\beta}$, $H_{\gamma}$, MgI b triplet region and the Calcium triplet (CaT). After this cut, we have obtained 31 galaxies, 29 star-forming and 2 passive, that have been analysed using the full spectral fitting code {\sc FADO} using the new models and the CHA IMF. We have run the analysis 10 times to avoid false minimum solutions and made the mean of all the results.   

We compared the current stellar masses and mean stellar metallicity, weighted by light, with other studies of the literature. Our results follow trends similar to other work that analysed star-forming galaxies \citep{Kudritzki+2016,Zahid+2017,Sextl+2023} and have slightly higher mean stellar metallicity weighted by light, but are within error bars of those studies \citep{Gallazzi+2005,Sextl+2023}. We have also made a direct comparison of our results with those obtained by \citet{Cano-Diaz+2022} for the same galaxies. In the case of the current stellar mass - mean stellar metallicity relation, our $\left<Z\right>_L$ is higher than CD22 for the same mass bin, but both models agree within error bars, except for the highest mass bin.

In the case of the current stellar mass - mean stellar age relation, our results in the lowest mass bin have higher $\left<\log (\tau)\right>_L$ than CD22 and have smaller $\left<\log (\tau)\right>_L$ than CD22 for higher stellar masses. We find a small trend of lower mean stellar age weighted by light for higher masses, while CD22 found no trend with stellar mass.

 Then, we compared the age-metallicity relation, we found two different stellar populations, one with low $\left<\log (\tau)\right>_L$ and high $\left<Z\right>_L$ and a second one with high $\left<\log (\tau)\right>_L$ and low $\left<Z\right>_L$. These two populations seem to have completely different star formation histories: the low $\left<\log (\tau)\right>_L$ and high $\left<Z\right>_L$ galaxies had a burst 100-500\, Myr ago followed by a very recent burst and have a high present day SF; while the high $\left<\log (\tau)\right>_L$ - low $\left<Z\right>_L$ galaxies had one or two bursts 3-4\,Gyr ago and their star formation rate is now very low, so the majority of the population is old age and low metallicity. However, the size of the sample is too small to have any definitive conclusion and it would be necessary to have a more complete and high S/N spectra sample of dwarf galaxies. 
 
Finally, we evaluated the possible causes of the different results between CD22 and our work. We have identified three possible causes: the difference in SSP models, the full-spectral fitting used code, and the method to analyze the data.

\section*{Data availability statement}

Tables 4 and 5 are only available in electronic form at the CDS via anonymous ftp to cdsarc.u-strasbg.fr (130.79.128.5) or via http://cdsweb.u-strasbg.fr/cgi-bin/qcat?J/A+A/. The computed SEDs and the tables with the magnitudes, are available on the web page: \url{http://www.pypopstar.com}. The user are available to download the complete set of models or a required subset (with all ages) just selecting by IMF and $Z$.

\begin{acknowledgements}

\\
I.M.-I. wants to thank Guinevere Kauffmann for the helpful comments and discussion that helped to improve the paper.
The authors would like to thank the anonymous referee who provided useful and detailed comments on an earlier version of the manuscript.
\\
This research has been funded by grant PID2019-107408GB-C41 and PID2022-136598NB-C33 funded by MCIN/AEI/10.13039/501100011033 and by “ERDF A way of making Europe”
\\
Funding for the Sloan Digital Sky Survey IV has been provided by the Alfred P. Sloan Foundation, the U.S. Department of Energy Office of Science, and the Participating Institutions. SDSS acknowledges support and resources from the Center for High-Performance Computing at the University of Utah. The SDSS web site is www.sdss4.org.
\\
SDSS is managed by the Astrophysical Research Consortium for the Participating Institutions of the SDSS Collaboration including the Brazilian Participation Group, the Carnegie Institution for Science, Carnegie Mellon University, Center for Astrophysics | Harvard \& Smithsonian (CfA), the Chilean Participation Group, the French Participation Group, Instituto de Astrofísica de Canarias, The Johns Hopkins University, Kavli Institute for the Physics and Mathematics of the Universe (IPMU) / University of Tokyo, the Korean Participation Group, Lawrence Berkeley National Laboratory, Leibniz Institut für Astrophysik Potsdam (AIP), Max-Planck-Institut für Astronomie (MPIA Heidelberg), Max-Planck-Institut für Astrophysik (MPA Garching), Max-Planck-Institut für Extraterrestrische Physik (MPE), National Astronomical Observatories of China, New Mexico State University, New York University, University of Notre Dame, Observatório Nacional / MCTI, The Ohio State University, Pennsylvania State University, Shanghai Astronomical Observatory, United Kingdom Participation Group, Universidad Nacional Autónoma de México, University of Arizona, University of Colorado Boulder, University of Oxford, University of Portsmouth, University of Utah, University of Virginia, University of Washington, University of Wisconsin, Vanderbilt University, and Yale University.

\end{acknowledgements}

\begin{appendix}
\label{appendix}
\section{Comparison of spectra and properties at $\rm Z=0.004$}

\begin{figure*}
    \centering    
\includegraphics[width=0.49\textwidth]{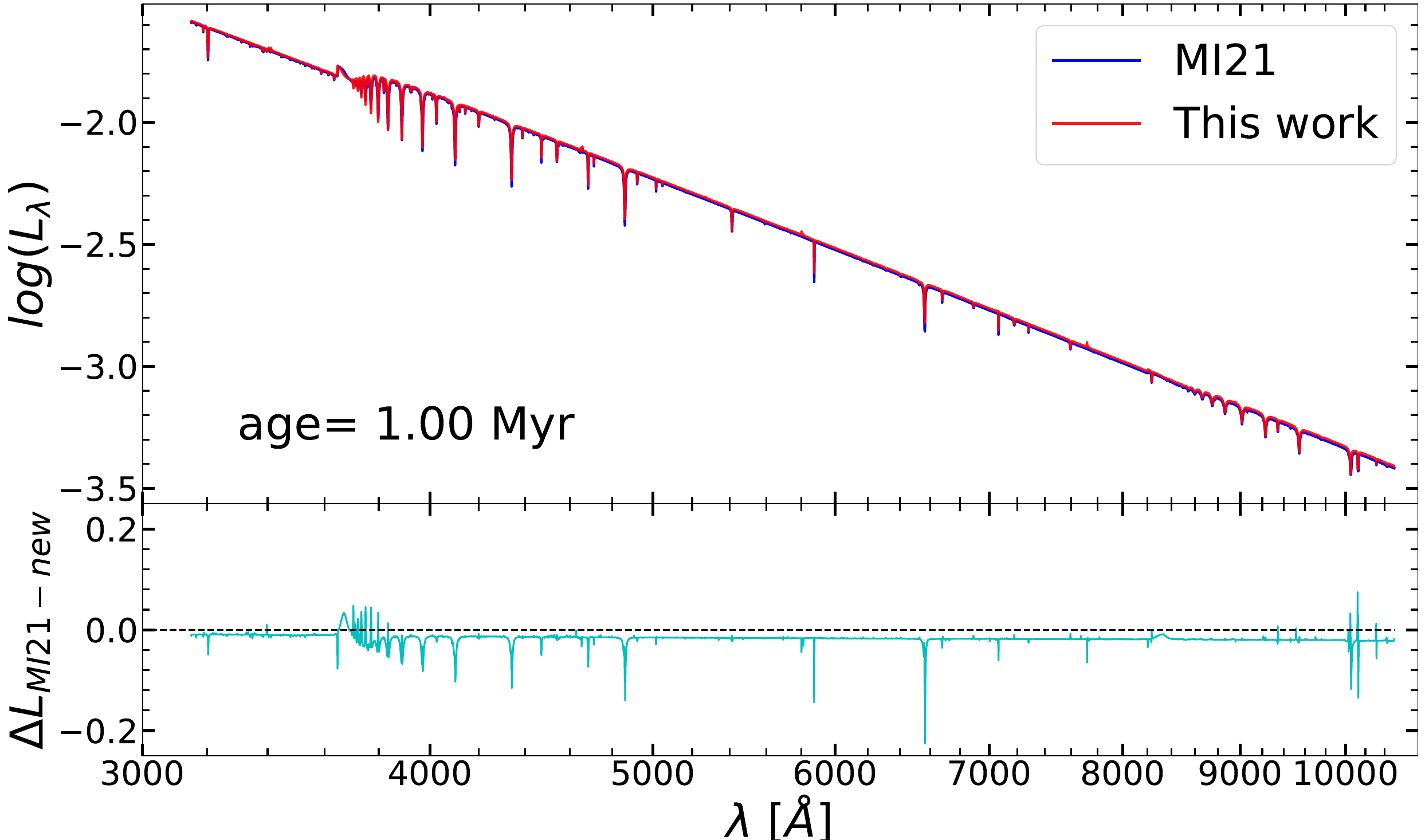}
\includegraphics[width=0.49\textwidth]{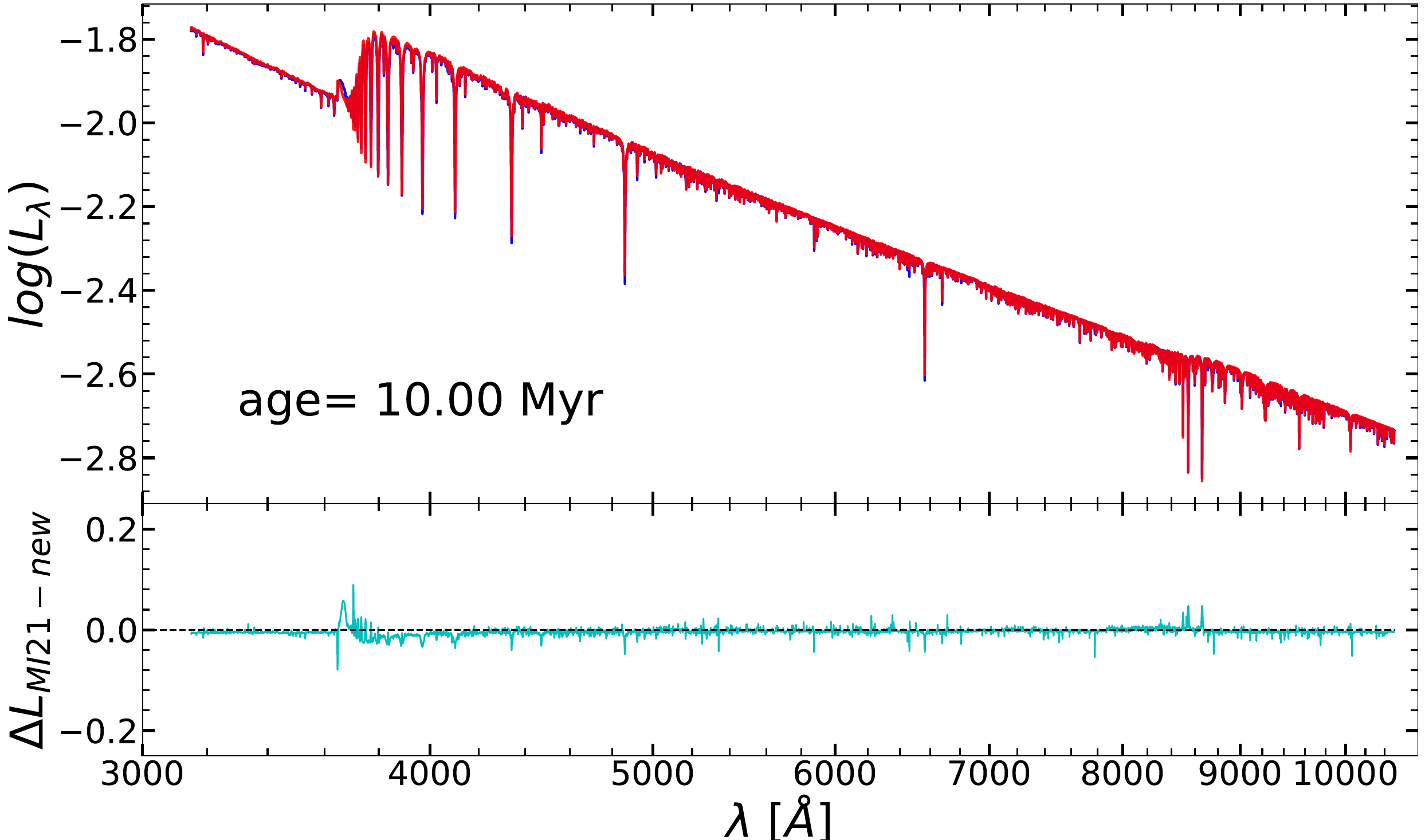}
\includegraphics[width=0.49\textwidth]{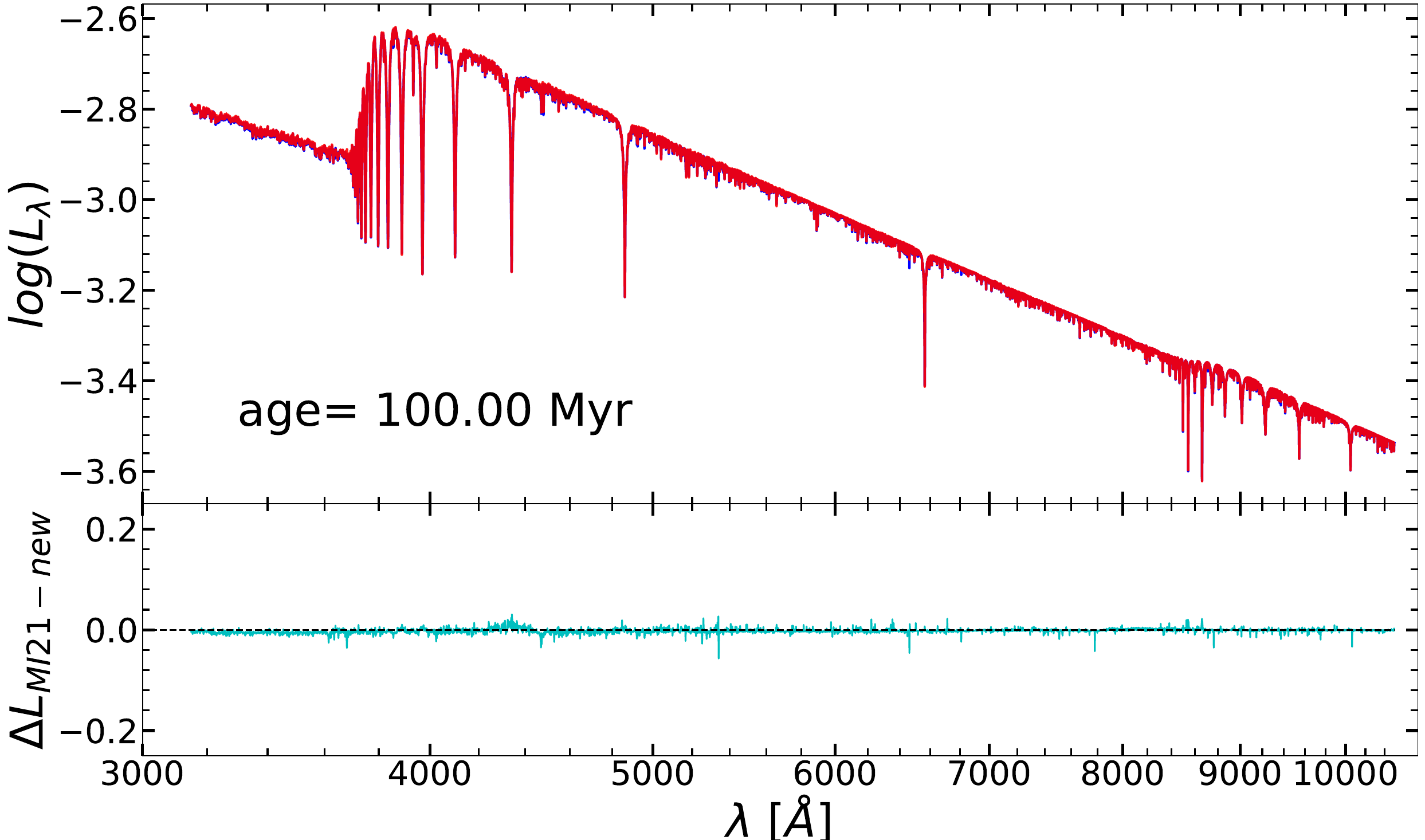}
\includegraphics[width=0.49\textwidth]{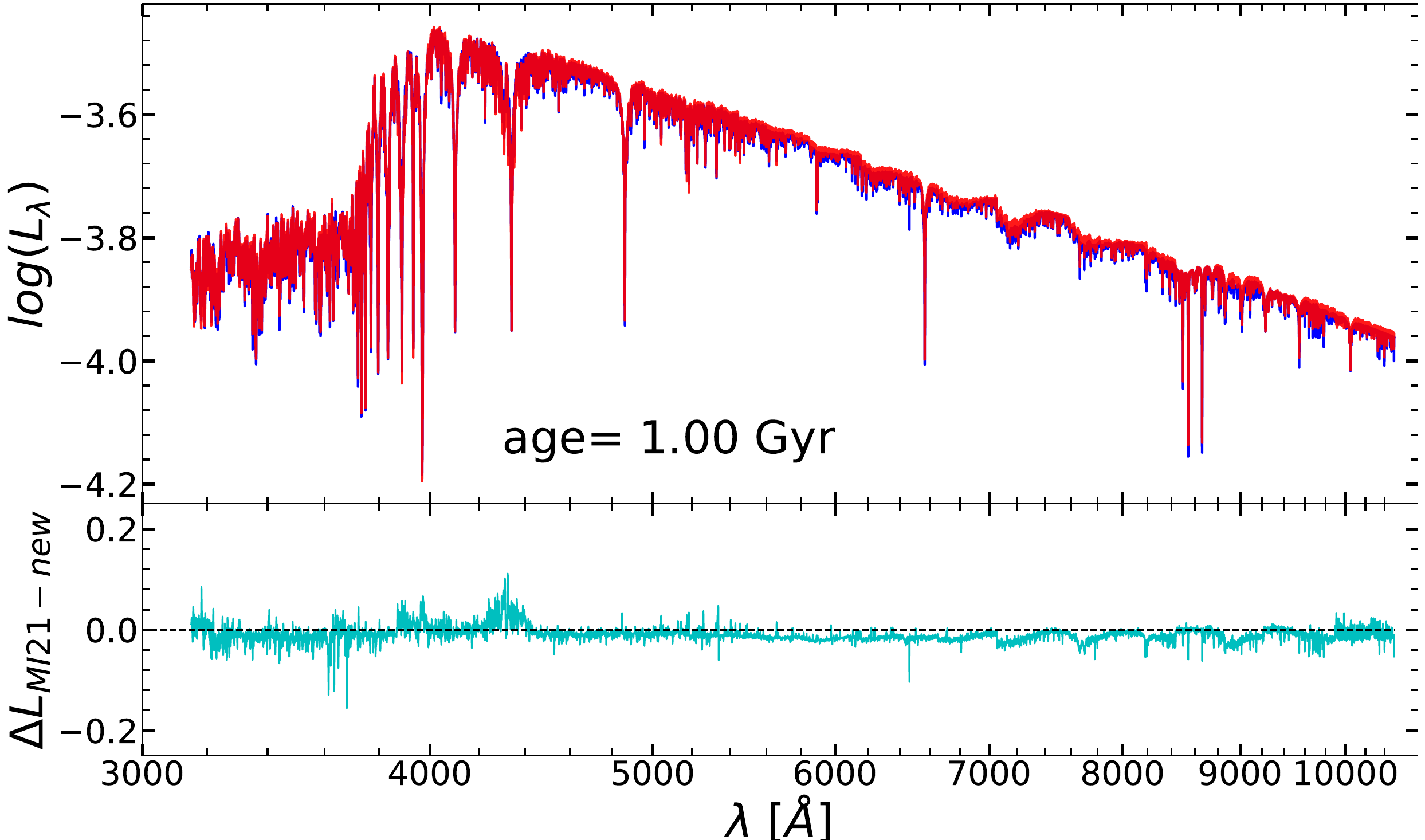}
\includegraphics[width=0.49\textwidth]{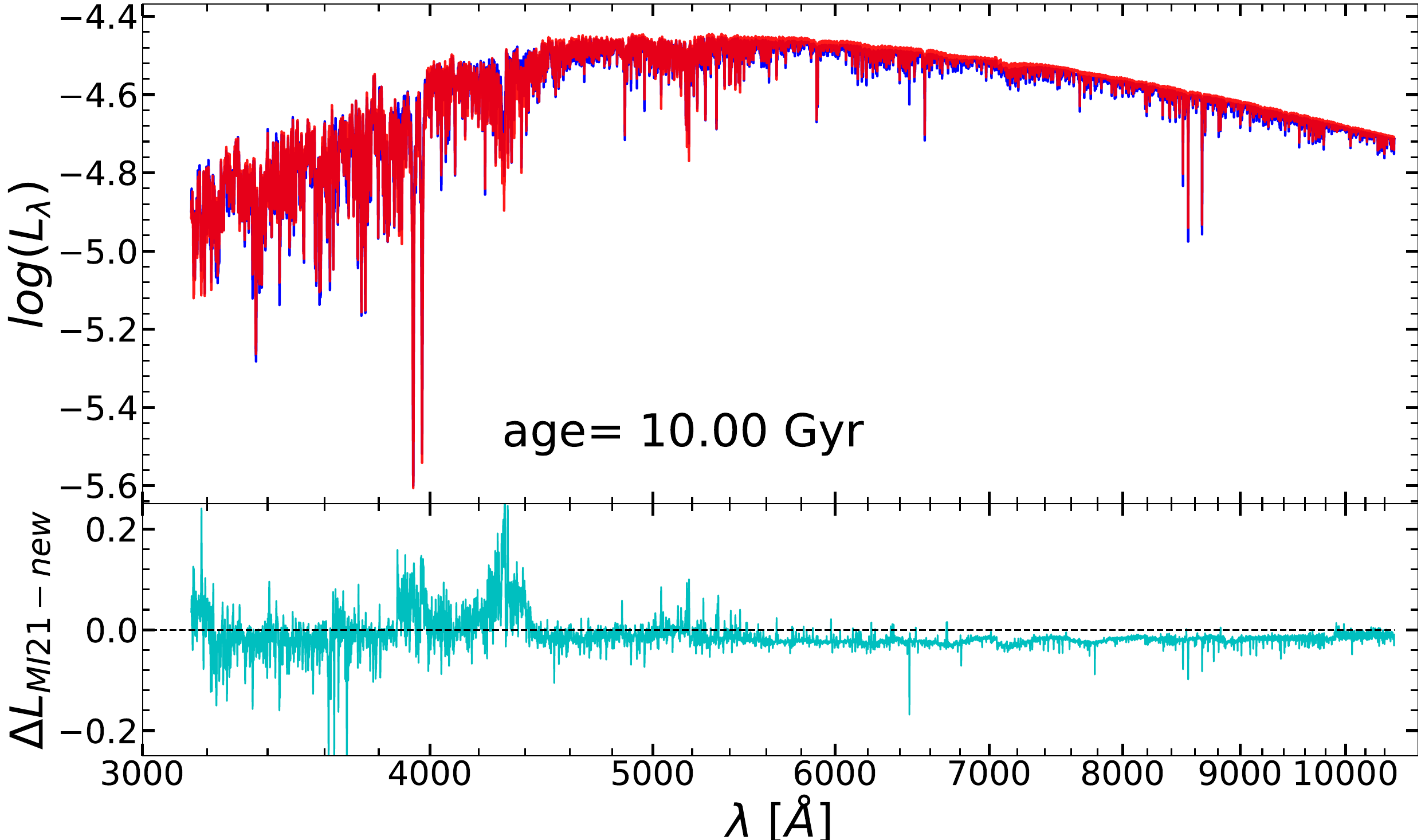}
\caption{Comparison between the old version of {\sc HR-pyPopStar} using the stellar library of C14 (MI21), in blue line, and the present models using MUN05 PHOENIX+ new PoWR stellar library, in red line, for $Z=0.004$ and CHA IMF. The bottom of part of every subfigure show the $\Delta L_{MI21-new}= \frac{L_{MI21}-L_{new}}{L_{MI21}}$.}
\label{fig:Munari_Coelho_Z004}
\end{figure*}

\subsection{Comparison of spectra }\label{subsec_app:}

We compared the spectra of MI21 and the new models for CHA IMF, $\rm Z=0.004$ and the ages 1, 10, 100, 1000 and 10000\,Myr in Fig. ~\ref{fig:Munari_Coelho_Z004}. 

In the case of the young ages 1 and 10 Myr, the new models have $1.5\%$ and $0.5\%$ more luminosity than the MI21 and have higher absorption in the lines of the Balmer series. As in $Z=0.02$, these changes are caused by the new models of the O and B stars from PoWR. For 100 Myr, the new models and MI21  are very similar. For 100 Myr the models are very similar with very small differences in some lines. In the case of 1 Gyr, the new models have stronger flux in the TiO and G-Band molecular bands. Finally, for the age of 10 Gyr, our new models have less flux in the G-Band than the previous models.  

The causes for the differences are the same as in Z=0.02 the change in the stellar libraries, specially the change of C14 to PHOENIX for cool stars and the update of the PoWR O and B stars. In general, the differences found at Z=0.02 are maintained at Z=0.004, but these differences are smaller at Z=0.004 than at Z=0.02.

\subsection{D4000 breaks}\label{subsec_app:D4000}

 We have made a comparison of the D4000 and Dn4000 for the lowest metallicity that both models share, $Z=0.004$. In the case of the D4000, the new results are 0.5\% smaller than the ones in MI21 for all ages. In the case of Dn4000, the new models have smaller Dn4000 than MI21 for young ages, but they have higher Dn4000s than MI21 for ages than 4 Gyr with a maximum difference of 2.8\%. The general trends of both D4000 and Dn4000 breaks observed for $Z=0.02$ are maintained for $Z=0.004$.

\begin{figure*}
\hspace{-0.5cm}
\begin{center}
\includegraphics[width=0.495\textwidth]{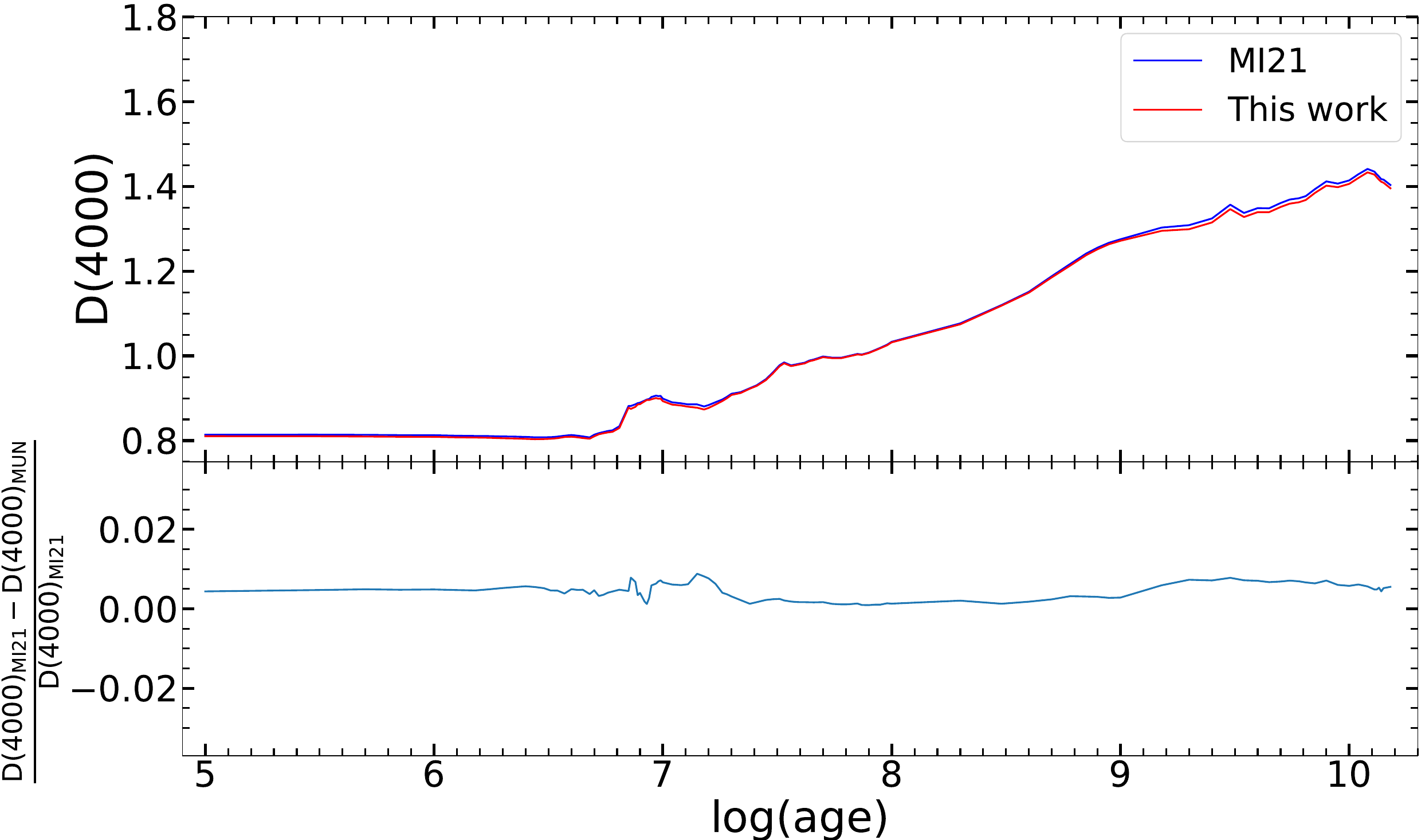}
\includegraphics[width=0.495\textwidth]{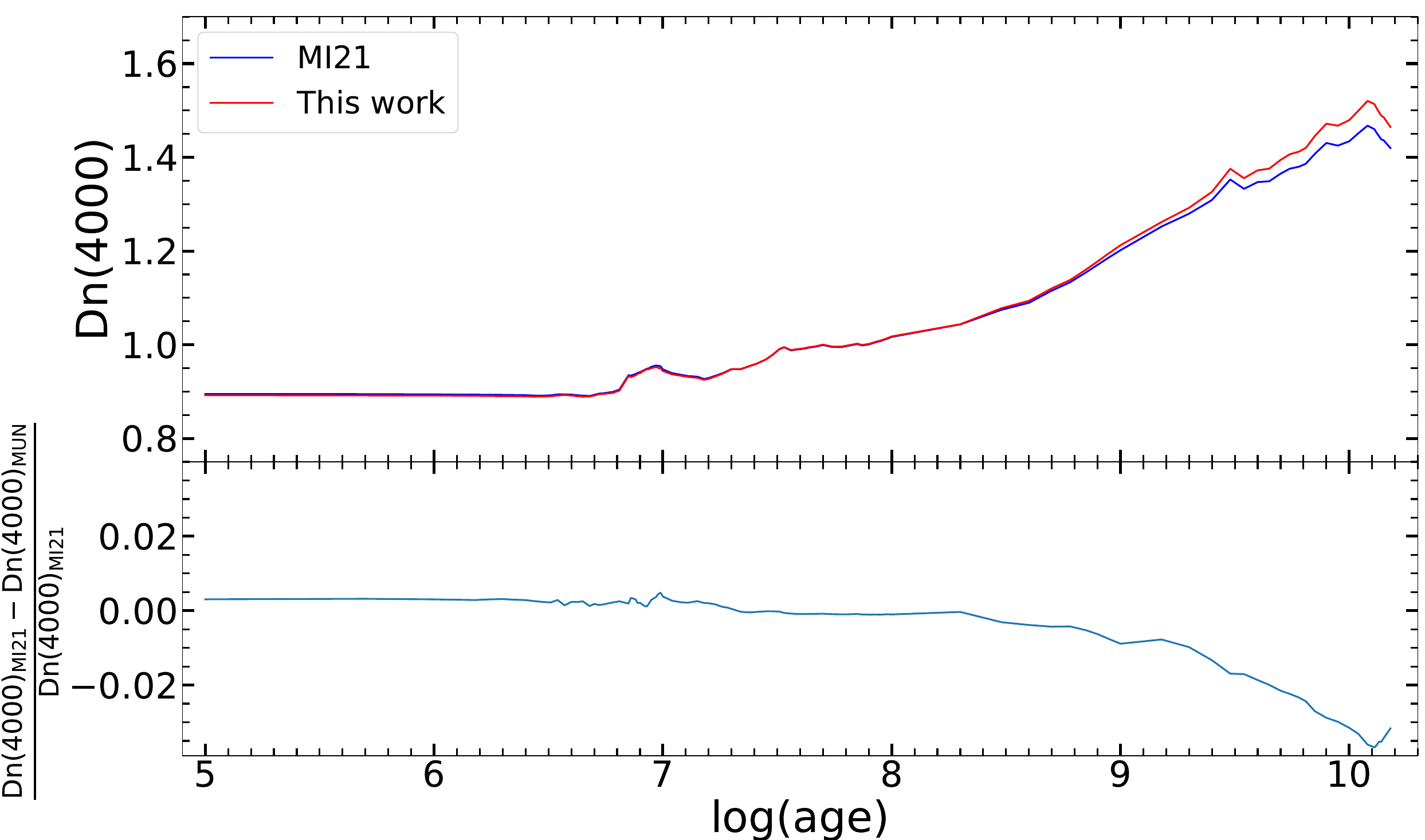}
\caption{D4000 and Dn4000 break time evolution comparison for between MI21 in blue line and this work red line for $Z=0.004$.}
\label{Fig:D4000_Z004}
\end{center}
\end{figure*}

\subsection{Calcium triplet}\label{subsec_app:CaT}

We have made a comparison of the calcium triplet (CaT) for the lowest metallicity that both models share Z = 0.004. The differences between the MI21 and the new models  for the CaT are significantly smaller than the solar metallicity case, but they still have differences. For young ages 0.1 Myr to 1 Myr the new models have smaller CaT than the MI21 models with approximately $4\%$. For 1 Myr to 500 Myr (log(age)=8.7), the new models have higher CaT than the old ones with a maximum peak of $\sim 20\%$ difference at 9 Myr. In the case of ages older than 500 Myr, the new models have smaller CaT than the ones from the MI21 models with a maximum difference of $\sim 7\%$ for the oldest age.

\begin{figure*}
\hspace{-0.5cm}
\begin{center}
\includegraphics[width=0.495\textwidth]{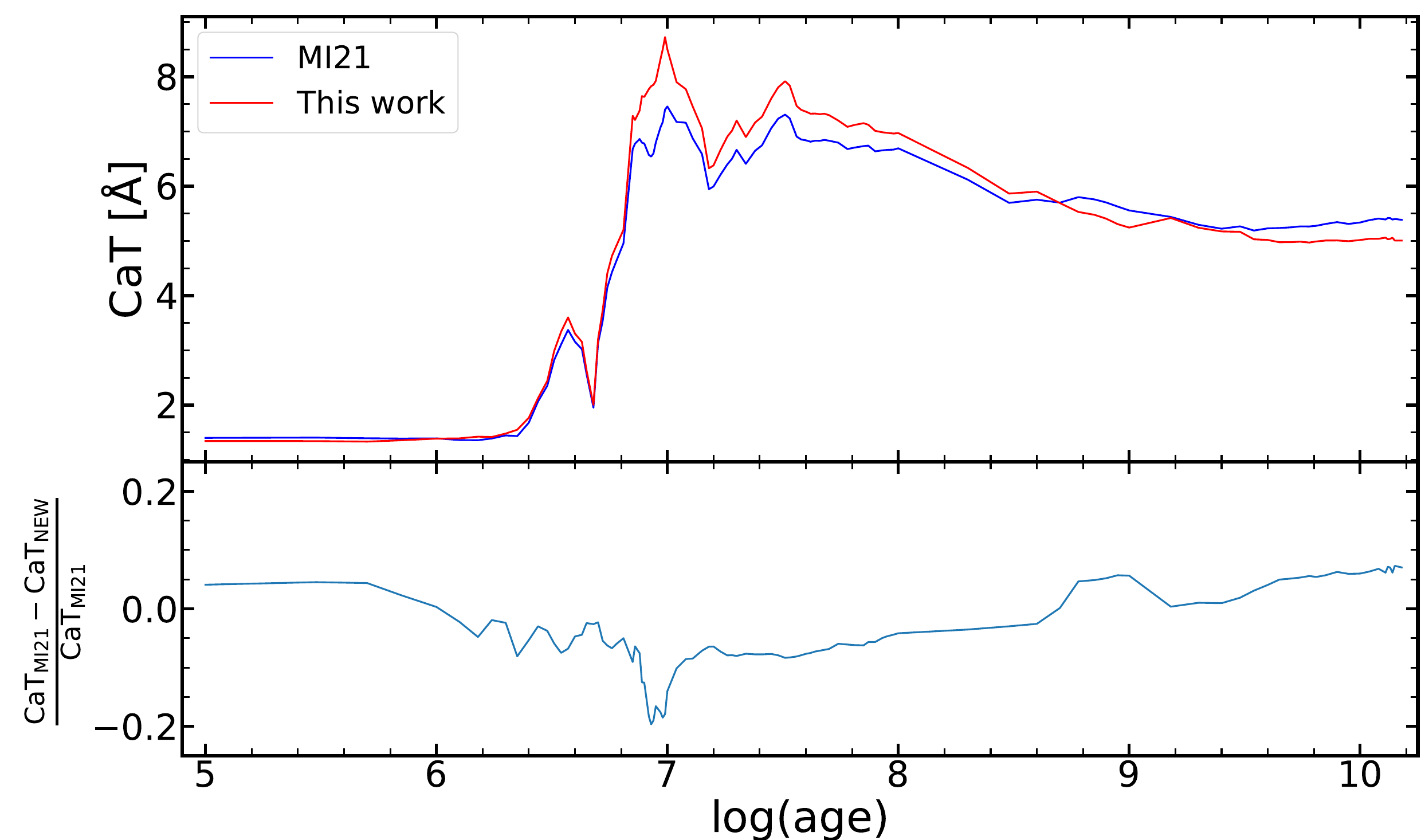}
\caption{Calcium triplet(CaT) break time evolution comparison for between MI21 in blue line and this work red line for $Z=0.004$.}
\label{Fig:CaT_Z004}
\end{center}
\end{figure*}

\subsection{Magnitudes}\label{subsec_app:magnitudes}

\begin{figure*}
\hspace{-0.5cm}
\begin{center}
\includegraphics[width=0.33\textwidth]{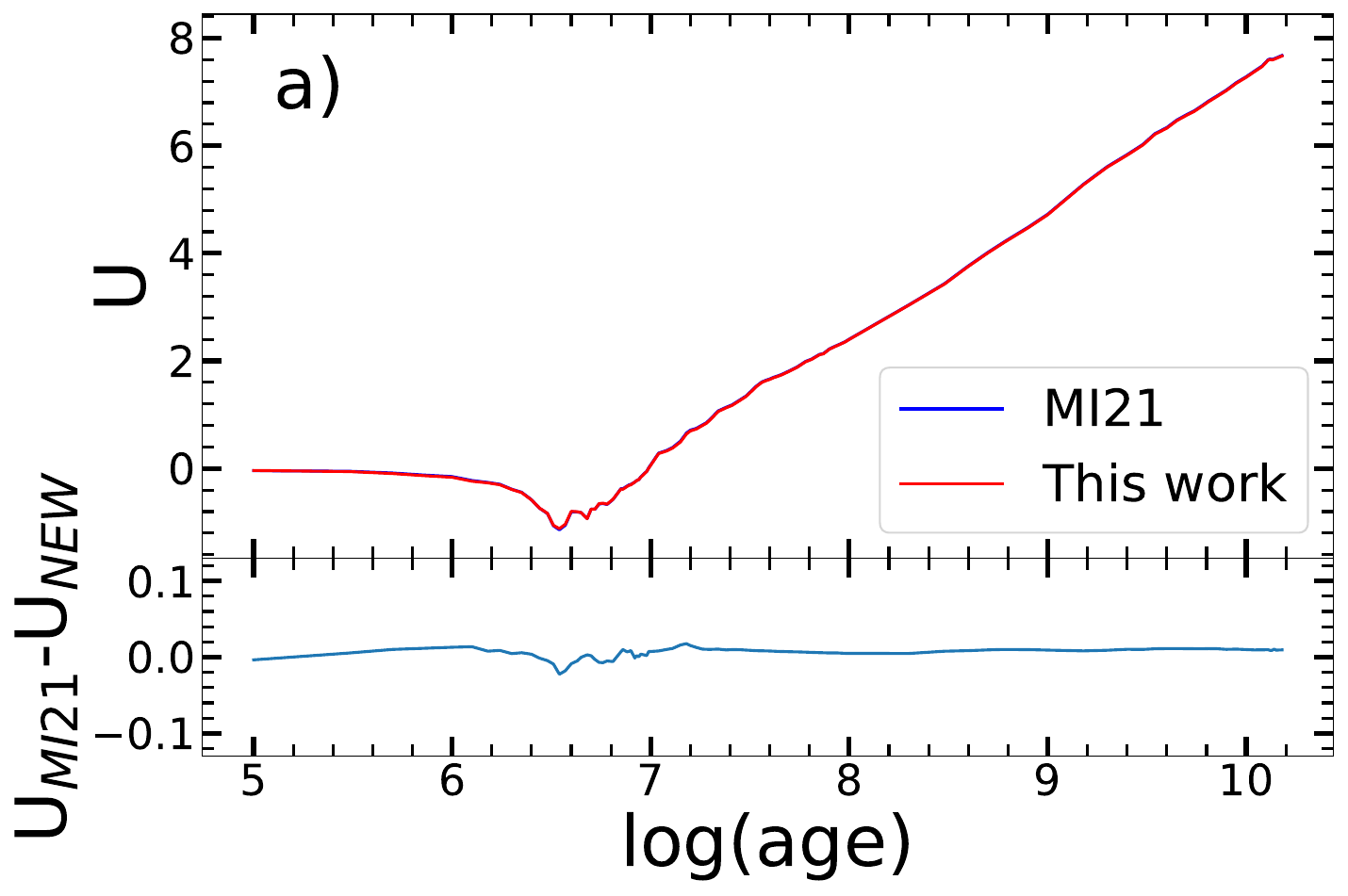}
\includegraphics[width=0.33\textwidth]{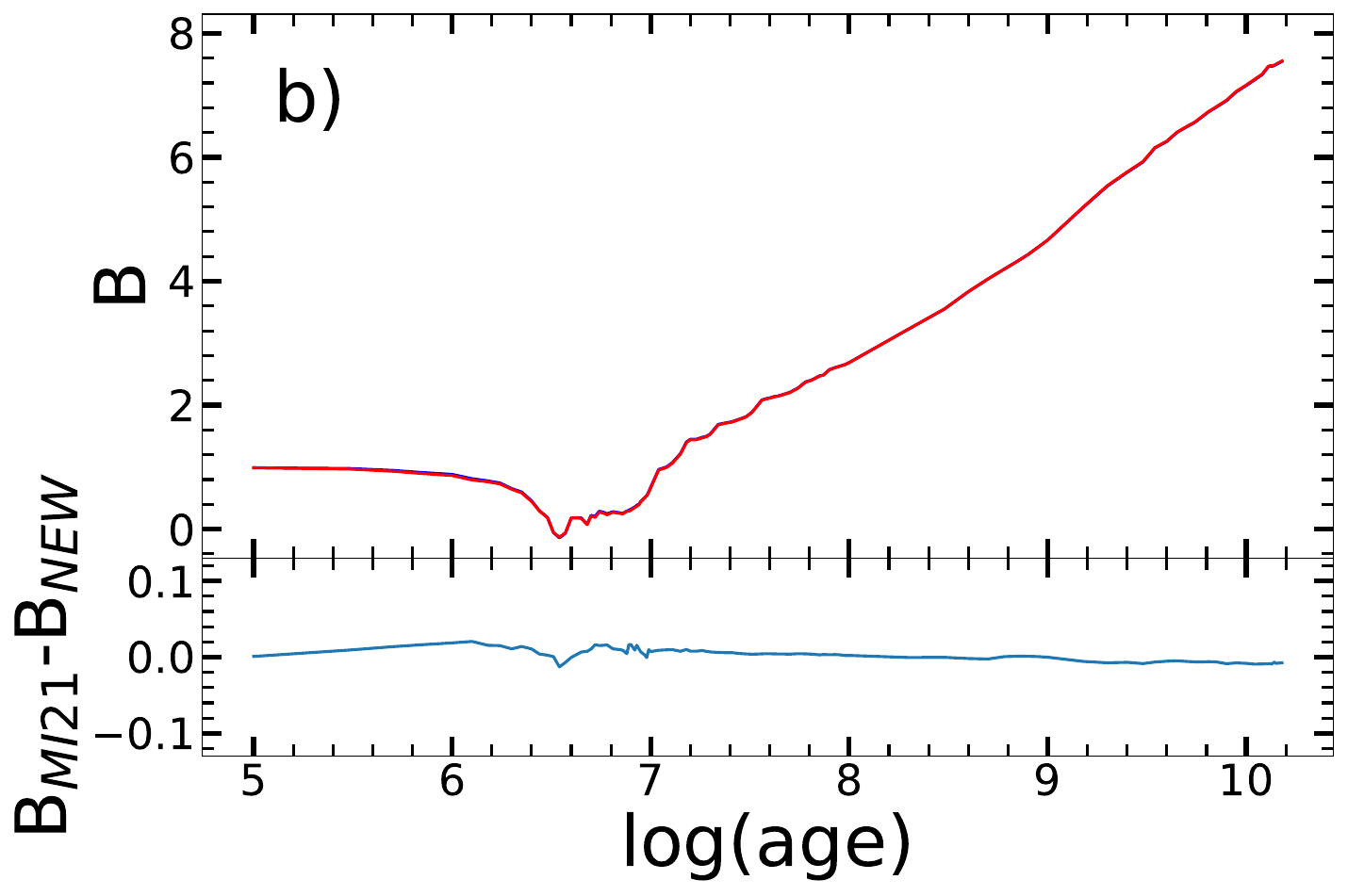}
\includegraphics[width=0.33\textwidth]{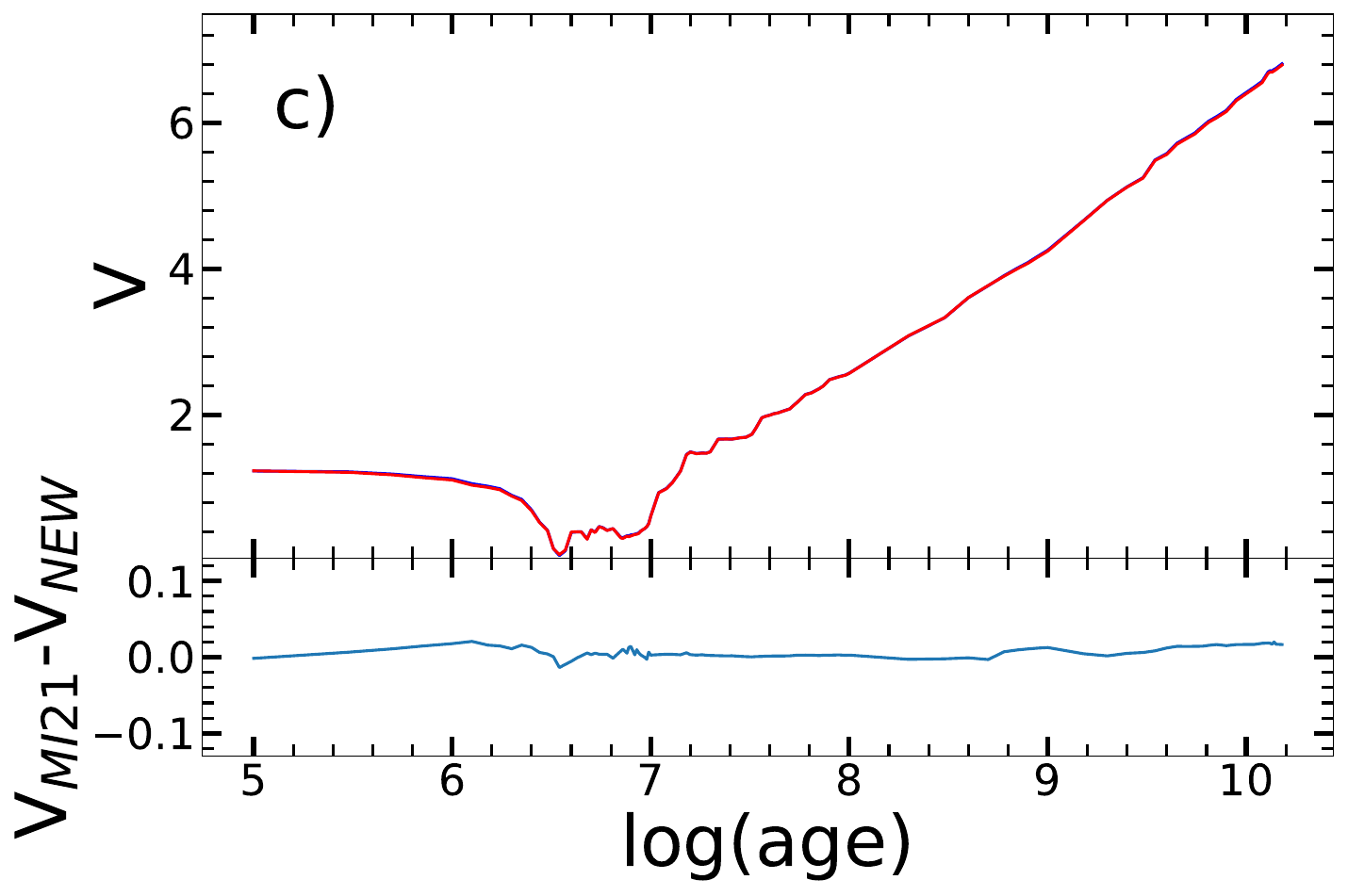}
\includegraphics[width=0.33\textwidth]{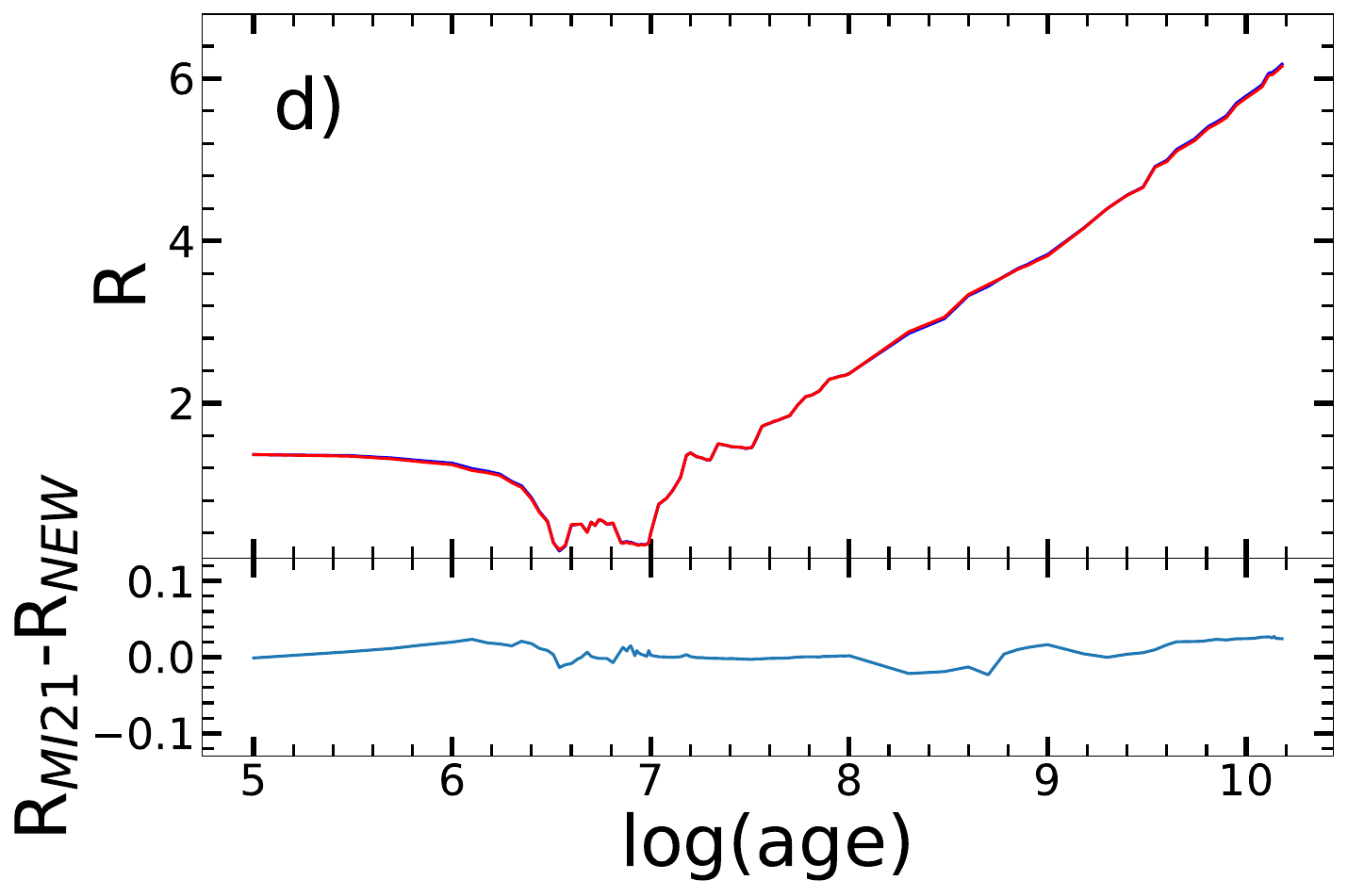}
\includegraphics[width=0.33\textwidth]{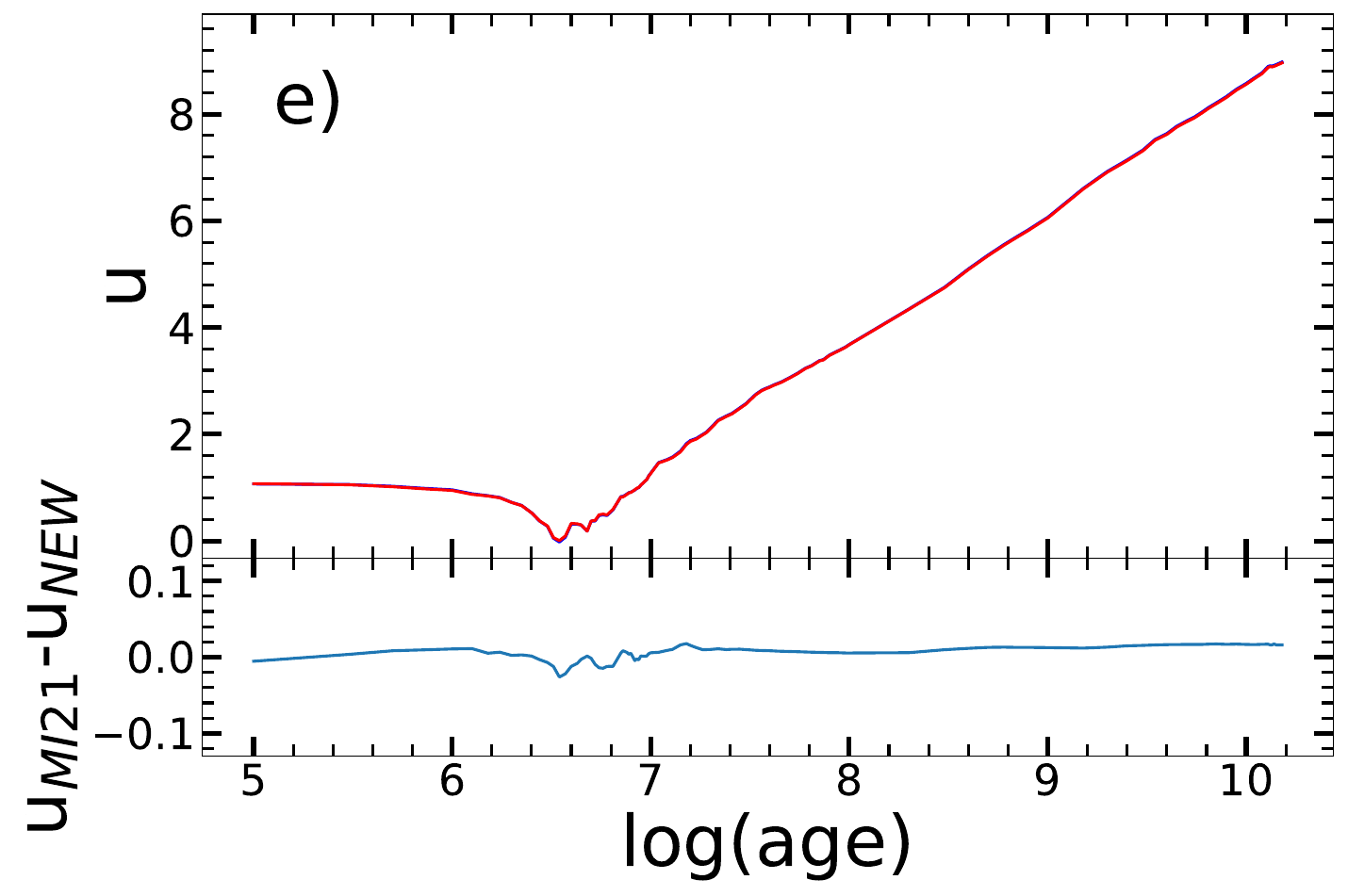}
\includegraphics[width=0.33\textwidth]{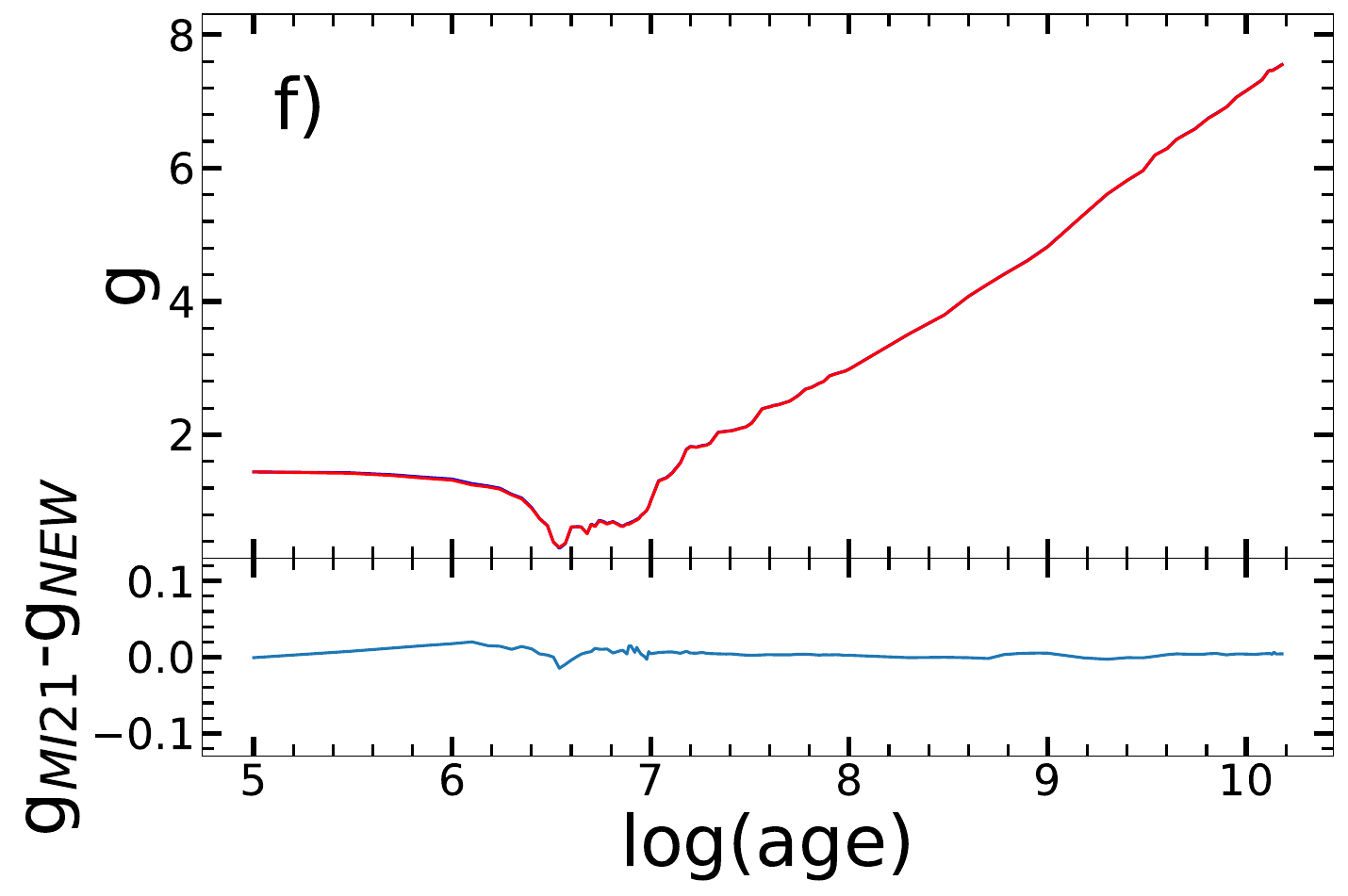}
\includegraphics[width=0.33\textwidth]{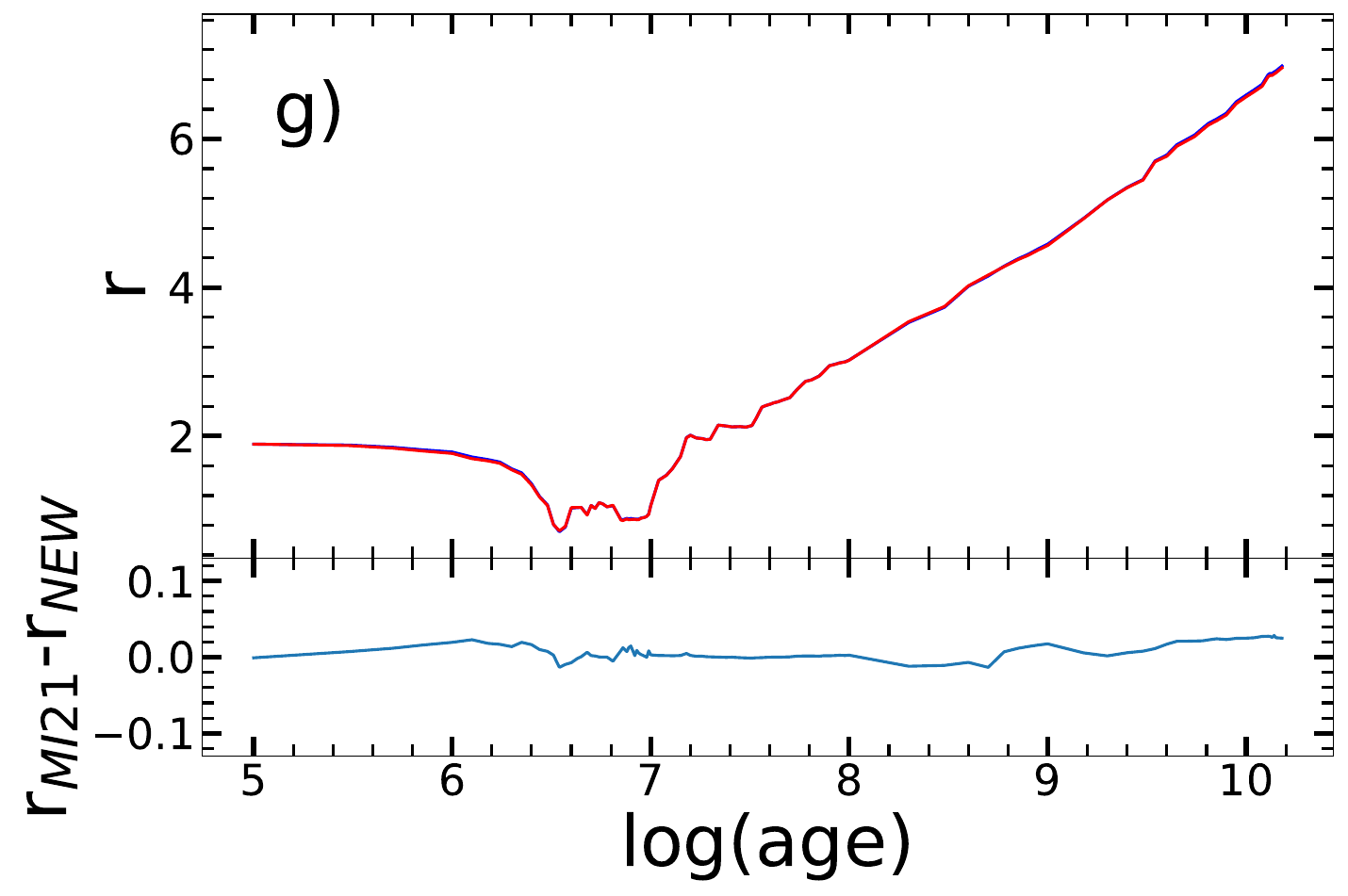}
\includegraphics[width=0.33\textwidth]{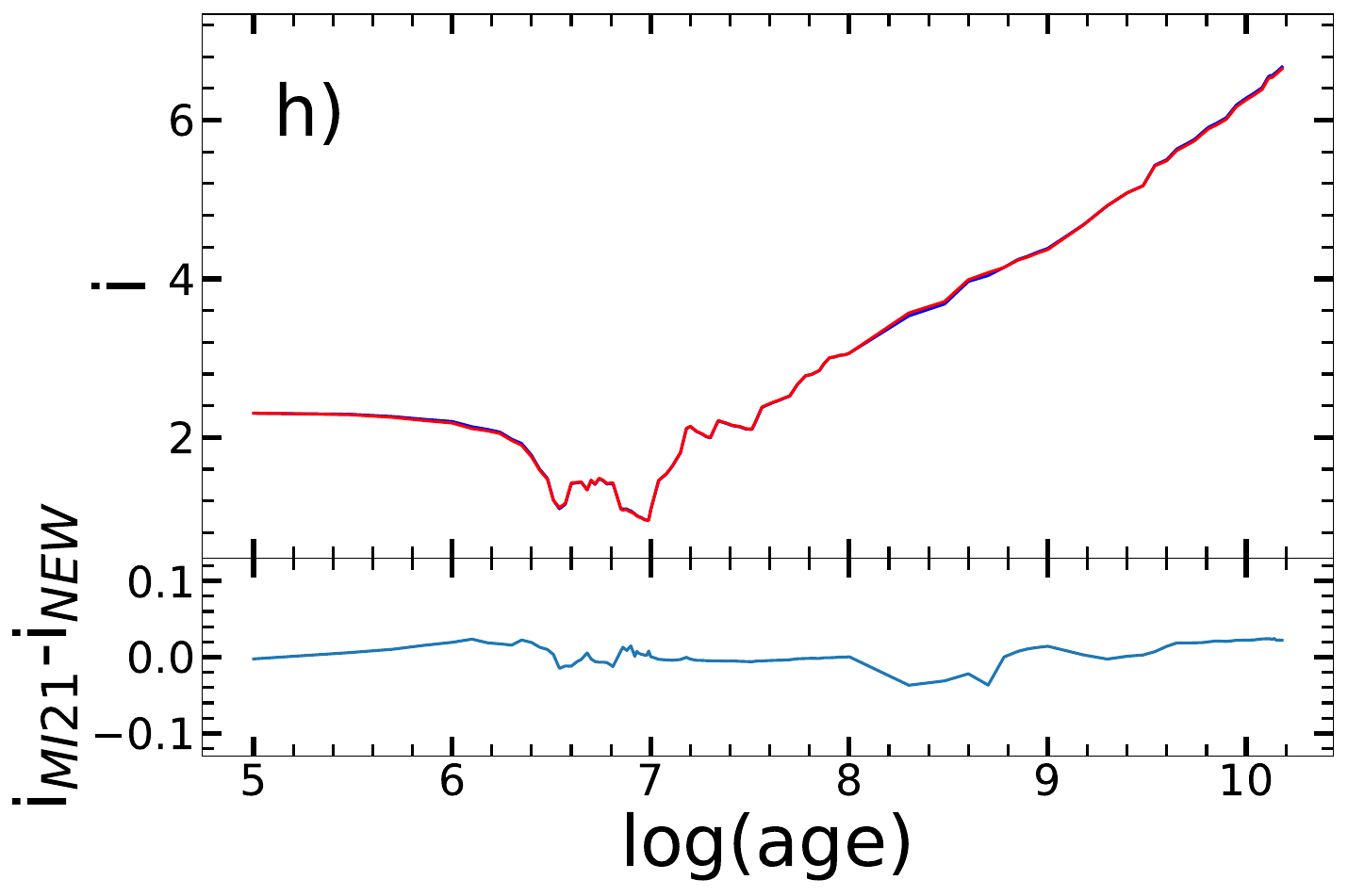}
\includegraphics[width=0.33\textwidth]{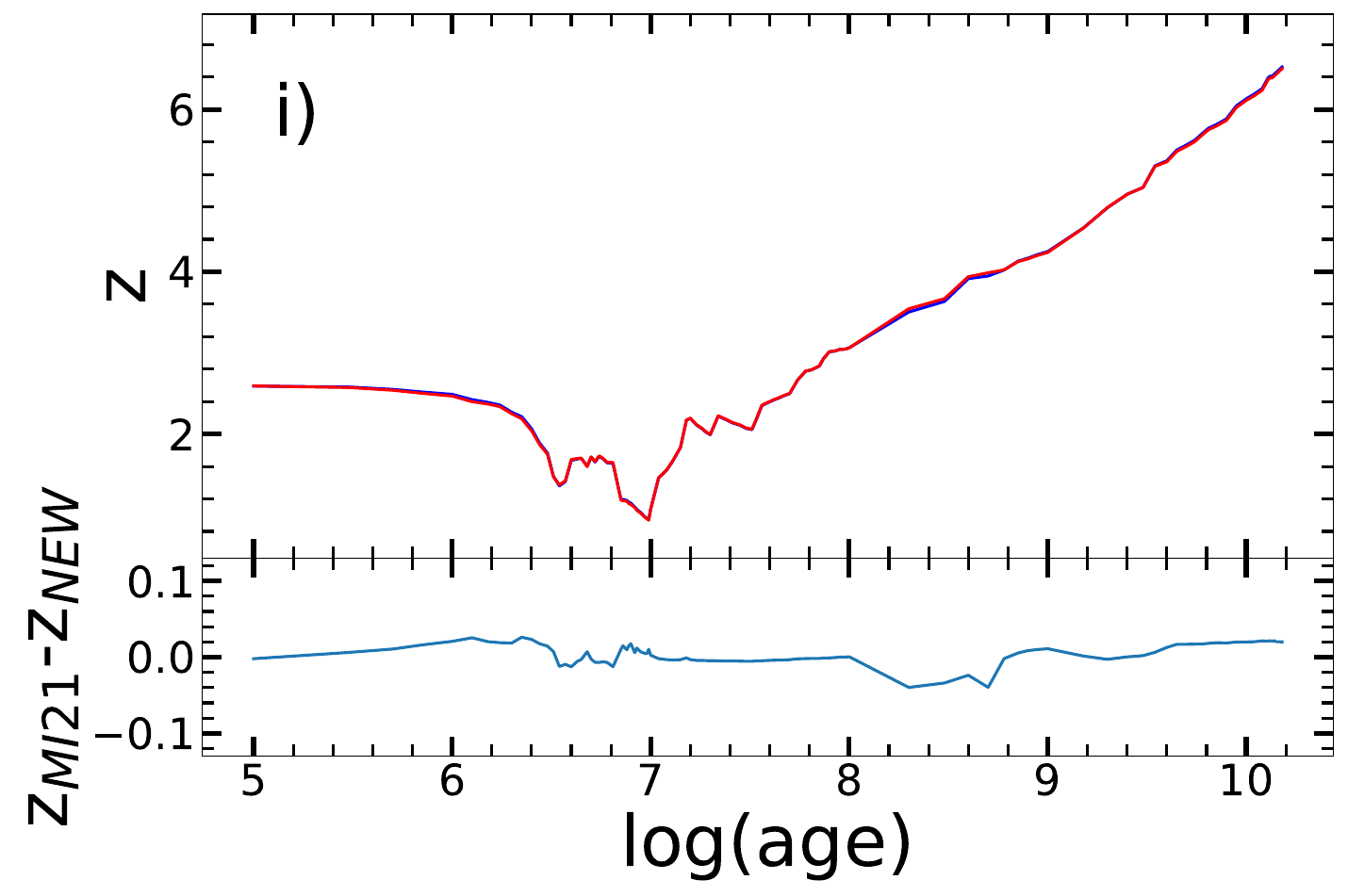}
\caption{Evolution of magnitudes  a) U, b) B, c) V, d) R, e) u, f) g, g) r, h) i, and i) z with age: comparison  of results between MI21, blue, and this work, red, with residuals in the bottom panel for CHA IMF and $Z=0.004$.
}

\label{Fig:magnitude_Z004}
\end{center}

\end{figure*}

We compared the magnitudes between MI21 and this work in the bands: U, B, V and R, in the Johnson system and u, g, r, i and z in the SDSS system. The evolution of the 9 magnitudes is shown in Fig. ~\ref{Fig:magnitude_Z004}.

In general, the differences between MI21 and the new models at Z=0.004 is very small around 0.02 dex except for i and z magnitudes where the new models have 0.04 dex higher than the MI21 for the 200 Myr old population.

\end{appendix}
\end{document}